\documentclass{aastex}

\usepackage{natbib,graphicx,latexsym}
\usepackage{emulateapj5}

\slugcomment{Astrophysical Journal, in press}

\shortauthors{Liu, Charlot \& Graham}
\shorttitle{SBFs and Stellar Populations of Ellipticals}

\newcommand{\HST}{{\sl HST}}
\newcommand{\SIRTF}{{\sl SIRTF}}
\newcommand{\eg}{e.g.}
\newcommand{\ie}{i.e.}

\newcommand{\etal}{et al.}
\newcommand{\Msun}{\mbox{$M_{\odot}$}}
\newcommand{\Zsun}{\mbox{$Z_{\odot}$}}
\newcommand{\Ho}{\mbox{$H_0$}}
\newcommand{\kms}{\hbox{km~s$^{-1}$}}

\newcommand{\Ic}{\mbox{${I_c}$}}
\newcommand{\Rc}{\mbox{${R_c}$}}
\newcommand{\Ks}{\mbox{$K_s$}}
\newcommand{\Kp}{\mbox{$K^{\prime}$}}
\newcommand{\Lp}{\mbox{$L^{\prime}$}}

\newcommand{\Bbar}{\mbox{$\overline{B}$}}
\newcommand{\Vbar}{\mbox{$\overline{V}$}}
\newcommand{\Rbar}{\mbox{$\overline{R_c}$}}
\newcommand{\Ibar}{\mbox{$\overline{I_c}$}}
\newcommand{\Ihstbar}{\mbox{$\overline{F814W}$}}

\newcommand{\Jbar}{\mbox{$\overline{J}$}}
\newcommand{\Hbar}{\mbox{$\overline{H}$}}

\newcommand{\Kbar}{\mbox{$\overline{K}$}}
\newcommand{\Kpbar}{\mbox{$\overline{K^{\prime}}$}}

\newcommand{\Lbar}{\mbox{$\overline{L}$}}
\newcommand{\Mbar}{\mbox{$\overline{M}$}}
\newcommand{\mbar}{\mbox{$\overline{m}$}}

\def\wig#1{\mathrel{\hbox{\hbox to 0pt{%
  \lower.5ex\hbox{$\sim$}\hss}\raise.4ex\hbox{$#1$}}}}

\begin{document}

\title{Theoretical Predictions for Surface Brightness Fluctuations\\
and Implications for Stellar Populations of Elliptical Galaxies}

\author{\sc Michael C. Liu}
\affil{Department of Astronomy, University of California, Berkeley, CA 94720}
\email{mliu@astro.berkeley.edu}

\author{\sc St\'ephane Charlot\altaffilmark{1}} 
\affil{Institut d'Astrophysique de Paris, CNRS, 98 bis boulevard
Arago, 75014 Paris, France}
\altaffiltext{1}{Also Max-Planck Institut f\"ur Astrophysik,
Karl-Schwarzschild-Strasse 1, 85748, Garching, Germany}

\author{\sc James R. Graham}
\affil{Department of Astronomy, University of California, Berkeley, CA 94720}

\begin{abstract}

We compute theoretical predictions for surface brightness fluctuations
(SBFs) of single-burst stellar populations (SSPs) using models optimized
for this purpose.  We present results over a wide range of ages (from 1
to 17~Gyr) and metallicities (from 1/200 to 2.5~times solar) and for a
comprehensive set of ground-based and space-based optical and infrared
bandpasses. Our models agree well with existing SBF observations of
Milky Way globular clusters and elliptical galaxies.

Our results provide refined theoretical calibrations and $k$-corrections
that are needed to use SBFs as standard candles.  We suggest that SBF
distance measurements can be improved by (1) using a filter around
1~\micron\ to minimize the influence of stellar population variations,
and (2) using the integrated $V-K$ galaxy color instead of $V-\Ic$ to
calibrate $I$-band SBF distances.

We show that available SBF observations set useful constraints on
current population synthesis models, and we suggest SBF-based tests for
future models.  The existing SBF data favor particular choices of
stellar evolutionary tracks and spectral libraries among the several
choices allowed by comparisons based on only the integrated properties
of galaxies.  Also, the tightness of the empirical $I$-band SBF
calibration as a function of $V-I_c$ galaxy color is a useful
constraint. It suggests that the model uncertainties in the lifetimes of
the post-main sequence evolutionary phases are probably less than
$\pm$50\% and that the initial mass function in elliptical galaxies is
probably not much steeper than that in the solar neighborhood.

Finally, we analyze the potential of SBFs for probing unresolved stellar
populations in elliptical galaxies. Since SBFs depend on the second
moment of the stellar luminosity function, they are sensitive to the
brightest giant stars and provide complementary information to
commonly-used integrated light and spectra. In particular, we find that
optical/near-infrared SBFs are much more sensitive to the metallicity
than the age of a stellar population.  Therefore, in combination with
age-sensitive observables, SBF magnitudes and colors are a valuable
complement to metal-line indices to break the age/metallicity degeneracy
in elliptical galaxy studies.  Our preliminary results suggest that the
most luminous stellar populations of bright galaxies in nearby clusters
have roughly solar metallicites and about a factor of three spread in
age.

\end{abstract}

% Astrophysical Journal Subject Headings
%       Authors should select subject headings for their manuscripts using
%       the list given below.  No more than six headings should be given
%       (and if headings are used for individual stars, galaxies, etc.,
%       each object normally counts as one heading). The overall
%       categories (which are indented) should not be included. Also, do
%       not include any part of a heading that is in parentheses: thus,
%       (stars:) binaries (including multiple): close should be given as
%       binaries: close.
\keywords{cosmology: distance scale --- Galaxy: globular clusters ---
	galaxies: elliptical and lenticular, stellar content, formation
	--- stars: late-type, AGB and post-AGB}

\section{Introduction}

When observing the inner regions of a nearby elliptical galaxy or the
bulge of a nearby spiral galaxy, there are two noticeable
characteristics of the surface brightness structure of these spheroidal
stellar systems.  The first characteristic is that the galaxy is
brightest in the center with the surface brightness falling off
gradually with increasing radial distance.  The second characteristic is
only apparent in good seeing conditions: on small scales, the galaxy has
a clumpy appearance on the spatial scale of the seeing disk.  The
clumpiness arises from Poisson statistical variations in the number of
stars within each resolution element.  This effect can be easily
recognized by visual inspection of images of nearby galaxies like M~31
and M~32, in which the small-scale clumpiness can be a few percent of the
mean surface brightness.  Historically, this effect was called
``incipient resolution.''  In the modern context, it is known as surface
brightness fluctuations (SBFs).

\citet{1988AJ.....96..807T} devised a technique to quantify SBFs for use
as an extragalactic distance indicator for undisturbed early-type
galaxies. (See also the reviews by \citealp{1992PASP..104..599J} and
\citealp{1999phcc.conf..181B}.) This method relies on using the ratio of
the second moment to the first moment of the stellar luminosity function
(LF) of the galaxy as a standard candle:
\begin{equation}
\Lbar \equiv \frac{\sum_i n_i L_i^2}{\sum_i n_i L_i}
\label{Lbar}
\end {equation}
where $n_i$ is the number of stars of type $i$ and luminosity $L_i$.
The quantity \Lbar\ has units of luminosity and is referred to as \Mbar\
when represented as an absolute magnitude.  The apparent SBF magnitude
\mbar\ can be determined observationally, and if the distance to the
galaxy is known, \Mbar\ can also be determined.  Of course, as in the
case of ordinary photometry, SBF colors are distance-independent
(provided that the $k$-correction is negligible).  Using SBFs as a
distance indicator requires that (1) the bright end of the stellar LF in
elliptical galaxies and spiral bulges is universal, or (2) variations in
the LF from galaxy to galaxy can be measured and corrected so that
\Mbar\ remains a standard candle.

SBFs are an intrinsic property of a stellar population as a whole.
Therefore, in addition to their utility as a distance indicator, SBFs
offer much promise in adding to our knowledge of the stellar content of
elliptical galaxies.  In fact, the use of SBFs for stellar population
studies arguably preceded its use as a distance indicator:
\citet{1955AJ.....60..247B} used the "count-brightness ratio," the ratio
of the number of resolved stars to the integrated light, to study the
populations of M~31 and M~32.  Hence, the idea of using observations
near or at the limit of resolution to explore stellar populations has a
long history.

Furthermore, SBFs provide information about stellar populations {\em
unique} from ordinary integrated light, which is the first moment of the
stellar LF.  Since SBFs also depend on the second moment of the stellar
luminosity function, they are especially sensitive to the most luminous
stars in elliptical galaxies, the evolved cool giant stars.  Thus, SBFs
can put stronger constraints on the evolution of these stars than
integrated light alone. Both the interior structure and emergent
spectral energy distributions of cool giant stars are poorly understood,
especially near the tip of the red giant branch (RGB) and the asymptotic
giant branch (AGB) populated by low and intermediate mass
($\lesssim5-7\Msun$) stars.  Ideally, we would study these stars using
Local Group star clusters, which comprise populations of homogenous
composition and age. However, because cool giants evolve rapidly, only a
handful are present in any cluster; therefore, small number statistics
and stochastic fluctuations are undesireable factors
\citep[e.g.,][]{1997ApJ...479..764S}.  In this context, SBF analyses of
entire galaxies can complement star cluster studies, since galaxian
light arises from several orders of magnitude more stars. Also, while
nearby globular clusters are mostly metal-poor, the dominant stellar
populations in ellipticals are thought to be generally old and
metal-rich. There is a dearth of such systems in the Local Group --- the
best examples are the bulges of the Milky Way and M~31, but these may be
imperfect analogs. In order to study metal-rich stellar evolution, one
naturally turns to elliptical galaxies.

There are two basic motivations for modeling SBFs: (1) as pointed out by
\citet{1988AJ.....96..807T}, one can derive the calibration of SBF
absolute magnitudes purely from models provided that stellar populations
in galaxies can be modeled accurately, and alternatively, (2) one can
use the observations of SBF magnitudes and colors to test and improve
the models.

The first attempt at deriving a purely theoretical SBF zeropoint was
that of \citet{1990AJ....100.1416T}. They used stellar evolutionary
models from the Revised Yale Isochrones \citep[][hereinafter RYI]{gre87}
supplemented with simple prescriptions for the horizontal branch and
AGB; the resulting \Ibar\ zeropoint and especially its dependence on the
integrated $V-\Ic$ galaxy color disagreed significantly with
observations \citep{1991ApJ...373L...1T}.  This was due to the fact that
the RYI giant branches failed to turn over in the optical at high
metallicity, probably because of inaccurate bolometric corrections for
the coolest giant stars \citep{1992IAUS..149..181M,
1994ApJ...429..557A}.  A subsequent study was made by
\citet{1993ApJ...409..530W}, who computed SBF magnitudes using his own
population synthesis models \citep{1994ApJS...95..107W}.  For the
main-sequence and RGB stars, these models used an amalgamation of
isochrones from Vandenberg and collaborators with the RYI; post-RGB
evolution was added using the fuel consumption theorem, including
``schematic'' treatments of the HB as a single red clump and of AGB
evolution using a variety of theoretical prescriptions. The resulting
SBF predictions agreed well with the observed optical SBF colors (\Vbar,
\Rbar, and \Ibar) and with the empirical calibration of \Ibar\ versus
$V-\Ic$ \citep[see also][]{1997ApJ...475..399T}.
\citet{1993A&A...275..433B} also computed predictions for SBF magnitudes
which were consistent with the existing optical SBF data at the time,
though there were some uncertainties in transforming from the Johnson
filters used in his models to the Kron-Cousins ones used for the
observations. Since his models used older theoretical spectra
\citep{1978A&AS...34..229B}, Buzzoni had to extrapolate the spectra for
wavelengths longward of 1.08~\micron\ and also for stars with
$T_{eff}<4000$~K.  For both of these reasons, the
\citet{1993A&A...275..433B} models are expected to be less accurate
for SBF predictions in the IR. For example, their predictions are at
least several tenths of a magnitude fainter in the $K$-band than the
observations.

The major observational effort on SBFs has been focused on $I$-band
measurements for distance determinations.  As mentioned above,
comparisons of stellar population models with observed SBF magnitudes
and colors can also help us calibrate the colors (\ie, stellar spectral
energy distributions [SEDs]) and numbers (\ie, evolutionary lifetimes)
of cool luminous giant stars in old stellar populations. SBF stellar
population studies have been less explored than distance measurements,
partly because of the lack of suitable datasets.  Multicolor optical
($VRI$) SBF measurements for Virgo cluster galaxies from
\citet{1990AJ....100.1416T} were analyzed by both
\citet{1993ApJ...409..530W} and \citet{1993A&A...275..433B}, who reached
opposite conclusions on whether the optical SBF colors indicated that
the galaxies contained a significant metal-poor component.
\citet{1994ApJ...429..557A} measured $I$-band SBFs for a sample of
Galactic globular clusters, the only observations to date for these
systems.  They sought to understand the empirical $I$-band SBF zeropoint
and its correlation with integrated $V-\Ic$ galaxy colors, as well as
the conflict between the data and the RYI models; they also addressed
the use of optical SBF colors for stellar population studies in galaxies
and for disentangling age from metallicity effects in globular
clusters. \citet{1995AJ....110..179S, 1996AJ....111..208S} performed a
detailed study of the optical SBF gradients within the nearby
ellipticals M~32 and NGC~3379. Finally, \citet{1998ApJ...505..111J}
found reasonable agreement between their \Kp-band SBF data for 11 nearby
galaxies and the \citet{1993ApJ...409..530W} models, with most of the
galaxies lying around the [Fe/H]~=~--0.25 models with a spread in ages.

Now is a ripe opportunity to revisit the issue of stellar population
modeling of surface brightness fluctuations.  There have been
significant recent improvements in the stellar evolution calculations
and spectral libraries used by population synthesis models. For example,
the latest stellar evolution calculations include updated input
radiative opacities \citep[e.g.,][]{1992ApJ...397..717I}.  There has
been even more progress on the observational front. The amount of
$I$-band SBF data has increased by nearly tenfold since the early
modeling of \citet{1993ApJ...409..530W}, and new data in the
near-infrared have extended the spectral range of SBF measurements
\citep{1994ApJ...433..567P, 1998ApJ...505..111J, liu2000, mei2000}.

In this paper, we present new models for optical/infrared SBFs of
intermediate-age and old single-burst stellar populations (SSPs) and
discuss their implications for SBF distance measurements and stellar
population studies.  Though interesting issues remain to be addressed by
blue/near-UV SBF measurements \citep{1993ApJ...415L..91W}, we focus on
the optical and near-infrared (\ie, $V$-band to $K$-band) SBFs, since
these constitute the bulk of past and ongoing observations. In \S~2, we
describe our models for computing SBFs. The models cover a wider range
of ages and metallicities than in previous SBF studies. Furthermore, we
have optimized the models for this work by refining the prescription for
the luminous cool stars, which are important contributors to the SBF
signal.  In \S~3, we present the predictions of our models, including
SBF magnitudes, integrated colors, and the fractional contribution of
different stellar evolutionary phases to the SBFs.  We also derive
theoretical calibrations and $k$-corrections that are needed to use SBFs
standard candles.  In \S~4, we compare our results with current SBF
observations and discuss implications for the stellar content of
elliptical galaxies.  In \S~5, we review the uncertainties in our
results and explore the potential of SBFs for breaking the
age-metallicity degeneracy in studies of elliptical galaxies.  Finally
in \S~6, we summarize our findings and offer some future directions for
SBF studies.

\section{Synthesizing SBFs of Stellar Populations}

In this section, we present our models for computing SBFs. We first
recall the origin of the SBF signal of stellar populations. We then
review current observations of the stars which dominate this signal in
order to establish a framework to interpret the model results.  Readers
interested only in the description of our models should skip to
\S~\ref{bc2000}.

\subsection{Origin of the SBF Signal}

\begin{figure}
\centering
\includegraphics[width=5in,angle=0]{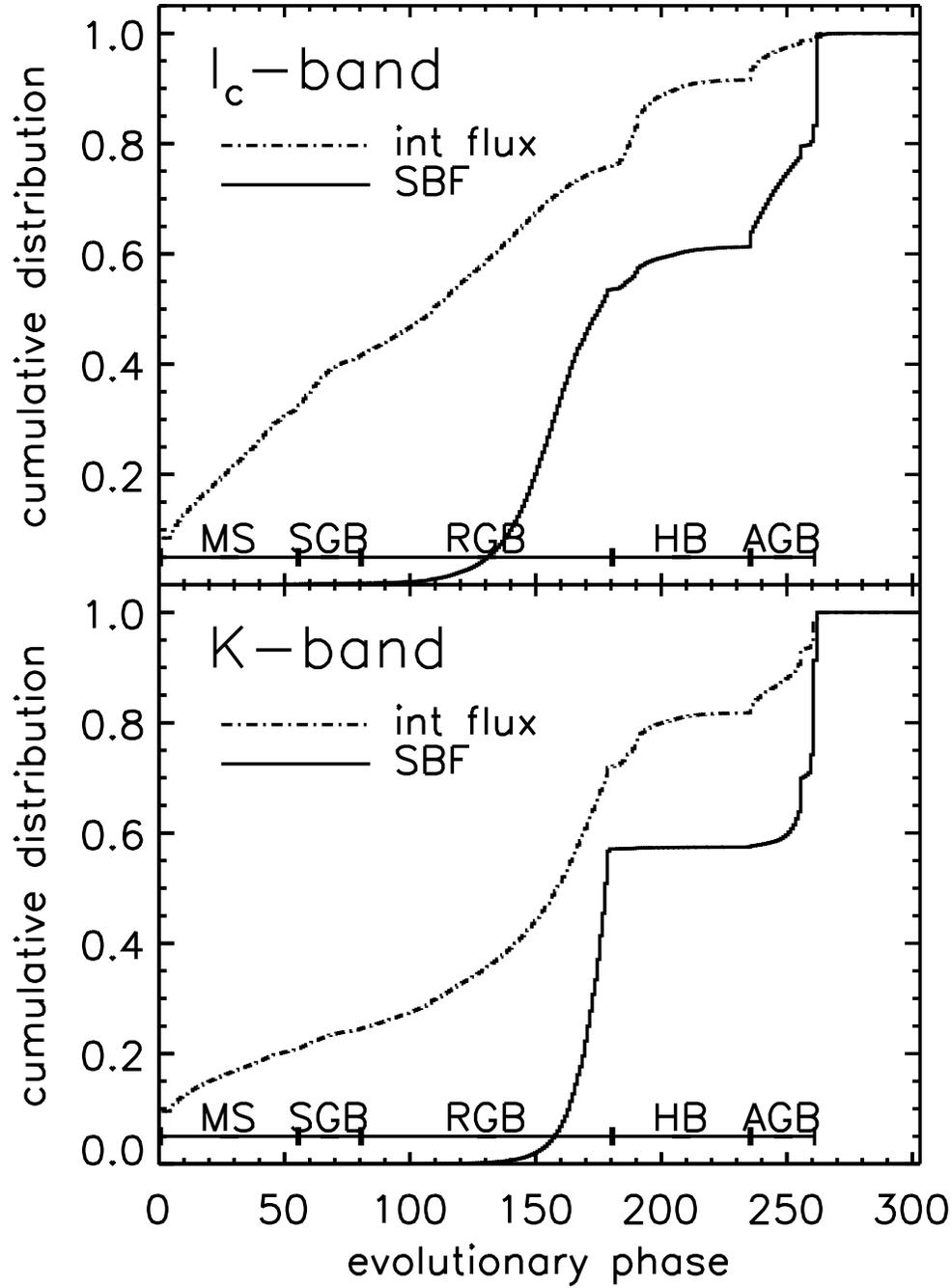}
\bigskip
\bigskip
\caption{\normalsize Cumulative distribution as a function of evolutionary
phase for the integrated light and SBFs of a 12~Gyr old
solar-metallicity single-burst stellar population from our standard
models (Padova tracks with semi-empirical SEDs). Evolutionary phases
range from the zero-age main sequence on the left to the end of the AGB
on the right using an arbitrary numerical index.  The integrated light
arises from stars of all phases. On the other hand, the SBFs originate
almost entirely from the RGB and AGB. \label{cumulative}}
\end{figure}

Since SBFs are weighted by the square of the stellar luminosity, they
are very sensitive probes of the most luminous cool giant stars. For
example, for the idealized case of a stellar population with a simple
power-law luminosity function, one can easily show via
equation~(\ref{Lbar}) that the fluctuation luminosity scales linearly
with the maximum luminosity of the stars. Figure~\ref{cumulative}
illustrates the greater sensitivity to luminous giant stars of SBFs
compared to ordinary integrated light for a more realistic model stellar
population (see \S~\ref{bc2000} below). For this 12~Gyr
solar-metallicity model, about 90\% of the SBF signal is contributed by
the brightest 2~mags of the stellar luminosity function in the near-IR
and the brightest 3--5~mags in the optical. Clearly, the
optical/infrared SBF signal is heavily weighted to the very brightest
stars in a stellar population.

\subsection{Observed Properties of Cool Giant Stars \label{obsframework}}

Since SBF magnitudes and colors are expected to closely follow the peak
magnitude and colors of the giant branch, it is insightful to review the
observed properties of cool giant stars, both on the RGB (H-shell
burning) and the AGB (He-shell burning), as a function of age and
metallicity for Milky Way and Magellanic Cloud star clusters.

\subsubsection{Giant Branches of Old Stellar Populations \label{obs-oldpops}}

\begin{figure}
\centering
\hbox{\hskip -0.75in
\includegraphics[width=5.5in,angle=90]{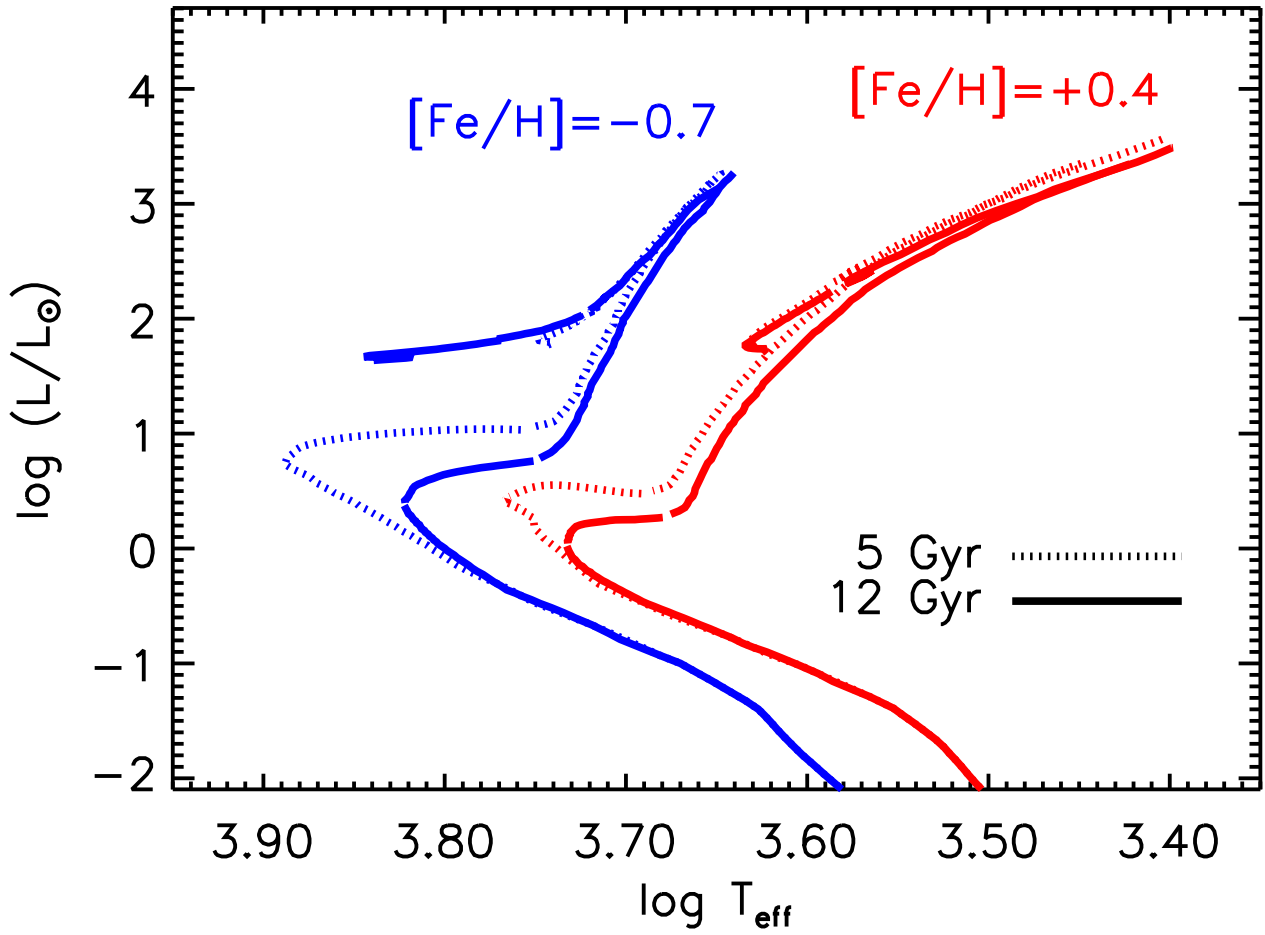}}
\caption{\normalsize Theoretical isochrones from the Padova models
illustrating the effect of age and metallicity variations for old
stellar populations. \label{hrdiagram}}
\end{figure}

The colors of the giant branches of old ($\gtrsim$5~Gyr) stellar
populations are primarily driven by metallicity, with more metal-rich
systems having redder giant branch colors
\citep[e.g.,][]{1983ApJ...275..773F}. This is illustrated in
Figure~\ref{hrdiagram}, which shows the change in the temperature of the
RGB and AGB for a change in metallicity from [Fe/H]~=~--0.7 to +0.4 with
ages of 5 and 12~Gyr.  This trend results from the increase in opacity
with metallicity from H$^-$ ions, which are the principal source of
continuum opacity for stars with $T_{eff}\approx3000-6000$~K. Metals
with low-ionization potential are the primary electron donors. Thus, as
the metallicity increases, H$^-$ opacity increases and the giant branch
temperature drops.  It is interesting to note that since Mg and Si are
also significant donors along with Fe \citep{ren77}, the giant branch
color traces the total metallicity [Z/H] and not just the Fe abundance
[Fe/H] (\eg, \citealp{1984ApJ...287L..85G}; see discussion in
\citealp{1996A&A...305..858S}). In stellar evolution models, the
absolute temperature of the RGB depends on the choice of the mixing
length used to parameterize the interior convection
\citep[e.g.,][]{1988ARA&A..26..199R}.

Similarly, the shape of the RGB, as measured from the RGB slope in
optical/IR color-magnitude diagrams, is also primarily driven by the
metal abundance for old populations, with more metal-rich clusters
having steeper slopes \citep{1995AJ....109.1131K, 1995AJ....110.2844K,
1997AJ....114..694T,ferraro99}.  This implies that giant stars which
contribute to the SBF signal will have a larger spread in color for
metal-rich populations than metal-poor ones.

The tip of the RGB (TRGB) is delineated by core-He ignition. The TRGB
bolometric luminosity in Milky Way globular clusters increases modestly
with metallicity \citep{1981ApJ...246..842F,
1983ApJ...275..773F}.\footnote{Observations actually measure the
brightest RGB stars in the clusters, rather than the true tip of the
RGB. Therefore, there is a potential systematic underestimate of the tip
luminosity due to statistical sampling.  Unpublished estimates by
\citet{rood97} show that this effect is probably not significant for the
Frogel \etal\ data \citep[see also][]{ferraro99}.}  The optical and IR
magnitudes for the TRGB have opposite dependences on metallicity. In the
optical, increasing metallicity leads to increased opacity from
molecular lines, especially from TiO, which reduces the optical flux.
Hence, the TRGB in the optical becomes cooler and fainter as metallicity
increases, first turning over in the $V$-band and then also in the
$I$-band for the most metal-rich Milky Way globular clusters
\citep[e.g.,][]{1977MNRAS.178..163L,1990AJ....100..162D,
1991A&A...249L..31O,1991ApJ...382L..15B}. Thus the brightest stars of
metal-rich globular clusters are fainter in the optical than those of
metal-poor clusters.  On the other hand, in the near-IR the $K$-band
magnitude of the TRGB of Milky Way globular clusters rises monotonically
with metallicity, with a roughly 1~mag increase over the range
$-2.2\lesssim{\rm [Fe/H]}\lesssim-0.2$ \citep{ferraro99}.

The AGB is expected to follow the behavior of the RGB in old populations
and reach cooler temperatures with increasing metal abundance.  AGB
studies in globular clusters are hampered by the scarcity of stars found
in this short evolutionary stage in any given cluster.  The AGB is
usually divided into two phases: (1) the initial early AGB (E-AGB),
where He-shell burning proceeds steadily, and (2) the thermally pulsing
AGB (TP-AGB), characterized by oscillations in luminosity due to
periodic flashes of the He shell. Mass loss is believed to occur during
both of these phases and to end in a superwind stage. In metal-poor
([Fe/H]~$\lesssim-1$) Milky Way globular clusters, the luminosities of
the brightest AGB stars do not exceed the tip of the RGB. In more
metal-rich systems, long-period variables are observed above the tip of
the RGB, which are presumably TP-AGB stars
\citep{1988ApJ...324..823F}. In fact, since AGB stars follow a core
mass-luminosity relation \citep{1971AcA....21..417P} and the
main-sequence turnoff mass increases with metallicity at a fixed age
\citep{1988ARA&A..26..199R}, higher metallicity leads to brighter AGB
stars.

\subsubsection{Extended Giant Branches of Intermediate-Age Populations}

Information on intermediate-age ($\approx0.5-5$~Gyr) stellar populations
largely comes from studies of star clusters in the Large and Small
Magellanic Clouds (LMC/SMC). The brightest stars in these clusters are
carbon AGB stars with bolometric magnitudes much brighter than the tip
of the RGB \citep{1979ApJ...232..421M, 1980ApJ...239..495F,
1980ApJ...240..464M, 1981ApJ...249..481C}.  These very cool stars
provide a significant fraction of the total luminosity of the clusters
\citep{1983ApJ...266..105P}.  They are redder and more luminous than the
M-type AGB stars. In contrast, the brightest stars in old LMC/SMC
clusters are M~giants no brighter than the brightest stars in Galactic
globular clusters \citep{1985ApJ...288..551A, 1990ApJ...352...96F,
1996A&A...316L...1M}.  As a result, the near-IR colors of
intermediate-age clusters (types IV to VI of
\citealp{1980ApJ...239..803S}) are much redder than those of older or
younger clusters, whereas the optical colors redden monotonically with
cluster age.

Carbon stars are believed to arise from carbon dredge-up during the
TP-AGB phase \citep{1983ARA&A..21..271I}. The observed ratio of C-stars
to M-stars in Local Group galaxies anti-correlates with metallicity,
C-stars being the dominant spectral type for [Fe/H]~$\lesssim-1$
\citep{1978Natur.271..638B, 1986ApJ...305..634C,
1987ApJ...323...79P}. This occurs because stars with lower metallicities
require fewer thermal pulses to change from an O-rich to a C-rich
star. In addition, dredge-up is more efficient at lower metallicities
because it begins at lower core masses, \ie, at earlier ages. There are
no examples of solar-metallicity intermediate-age clusters in the
LMC/SMC which would be most directly comparable to elliptical
galaxies. However, models by \citet{1993ApJ...413..641V} predict that
the bolometric luminosity of the tip of the AGB should remain much
brighter than the tip of the RGB and have even cooler temperatures and
redder colors than in LMC/SMC clusters \citep[see also discussion
in][]{1998AJ....116...85S}.

Since the giant branches of intermediate-age Magellanic Cloud clusters
reach higher luminosities than those of old Galactic clusters, they are
often referred to as ``extended'' giant branches and are considered the
hallmark of intermediate-age populations
\citep[e.g.,][]{1997ApJ...477L..21G}. This occurs because the core
mass-luminosity relation for AGB stars implies that younger populations
have more luminous AGB stars. In the context of SBF studies, elliptical
galaxies are generally thought to be composed entirely of old stellar
populations. However, any recent episodes of star formation would lead
to extended giant branches. The observable consequence of this would be
brighter fluctuation magnitudes (and perhaps somewhat redder fluctuation
colors), especially in the infrared, without significant changes in the
integrated colors of the galaxies.\footnote{Extended giant branches have
been claimed to be detected in optical and near-IR images of M~32 and
the bulge of M~31 \citep{1991AJ....101.1286R, 1992AJ....104.1349F,
1992AJ....104.1360E, 1993AJ....106.2252R}, which would mean recent star
formation in these supposedly canonical examples of old stellar
populations.  However, more recent work has suggested this conclusion is
erroneous and due to severe image crowding \citep{1993AJ....105.2121D,
1996AJ....112.1975G, 1998AJ....115.2459R, 1998A&AS..127..327S,
1999ApJ...518..627J}. In addition, as pointed out by
\citet{1998AJ....115.2459R}, the presence of objects brighter than the
tip of the RGB (\hbox{$M_{bol}\approx-4$} for solar-metallicity) cannot
be readily interpreted as the sign of an intermediate-age
population. Long-period variables in metal-rich globular clusters do
reach $M_{bol}\approx-5$, so the only unambiguous signs of an
intermediate-age population would be either the presence of (1) stars
with even larger bolometric luminosities or (2) many more stars with
$M_{bol}=-4$~to~$-$5 than expected in a metal-rich population.} We
return to this point in \S~\ref{section-kbar-compare} and
\S~6.

\subsection{Stellar Population Synthesis Models for SBFs \label{bc2000}}

We use the latest version of the Bruzual-Charlot population synthesis
models \citep[][hereinafter BC2000]{bc2000}. These models allow us to
predict the spectral evolution of stellar populations with arbitrary
star formation rates and initial mass functions (IMFs) for a wide range
of metallicities. The main ingredients of the models are the stellar
evolution theory used to predict the distribution of stars in the
theoretical Hertzsprung-Russell diagram and the library of spectra
assigned to stars as a function of effective temperature and
luminosity. The integrated spectrophotometric properties of an entire
stellar population are obtained by summing the spectra of its component
stars. We now describe the model features most relevant to the present
work.

\subsubsection{Model Inputs} 

The BC2000 models present several major improvements over the earlier
models of \citet{1993ApJ...405..538B}. They also offer several choices
of input stellar evolution theory and atmospheres
(Table~\ref{table-models}). Stellar evolution can be followed according
to the prescription of the Geneva school for solar metallicity or that
of the Padova school for metallicites ranging from 1/200 to 2.5~times
solar. (See \citealp{1996ApJ...457..625C} for a detailed description of
the two prescriptions, and \citealp{1997AJ....114.1531B} for a complete
list of references.) These prescriptions include all phases of stellar
evolution from the zero-age main sequence to the beginning of the TP-AGB
(for low- and intermediate-mass stars) and core-carbon ignition (for
massive stars). In both sets of tracks, solar metallicity corresponds to
$Z_\odot =0.020$, and in the Padova tracks, the metallicity scales as
\hbox{[Fe/H] = 1.024 log$Z$ + 1.739} \citep{1994A&AS..106..275B}. We
note that the Padova and Geneva models predict very different fractional
contributions from RGB and AGB stars to the integrated light of
solar-metallicity SSPs \citep{1996fstg.conf...14B}; this has strong
implications for SBF predictions.\footnote{\citet{2000A&AS..141..371G}
have recently updated the Padova library. The new library covers only a
limited range of initial stellar masses and metallicities and cannot be
combined with the library used here because of the slightly different
chemical composition used at a fixed metallicity. The new tracks are
almost identical to the previous ones, the main purpose of the new
library being a finer sampling of initial stellar masses (Girardi \etal\
2000; A.~Bressan 2000, private communication).}

\citet{bc2000} supplemented these tracks with a new prescription for the
thermally-pulsing regime at the tip of the AGB, which improves over that
adopted by \citet{1993ApJ...405..538B}. This prescription was motivated
by the inability of that proposed by \citet{1991ApJ...367..126C} to
account for the SBF properties of observed galaxies when we started the
present study.\footnote{Preliminary results for the $K$-band SBF
magnitudes in \citet{liu99b,liu99c} used the old TP-AGB prescription and
are supplanted by the results in this paper.}  In the new prescription,
the effective temperatures, bolometric luminosities, and lifetimes (of
both the optically-visible and superwind phases) of TP-AGB stars of
various metallicities are taken from the models of
\citet{1993ApJ...413..641V}.  For stars undergoing a transition from
oxygen-rich (M-type) to carbon-rich (C-type) on the TP-AGB, the relative
lifetimes of the two phases as a function of metallicity are taken from
the models of \citet{1993A&A...267..410G} and
\citet{1995A&A...293..381G}.  These models reproduce the observed ratios
of C to M stars in the LMC and the Galaxy; however, since they do not
extend to sub-Magellanic ($Z\lesssim0.004$) nor super-solar ($Z>0.2$)
metallicities, the same relative durations of the M-type and C-type
phases are applied to the Padova tracks at the more extreme
metallicities.  Finally, the tracks are supplemented with a prescription
for post-AGB evolution and with unevolving main-sequence stars in the
mass range 0.1--0.6~$M_\odot$.

The BC2000 models also offer a choice of spectral libraries, summarized
in Table~\ref{table-models}. For solar metallicity, a quasi-empirical
set of spectra is available from a compilation by
\citet{1998PASP..110..863P}.  The spectra of M0--M10 giant stars are the
only non-empirical ones in this library, as they are based on the
synthetic M-giant spectra computed by \citet{1994A&AS..105..311F}. These
models were constructed to reproduce period-averaged spectra from
observations of long-period variable TP-AGB stars. In the remainder of
this paper, we refer to this spectral library as the {\em empirical}
SEDs.

For stars of all metallicities, a comprehensive spectral library has
been assembled by \citet[][hereinafter LCB97]{1997A&AS..125..229L,
1998A&AS..130...65L}. It is a combination of model atmospheres by Kurucz
(1995; private communication to Lejeune), 
\citet{1989A&AS...77....1B,1991A&AS...89..335B}, and the same
\citet{1994A&AS..105..311F} spectra as above, rebinned
onto homogeneous scales in wavelength and physical parameters (effective
temperature, surface gravity, and metallicity).  Hereinafter, we refer
to this spectral library as the {\em theoretical} SEDs.

LCB97 further corrected the synthetic spectra in their library by
requiring that these reproduce the color-temperature relations observed
for solar-metallicity stars (see also
\citealp{1996fstg.conf...94L}). This adjustment is especially important
for the M-star spectra.  Lacking empirical calibration for non-solar
metallicities, LCB97 applied the corrections derived at solar
metallicities to all other metallicities.  We refer to this spectral
library as the {\em semi-empirical} SEDs.

\citet{bc2000} supplemented all three of these spectral libraries with
period-averaged spectra for C-type stars on the TP-AGB based on model
atmospheres from Loidl et al. (1999, private communication; see
\citealp{2000ibp..confE..39H}). BC2000 applied empirical corrections to
these spectra to match observed color-color calibrations by
\citet{1965ApJ...141...161M}. The spectra of stars in the superwind
phase at the end of the TP-AGB are based on observations by
\citet{1996A&A...314..896L} and \citet[][and references
therein]{1997A&A...324.1059L}.

The models used here have been tested successfully against observed
spectra of star clusters and galaxies
\citep{1997AJ....114.1531B,bc2000}.  Complementary descriptions and
applications of previous versions of these models can also be found in
\citet{1996ApJ...457..625C}, \citet{1997AJ....114.1531B}, and
\citet{1998MNRAS.294..705K}.  \citet{1996fstg.conf...14B} has compared
the results of different population synthesis codes and examined the
dependence of integrated galaxy colors, SEDs, and spectral indices
predicted by the \citet{bc98} models (very similar to BC2000 for his
purposes) on the choice of input evolutionary tracks and spectra.

Unless otherwise indicated, in all calculations below we adopt a
\citet{1955ApJ...121..161S} IMF truncated at 0.1 and 125~\Msun.  For old
($\ge$5~Gyr) populations, adopting the \citet{1986FCPh...11....1S} IMF
would lead to fluctuation magnitudes only $\le$0.05~mag brighter and
integrated colors only $\le$0.02~mag bluer. (The changes are somewhat
larger at younger ages; at 1~Gyr, the Scalo IMF leads to \Mbar's fainter
by about 0.1--0.2~mag.)  These changes would have a negligible effect on
our results and the agreement of our models with SBF observations. (See
\S~\ref{imf+lifetimes} for a discussion of alternative IMFs.)
Furthermore, in this paper, we approximate star clusters and elliptical
galaxies as single-burst stellar populations.  While this approximation
is adequate for our present purposes, it limits investigations of the
star formation history of elliptical galaxies. This is our goal in a
complementary study (M. Liu et al., in preparation; see also
\S~6).

\subsubsection{Model Outputs \label{output}}

We compute SBF magnitudes of single-burst stellar populations for ages
of 1--17~Gyr and metallicities $Z=0.0001 -0.05$ ($Z/Z_{\odot} = 0.005 -
2.5$, \hbox{[Fe/H] = --2.4 to +0.4}).  We generate results for the
$BV{\Rc\Ic}JH{\Kp\Ks}KL{\Lp}M$ bandpasses used by ground-based
observatories and the $F814W,$ $F110M,$ $F110W,$ $F160W,$ and $F222M$
bandpasses used by the WFPC2 and NICMOS instruments aboard \HST.  The
zeropoints for our magnitudes are computed on the Vega magnitude system.

The response functions for the Johnson-Morgan $BV$ bandpasses are from
\citet{1978A&A....62..411B}.  The \Rc\ and \Ic\ bandpasses are from
\citet{1990PASP..102.1181B}.  For the ground-based near-infrared
$JHKL{\Lp}M$ bandpasses, we use traces from
\citet{1988PASP..100.1134B}.\footnote{There is potential ambiguity in
the notation of \Lbar\ and \Mbar.  In the case of \Lbar, this can be
either the fluctuation luminosity or the SBF magnitude in the $L$-band
filter ($\lambda_{eff}=3.5$~\micron); we discuss the filter only in
\S~\ref{section-age-z} and Table~\ref{table-salp}. In the case of \Mbar,
we use this to refer generically to the SBF absolute magnitude in any
filter bandpass.  This could also represent the SBF absolute magnitude
in the $M$-band filter ($\lambda_{eff}=4.8~\micron$), but we avoid using
this representation except in Table~\ref{table-salp}.}  These are
similiar to the filter system used at the UK Infrared Telescope (UKIRT),
which is described in \citet{1992ApJS...82..351L} and \citet{cas92}.  We
choose not to use the CIT bandpasses \citep{1982AJ.....87.1029E} as the
effective wavelength of the CIT $J$-band is noticeably redder than that
of most other IR systems
\citep[e.g.,][]{1992ApJS...82..351L,1988PASP..100.1134B} due to a Si
lens in the original CIT dewar which defined the blue end of the
bandpass.

We also include calculations for popular alternatives to the $K$ filter:
the \Kp\ \citep[1.9--2.3~micron;][]{1992AJ....103..332W} and \Ks\
\citep[2.0--2.3~\micron;][]{1995ApJS...96..117M} bandpasses. The \Ks\
filter trace is from the IR imager CIRIM used at Cerro Tololo
Interamerican Observatory, and the \Kp\ filter is from the UCLA Gemini
instrument at Lick Observatory \citep{1994SPIE.2198..457M}.  Our \Kp\
filter is virtually identical to that in the Hawaii QUIRC infrared
camera \citep{1996NewA....1..177H} used by \citet{1998ApJ...505..111J,
1999ApJ...510...71J} for their SBF measurements. However, it is slightly
different than the trace originally given by \citet{1992AJ....103..332W}
in that the peak transmission of our filter is slightly redder.
Transmission data for these IR filters were measured at 77~K.  The
traces were then multiplied with a theoretical model for atmospheric
transmission generated by the ATRAN program \citep{atran} as atmospheric
absorption can be significant at the edges of the bandpasses. We also
included the (negligible) effect of the quantum efficiency (QE) of the
IR detectors in these bandpasses.

The \HST\ bandpasses include the filter transmission profiles, the
spectral response of the \HST\ mirror and instrument optics, and the QE
of the detectors. The latter is especially important for the NICMOS
filters as the QE rises sharply from 1--2 \micron\ \citep{nicmos} due to
the low operating temperature of the instrument's HgCdTe arrays.

Also, we compute the ordinary $V-\Ic$, $V-K$, and $J-K$ integrated
colors of the SSPs and the strengths of several absorption lines
($H\beta$, Mg$_2$, Mgb, H$\gamma_A$, and C4668) defined on the standard
Lick/IDS system \citep{1994ApJS...95..107W,1994ApJS...94..687W,
1997ApJS..111..377W}.  We use the analytic fitting functions derived in
these studies for index strength as a function of stellar temperature,
gravity, and metallicity.  We note that $\alpha$-element enhancement in
massive elliptical galaxies leads to [Mg/Fe] ratios above that of the
most metal-rich stars in the solar neighborhood
\citep{1992ApJ...398...69W}. Since our models use scaled-solar
metallicity, we expect our predictions for the Mg$_2$ indices to be
inaccurate and tabulate them only for completeness.  Elemental
enhancements, however, should have little effect on the other
spectrophotometric properties of the model stellar populations at fixed
total metallicity (A. Bressan \etal, in preparation).

\section{Results}

\subsection{SBF Magnitudes \label{sbfmags}}

We compute \Mbar\ for SSPs as a function of age and metallicity using
several combinations of evolutionary tracks and SEDs.  In rough terms,
models with \hbox{$Z\gtrsim0.008$} and \hbox{ages $\gtrsim5$~Gyr} can be
thought of as corresponding to the classical picture of old, metal-rich
stellar populations in giant elliptical galaxies, while lower
metallicity but comparably old SSPs would correspond to Galactic
globular clusters.

Table~\ref{table-salp} presents the SBF magnitudes computed using the
Padova evolutionary tracks and the semi-empirical SEDs.  Given the wide
metallicity range spanned by the Padova tracks and the corrections
applied by LCB97 to make the semi-empirical SEDs agree with
observations, we consider this combination of tracks and SEDs as our
``standard models.''

\begin{figure}
\centering 
\includegraphics[width=5.5in]{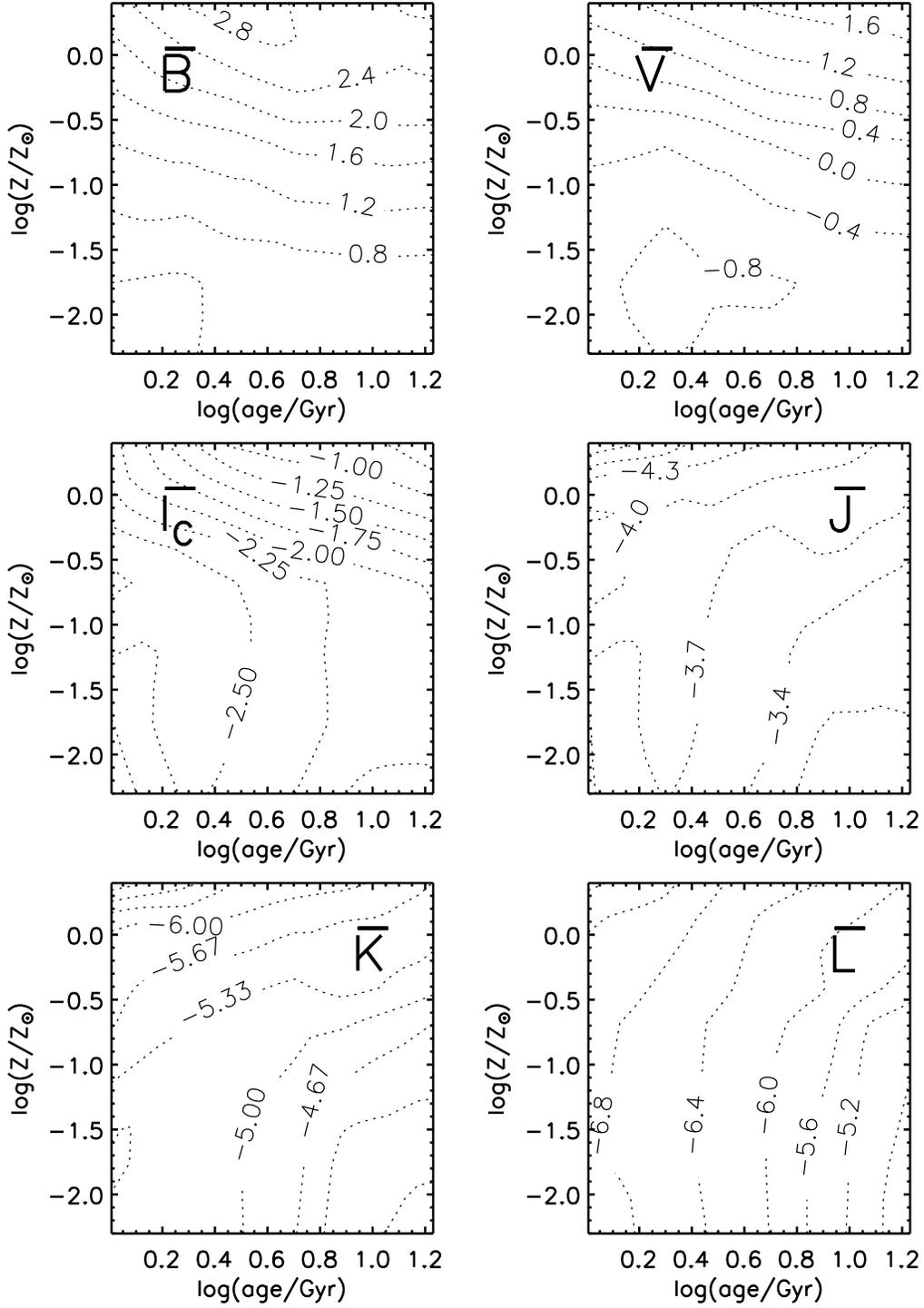}
\vskip 0.3in
\caption{\normalsize Contours of SBF magnitudes for the $BV{\Ic}JKL$
bandpasses in the parameter space of age and metallicity from our
standard models (Padova tracks with semi-empirical SEDs).  Horizontal
contours would indicate that a quantity depends only on metallicity, and
vertical contours that it depends only on age. \label{contours-sbf}}
\end{figure}

\begin{figure}
\centering 
\includegraphics[width=5.5in]{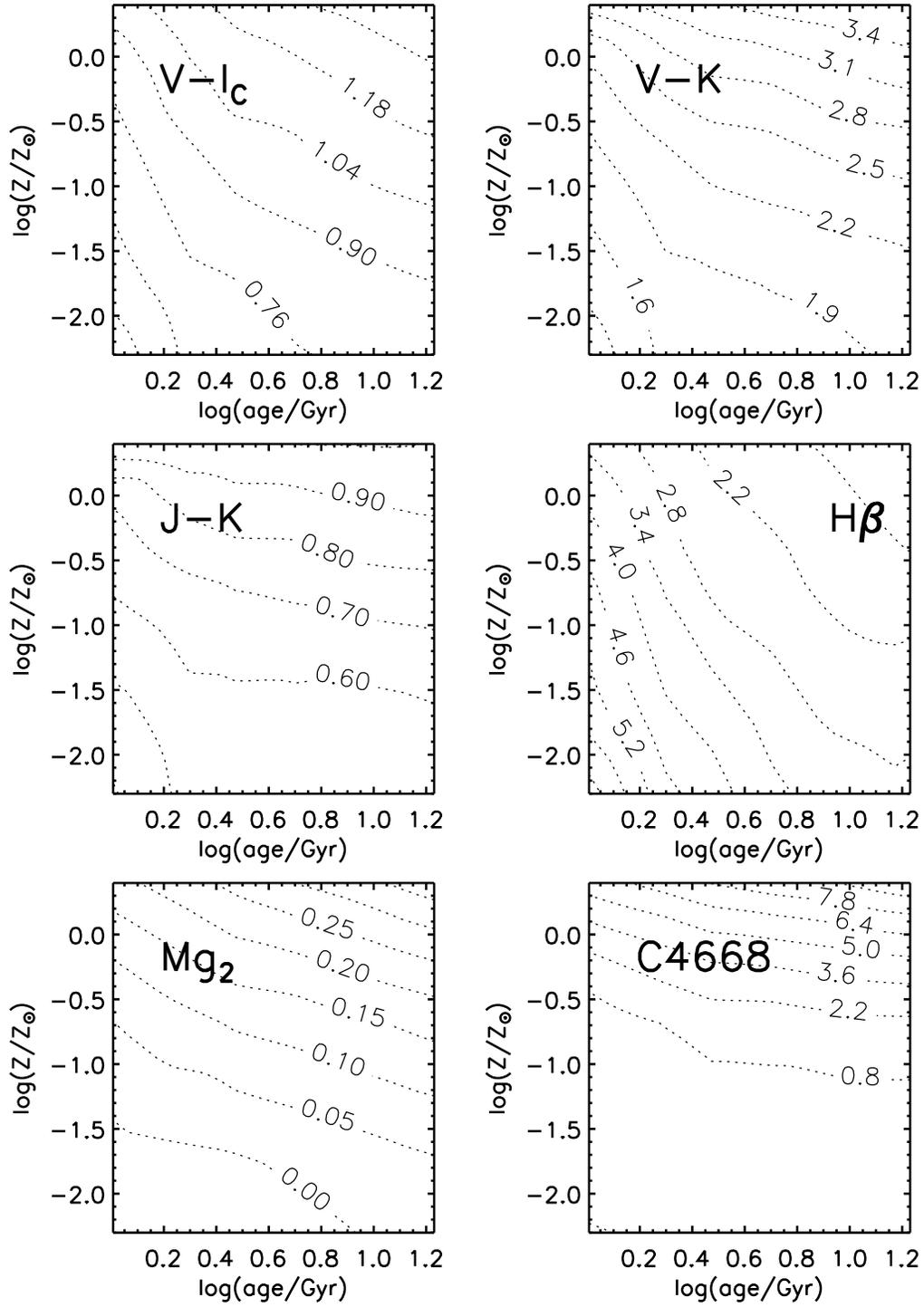}
\vskip 0.3in
\caption{\normalsize Contours of predicted galaxy colors and absorption
line indices in the parameter space of age and metallicity for the same
models as in Figure~\ref{contours-sbf}.  Horizontal contours would
indicate that a quantity depends only on metallicity, and vertical
contours that it depends only on age.
\label{contours-colors}}
\end{figure}

A more enlightening representation of our results is given in
Figure~\ref{contours-sbf}, which shows contour plots of \Mbar\ as a
function of age and metallicity for a subset of bandpasses computed with
our standard models.  For comparison, Figure~\ref{contours-colors} shows
contour plots for a set of integrated galaxy colors and absorption
indices as predicted by the same models.  Any quantity (SBF magnitude,
galaxy color, or absorption index) with all vertical contours would be
completely age-dependent, whereas any quantity with all horizontal
contours would be completely metallicity-dependent.
Figure~\ref{contours-colors} shows that the broad-band colors and
absorption indices have contours which are nearly straight lines. This
is the origin of the ``3/2 rule'' of \citet{1994ApJS...95..107W}, which
states that changes by a factor of two in metallicity roughly mimic
changes by a factor of three in age for the integrated colors and line
strengths of old stellar populations. Figure~\ref{contours-colors} does
reveal some differences in the slopes of the contours for different
integrated colors and line strengths, which reflects the varying age and
metallicity dependences. For example, H$\beta$ is mostly age-sensitive
while C4668 is mostly metallicity-sensitive.

Although the contours of the fluctuation magnitudes are not as regular
as those of the integrated colors and line strengths,
Figure~\ref{contours-sbf} does reveal regions in age and metallicity
where the behavior of \Mbar\ is expected to be relatively simple. The
optical SBFs of more metal-rich ($Z\gtrsim0.004$) populations are
largely age-independent, as are near-IR ($JHK$) SBFs for
$t\gtrsim5$~Gyr.  This is the age/metallicity regime expected to be
relevant for bright elliptical galaxies.  At longer IR wavelengths
($L$-band and redward), the SBFs are mostly age-dependent. However, we
caution that these predictions rely more heavily on our prescription for
stars in the superwind phase on the TP-AGB (which have $K-L\ga1$), and
the stellar spectral libraries are less robust at these wavelengths ---
our models have not been well-tested at wavelengths
$\lambda>2.4~\micron$ due to the lack of strong observational
constraints.

\subsection{Fractional Contributions of Different Evolutionary Phases} 

It is instructive to examine the fractional contributions to \Mbar\ from
different phases of stellar evolution (main sequence, sub-giant branch
[SGB], RGB, horizontal branch [HB], AGB, bare planetary nebula nucleus,
and white dwarf). These can be computed from the corresponding terms in
the numerator of equation~(\ref{Lbar}).  Such valuable information
cannot be currently acquired from observations given the challenges of
resolving, let alone categorizing, individual stars in galaxies. This is
especially true for the old metal-rich populations of ellipticals
because there are few nearby examples.  Theoretical predictions can
indicate the sensitivity of \Mbar\ to different stages of stellar
evolution and, consequently, help to identify which aspects of stellar
evolution theory can be tested via SBF observations.

\begin{figure}
\centering \includegraphics[width=5.5in]{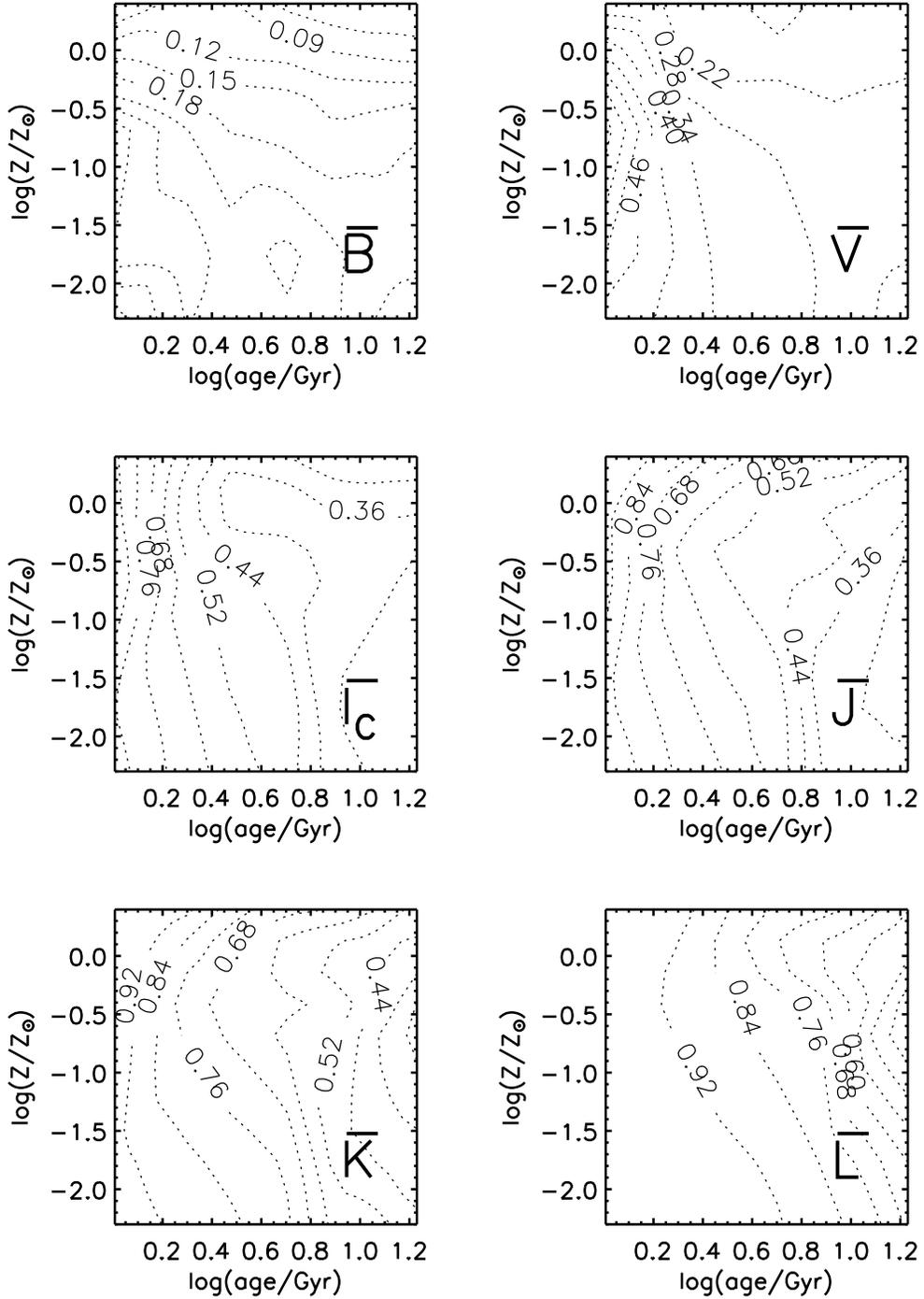}
\vskip 0.3in
\caption{\normalsize Fractional contribution of all AGB stars (including
carbon stars) to the SBF magnitudes in the $BV{\Ic}JKL$ bandpasses as a
function of age and metallicity for our standard models (Padova tracks
with semi-empirical SEDs). For the IR ($\lambda>1~\micron$) bandpasses,
the remaining contribution is almost exclusively from RGB
stars. \label{contours-agb}}
\end{figure}

The relative contribution of each evolutionary phase to the fluctuation
magnitude in a given bandpass is a function of both the age and
metallicity of the stellar population. For $t\wig{>}3$~Gyr, the relative
contributions of the different phases depend mostly on metallicity,
staying roughly constant as the population ages.
Figure~\ref{contours-agb} gives a contour plot of the fractional
contribution of AGB stars to \Bbar~\Vbar~\Ibar~\Jbar~\Kbar~\Lbar\ as a
function of age and metallicity for our standard models.  The IR
($\lambda>1~\micron$) SBFs arise almost exclusively from the RGB and AGB
stars, with the AGB stars dominating at all metallicities for
$t\wig{<}3$~Gyr.

Optical SBFs depend more significantly on evolutionary phases other than
the RGB and AGB. In the \Ic-band, the HB contribution rises with
metallicity, from about 1\% at $Z=0.0001$ (\Zsun/200) to about 20\% at
$Z=0.05$ (2.5~\Zsun) for a $>5$~Gyr old population.  It is worth noting
that in our models, AGB stars are important contributors to even the
$I$-band SBFs, unlike in the \citet{1993ApJ...409..530W} models. In the
$V$ and \Rc-band, the HB contribution is higher still at the expense of
the RGB and AGB, and in the $B$-band, planetary nebula nuclei become
important for $Z\wig{>}0.02$ (\Zsun).  For comparison, a decomposition
of the integrated light of a solar-metallicity SSP from each
evolutionary phase is given in Figure~2 of \citet{1993ApJ...405..538B}
and in \citet{1996fstg.conf...14B}.

\subsection{Calibration of SBFs for Distance Determinations
\label{distance-calibration}}

We now use our models to calibrate \Mbar\ as a distance indicator via
equation~(\ref{Lbar}).  Since the ages and metallicities of the observed
galaxies are not known a priori, the hope is to calibrate variations in
\Mbar\ arising from variations in these parameters by using another
observable quantity, such as the integrated galaxy color or spectral
index.  \citet{1997ApJ...475..399T} have derived an accurate calibration
for \Ibar; after applying a correction based on a galaxy's $V-\Ic$
color, they find the intrinsic scatter in \Ibar\ to be only 0.05~mag.
Preliminary data for \Kpbar\ reveal no obvious correlation with $V-\Ic$
color, though the \Kpbar\ sample is still too small for a firm
conclusion \citep{1998ApJ...505..111J}. Given the difficulties involved
in tying the SBF observations to the Cepheid distance scale \citep[see
discussion in][]{ton99}, it is important to determine the theoretical
zeropoint of the \Ibar\ versus $V-\Ic$ relation.

\begin{figure}
\centering
\vskip -0.3in
\includegraphics[width=3.5in,angle=90]{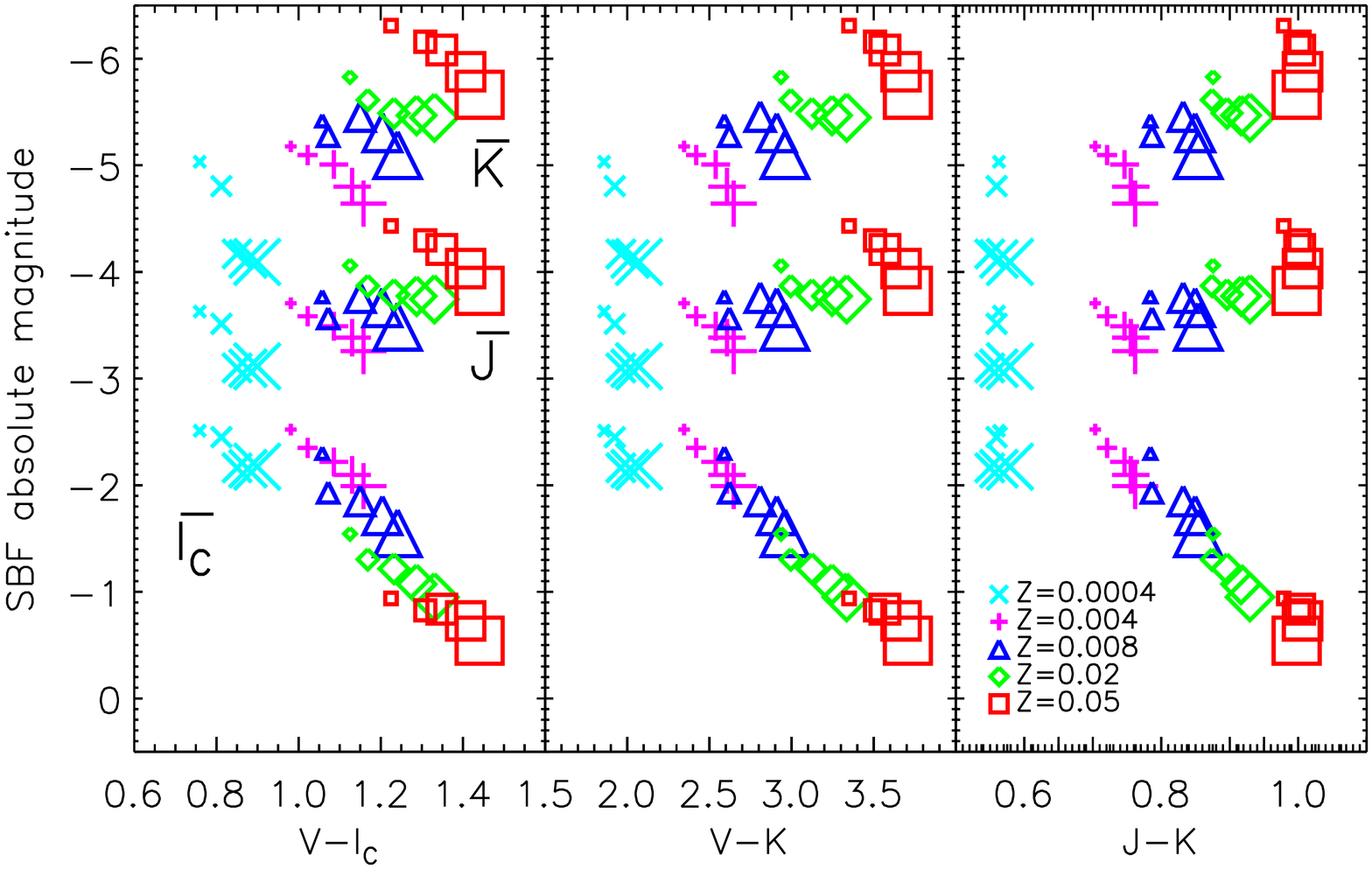}
\vskip 0.3in
\includegraphics[width=3.5in,angle=90]{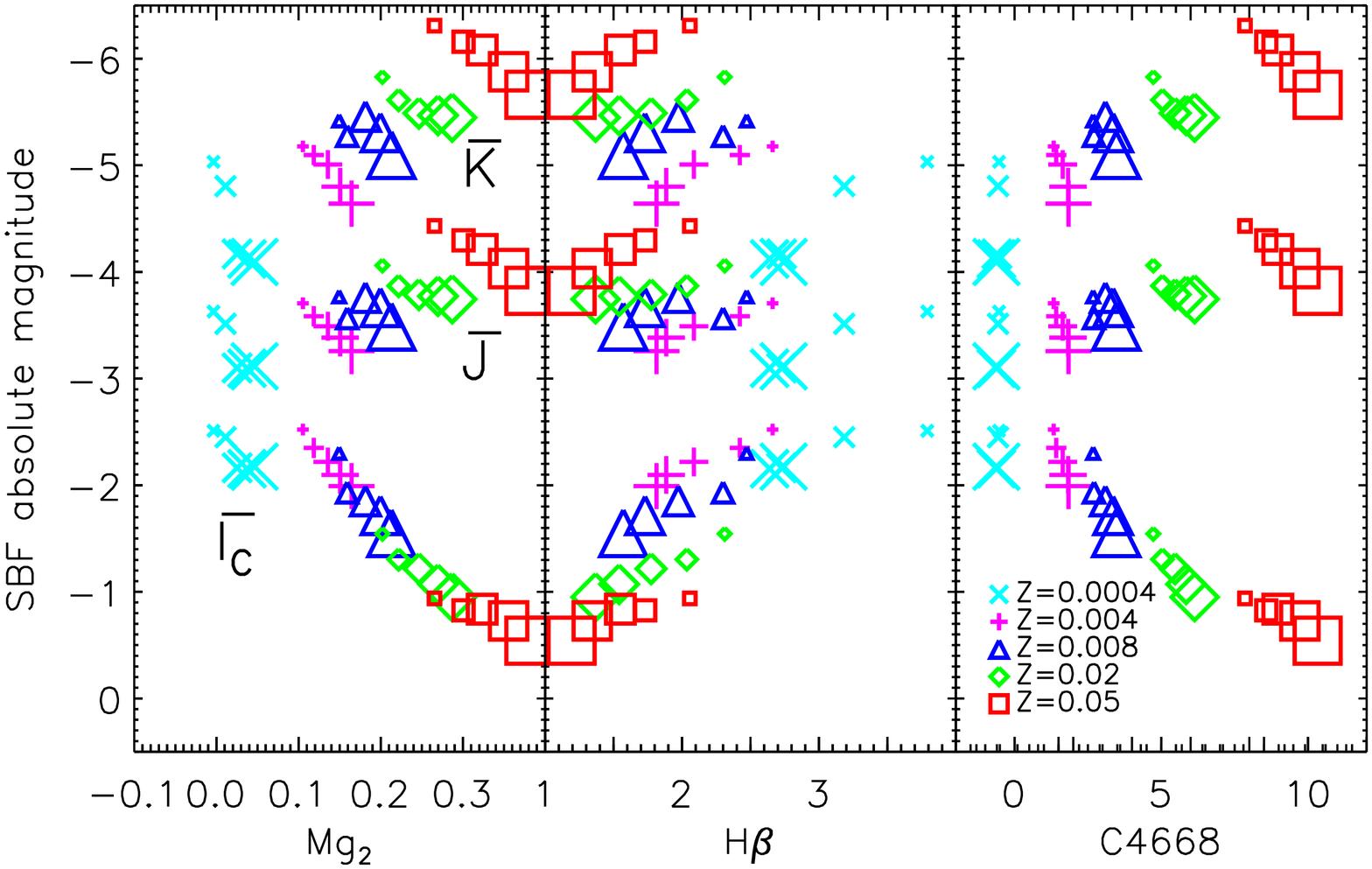}
\vskip 0.3in
\caption{\normalsize Results for $IJK$ SBF magnitudes as a function of
integrated light properties from our standard models. Ages of 3, 5, 8,
12, 17~Gyr are plotted for a range of metallicities, as indicated.
Models of a given metallicity have the same symbol, with increasing
symbol size representing increasing age.
\label{observables}}
\end{figure}

Figure~\ref{observables} shows \Ibar~\Jbar~\Kbar\ as a function of some
integrated colors and absorption-line indices for models with ages of
3--17 Gyr and $Z$~=~0.0004, 0.004, 0.008, 0.02, and 0.05.  Fluctuation
magnitudes appear to be the best behaved in the \Ic-band, \ie,
variations in \Ibar\ with age and metallicity are well-correlated with
the integrated properties.  To gauge how different observables are
correlated with \Mbar\ in different bandpasses, we performed a robust
linear fit to the fluctuation magnitudes as a function of the various
colors and line indices for the 5--17~Gyr and $Z=0.004-0.05$ models. We
then calculated the rms scatter of the residuals to these fits.  The rms
scatter indicates the effectiveness of a particular color or index in
compensating for stellar population variations. The results of these
fits are given in Table~\ref{table-calcobs}. We compare them with
observations in \S~\ref{section-ibar-compare}.

The scatter in $\Ibar$ in Figure~\ref{observables} is only
$\approx0.1$~mag (\ie, 5\% in distance) as a function of $V-K$ or $J-K$
color, while it is 0.19~mag as a function of $V-\Ic$. This suggests that
the accuracy of $\Ibar$ measurements could be improved by using $V-K$ as
a calibrator instead of $V-\Ic$ as is currently done. Moreover, $\Ibar$
as a function of $V-K$ has a shallower slope (1.4, as compared to 4.6
for \Ibar\ as a function of $V-\Ic$) meaning a greater tolerance of
photometric and reddening measurement errors.  Finally, the
model-predicted scatter for \Ibar\ is also typically smaller than for
that for \Jbar\ and \Kbar.

Spectral features can also be used to calibrate SBF magnitudes
\citep{1997ApJ...483L..37T}, though uncertainties in the effects of
$\alpha$-element enhancement are a concern for the models
(\S~\ref{output}). The metallicity-sensitive indices Mg$_2$ and C4668
generally have the smallest residuals ($\approx0.1-0.2$~mag) with
H$\beta$ having the largest residuals. This is expected since SBFs of
$\gtrsim5$~Gyr populations are mostly driven by metallicity.

Finally, the results of Figure~\ref{observables} suggest that the SBFs
measured with a filter around 1~\micron, \ie, between the \Ic\ and $J$
bandpasses, should be nearly independent of both age and metallicity
(leading to horizontal lines in these plots). This was also found by
\citet{1993ApJ...409..530W} who used very different models, suggesting
the conclusion is robust.

\subsection{$k$-Corrections}\label{kcorr}

Optical and near-infrared SBFs can be detected to $cz\sim10,000$~\kms\
using both ground-based telescopes and \HST\
\citep[e.g.,][]{1998ApJ...499..577L}. To derive accurate distances to
galaxies, and hence an accurate measure of \Ho, one first needs to apply
$k$-corrections to the observed SBF magnitudes and integrated colors of
the galaxies. Such corrections can be derived only from models.

We compute $k$-corrections for the integrated galaxy colors and for
\Mbar\ in various bandpasses useful for observing distant galaxies.  We
calculate corrections out to $cz=15,000$~\kms\ ($z=0.04$) for models
with $Z/\Zsun = \{0.2, 0.4, 1.0, 2.5\}$ and ages of 5, 8, 12, and
17~Gyr.  The corrections for the SBF magnitudes depend mostly on the
metallicity and very little on the age.  They correlate almost linearly
with redshift over the range considered. Table~\ref{table-kcorr} lists
the resulting $k$-corrections as a linear function of redshift for each
metallicity.  We calculated the slope of $k(z)$ for each of the four
model ages, and the average and rms of these slopes are tabulated.  The
sign convention used for the $k$-correction is $k(z) = \Mbar(z) -
\Mbar(z=0)$, \ie, the term is subtracted from observations at $z>0$ in
order to compare to local observations \citep{1956AJ.....61...97H,
row85}.

The amplitudes of the $k$-corrections can be significant.  For instance,
at distances of \hbox{$cz\approx 10,000$~\kms}, the $k$-correction
computed for $\Ibar$ is $\approx$0.2 mag with little dependence on the
choice of age and metallicity. This corresponds to a 10\% correction in
distance (and hence \Ho).  Also, our $k$-corrections for \Ibar\ and
$V-\Ic$ agree well with those of \citet{1997ApJ...475..399T}. For the IR
bandpasses, the $k$-corrections have a stronger dependence on
metallicity, but the amplitudes of the corrections are generally smaller
than in the optical.  There are some small irregularities seen in the
$k$-corrections for the 2~\micron\ filters: for \Kpbar\ and \Kbar, the
$k$-corrections can have different signs depending on the choice of
metallicity, though the amplitudes are small.  Also, the $k$-corrections
for the $K$ and $K^\prime$ filters differ significantly despite their
small wavelength separation. This is due to the CO bandhead at
2.3~\micron, which appears in absorption in the spectra of early-type
galaxies \citep{1975ApJ...195L..15F,1978ApJ...220...75F}. Our $K$-band
$k$-corrections are significantly larger than those used by
\citet{1998ApJ...505..111J}, which were computed from the
\citet{1994ApJS...95..107W} models, though the actual amplitudes of our
corrections are still relatively small. (Note that the Jensen \etal\
corrections have the opposite sign convention.)

We recommend using the $k$-corrections derived from the $Z=0.008$ and/or
$Z=0.02$ models, since these models roughly span the SBF observational
constraints discussed below.

\section{Comparison with Observations and Constraints on Stellar Populations}

\subsection{SBFs of Globular Clusters \label{globs}}

\begin{figure}
\centering
\vskip -0.3in
\includegraphics[width=5in,angle=0]{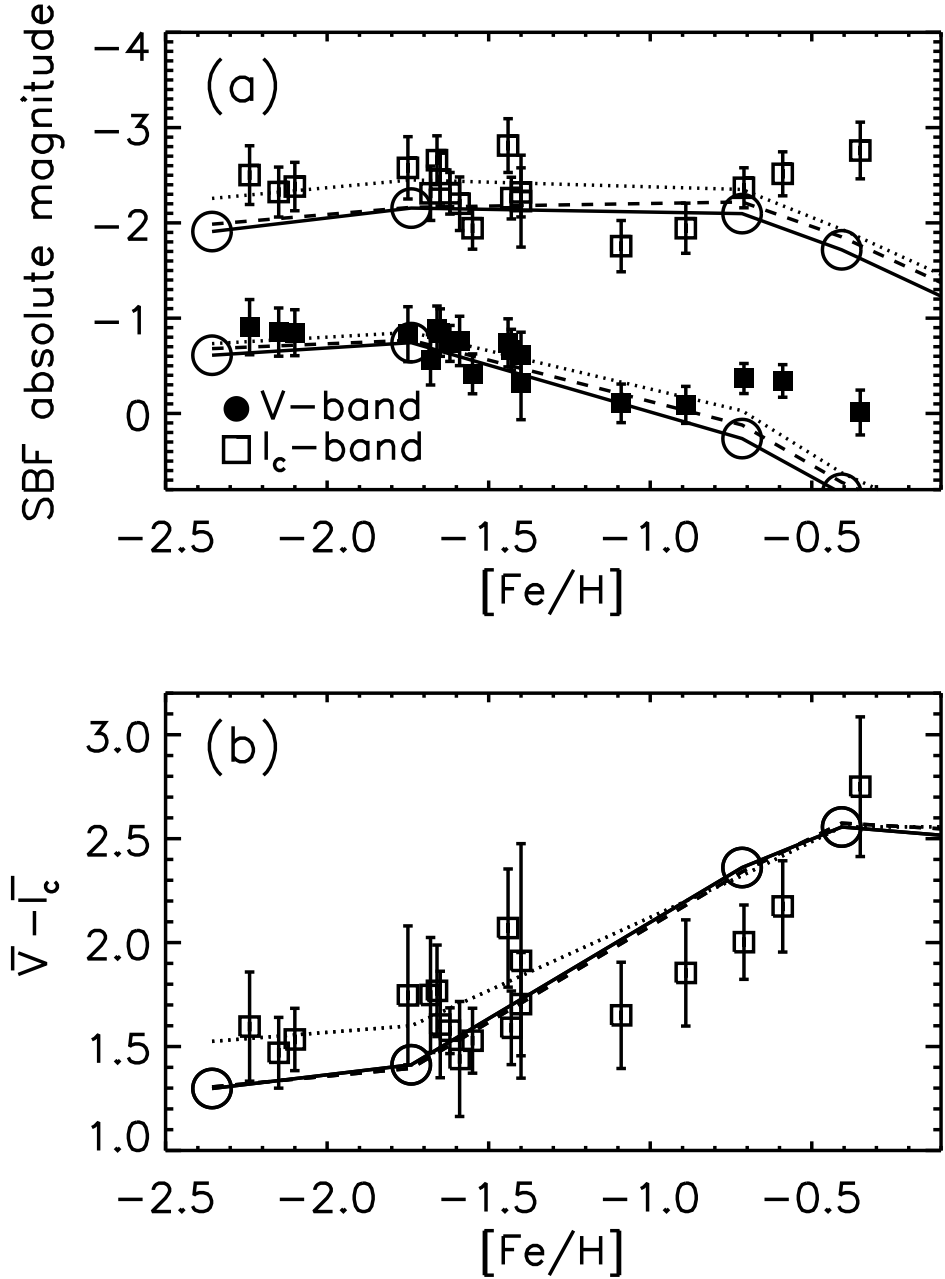}
\bigskip
\caption{\normalsize Comparison of our standard models with the observed
$V$ and \Ic\ fluctuation magnitudes ({\sl top}) and colors ({\sl
bottom}) of Milky Way globular clusters. The data are from Ajhar \etal\
(1994), adjusted to the Hipparcos distance scale from Carretta \etal\
(1999).  The models are defined at [Fe/H] = --2.4, --1.7, --0.7, and
--0.4, with the 12~Gyr models plotted as open circles ($\circ$).  We
have drawn lines connecting the 5~Gyr (dotted line), 8~Gyr (dashed line)
and 12~Gyr (solid line) models to guide the eye.
\label{Mbar-v-feh}}
\end{figure}

SBF measurements of globular clusters are useful tests of the models for
two main reasons: (1) these systems are believed to comprise stellar
populations with internally homogenous metallicity and age, and (2) they
extend to lower metallicities than the available galaxy sample.  The
only observations available are those of \citet{1994ApJ...429..557A}.
Figure~\ref{Mbar-v-feh} shows their $\Vbar$ and $\Ibar$ data as a
function of [Fe/H] compared to our standard models.\footnote{The values
of [Fe/H] for the data in Figure~\ref{Mbar-v-feh} are those tabulated in
\citet{1994ApJ...429..557A}, which are from
\citet{1985ApJ...293..424Z}. Recently \citet[][hereinafter
CG97]{1997A&AS..121...95C} have obtained high-dispersion spectra for a
subset of the \citet{1985ApJ...293..424Z} clusters to derive more
accurate metallicities. They offer a quadratic equation for transforming
from the [Fe/H] of \citet{1985ApJ...293..424Z} to their measurements.
The effect of changing to the CG97 scale is to slightly increase
($\delta$[Fe/H]~$\approx0.1-0.2$) the metallicity of clusters with
[Fe/H]~$\lesssim$~--1.0 and to slightly decrease
($\delta$[Fe/H]~$\approx$~0.1) the metallicity of more metal-rich
clusters.  The net effect on Figure~\ref{Mbar-v-feh} is small: the
[Fe/H] range of the data is slightly compressed, and the bulk of the
points are shifted by $\approx0.15-0.20$~dex to higher [Fe/H].}  We have
updated the distances used by \citet{1994ApJ...429..557A} to the {\sl
Hipparcos} distance scale from \citet{car00}; the net result is to make
the SBF magnitudes $\approx$0.35~mag brighter than reported in the
original paper.  The agreement between models and data is generally
good, though the models are perhaps too faint by $\approx0.2$~mag for
[Fe/H]~$\approx$~--0.7 ($Z=0.004$).  At higher metallicities, the models
also seem to be too red in \Vbar--\Ibar\ by about the same amount. (See
also \S~\ref{section-sbfcolors}.)

\subsection{\Ibar\ Measurements of Galaxies \label{section-ibar-compare}}

\begin{figure}
\centering
\hbox{\hskip -0.2in
\includegraphics[width=3.7in,angle=0]{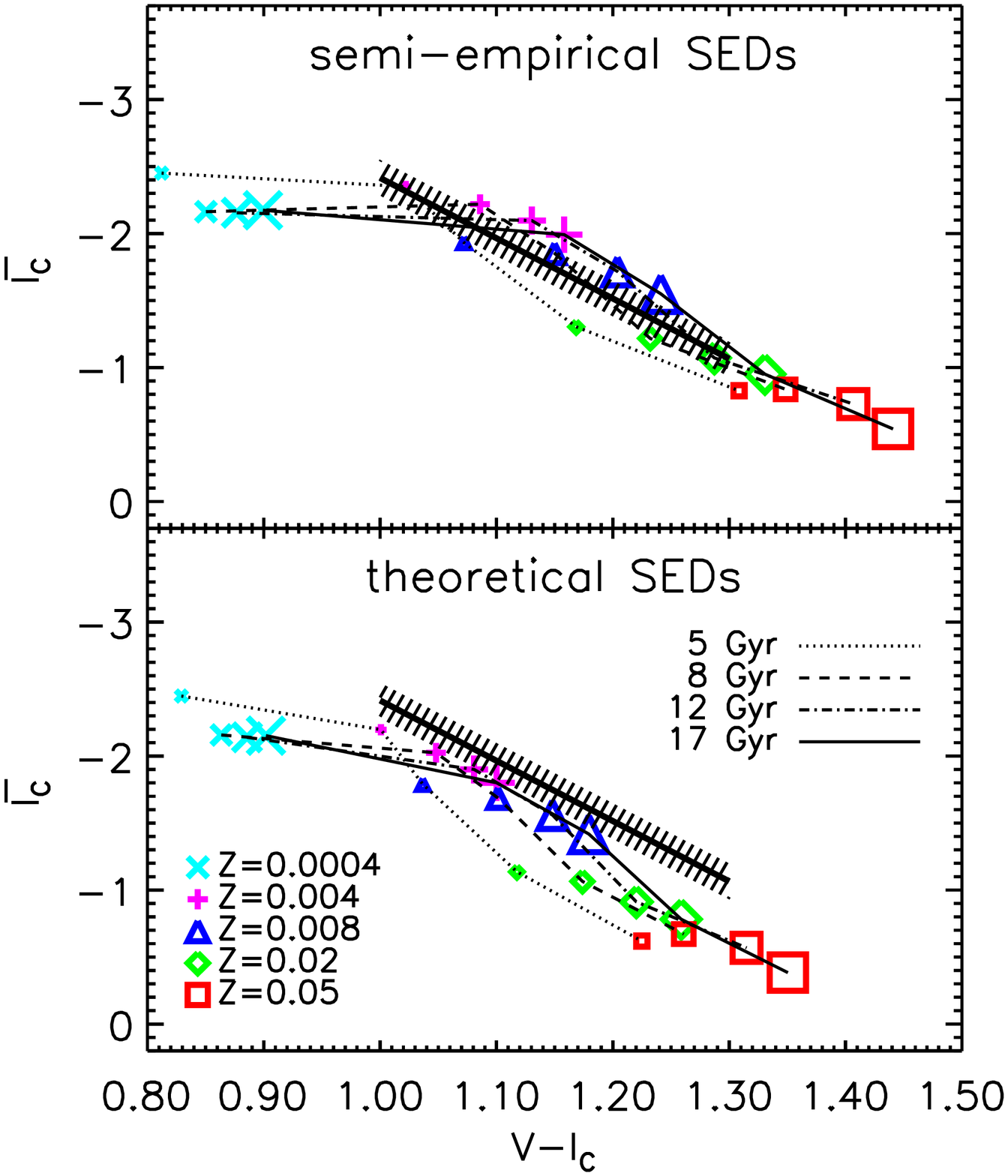}
\hskip 0.2in
\includegraphics[width=3.7in,angle=0]{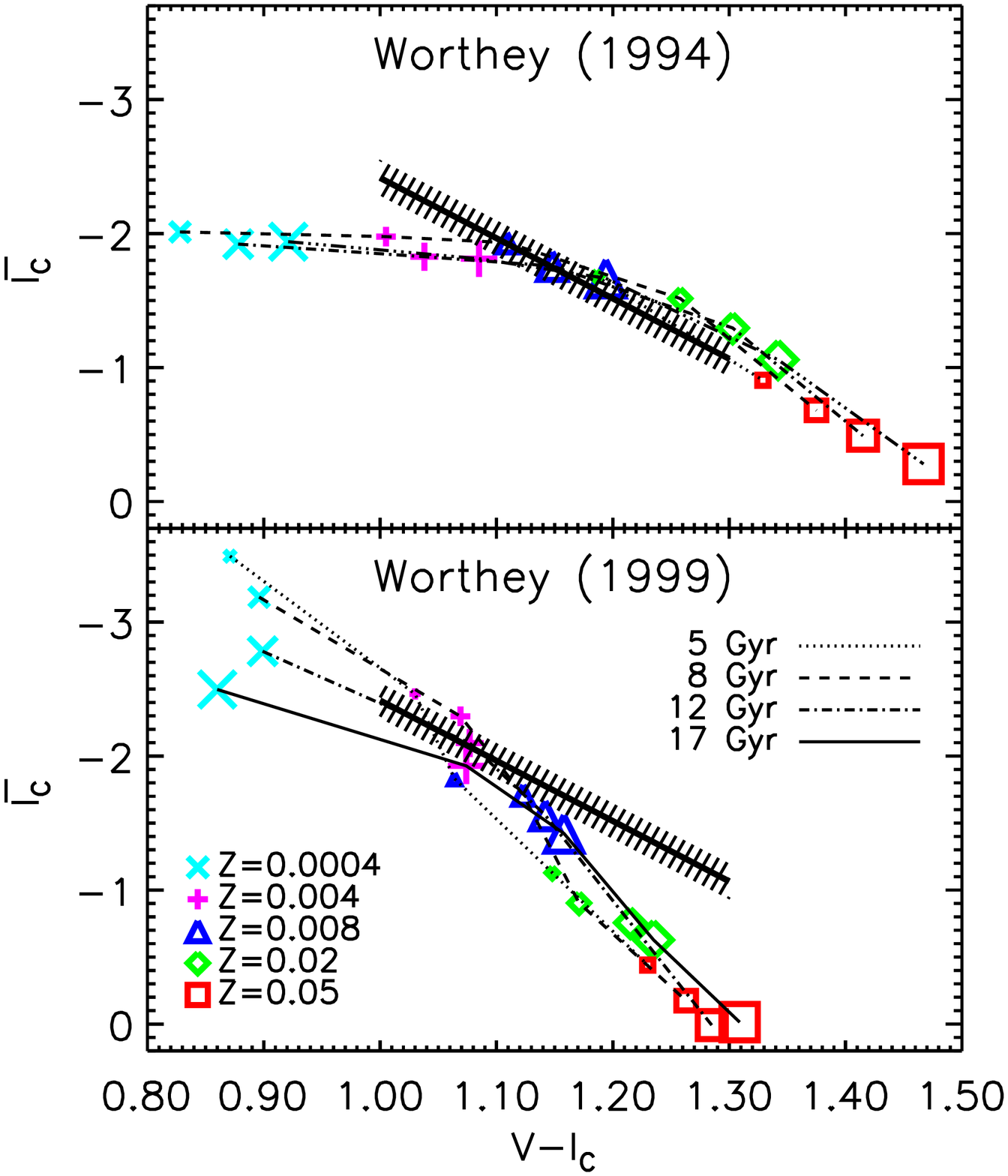}}
\vskip 0.5in
\caption{\normalsize Comparison of different models with the empirical
calibration of \Ibar\ as a function of $V-\Ic$ color (heavy line; Tonry
\etal\ 1997, 1999). The hatched region represents the 1$\sigma$ spread
in the empirical calibration. Models of a given metallicity have the
same symbol, with increasing symbol size representing increasing age.
Lines connect models with the same age. {\bf Left:}~Our model
predictions using Padova tracks with semi-empirical ({\sl top}) and
theoretical ({\sl bottom}) SEDs.  {\bf Right: }~Worthey (1994) models
({\sl top)} and Worthey (1999) models ({\sl bottom}). The Worthey (1999)
models use the same evolutionary tracks as our models for stars from the
main-sequence through the early AGB. The same symbols refer to models
with the same ages and metallicities in all the diagrams.
\label{ibar-compare}}
\end{figure}

Since the $I$-band SBF data comprise by far the largest set of
measurements to date, they serve as the most stringent test of the
models.  The calibration of the \Ibar\ zeropoint by Tonry \etal\ (1997)
relied on observations of early-type galaxies belonging to seven galaxy
groups with \HST\ Cepheid distances, and a much larger sample of
galaxies was used to determine the slope of the \Ibar\ versus $V-\Ic$
relation. In Figure~\ref{ibar-compare}, we show the empirical
calibration for \Ibar\ as a function of $V-\Ic$ galaxy color derived by
\citet{1997ApJ...475..399T}. This relation has been re-calibrated by
\citet{ton99} based on the bulges of six spiral galaxies with both SBF
and \HST\ Cepheid distances; the numerical results are the same as their
initial determination. A slightly different calibration of the same
\Ibar\ data was derived by \citet{fer99a}, leading to a 0.05~mag
brighter zeropoint.

Also shown in Figure~\ref{ibar-compare} are the results of our standard
models (Padova tracks with semi-empirical SEDs) and those of models with
the Padova tracks but with the theoreteical SEDs. The results for
$\Ibar$ from our standard models are in excellent agreement with the
empirical calibration. The slope fitted to the models
(\S~\ref{distance-calibration} and Table~\ref{table-calcobs}) is 4.56,
to be compared with the value $4.50\pm0.25$ measured. The
model-predicted zeropoint at $V-\Ic=1.15$ is --1.79~mag, to be compared
with the value of --1.74$\pm$0.07~mag derived empirically by
Tonry~et~al. In contrast, the models with theoretical SEDs miss the
observational slope by $\approx0.05-0.10$~mag in $V-\Ic$ color.
Therefore, it appears that the corrections introduced by LCB97 to match
the observed color-temperature relations of local stars are also useful
in improving the agreement with \Ibar\ observations.

We note, however, that the slope predicted from our standard models for
$\overline{F814W}$ as a function of $V-\Ic$ is 4.42, while the slope
measured by \citet{1997AJ....114..626A} from 16 galaxies is
$6.5\pm0.7$. (See discussion in \citealp{fer99a} about this point.)  A
comparison (not plotted) with the \Ihstbar\ measurements also shows our
standard models match better than models using the theoretical SEDs.

Figure~\ref{ibar-compare} shows that both \Ibar\ and $V-\Ic$ vary with
age and metallicity, although \Ibar\ depends more strongly on
metallicity than age.  At the very lowest metallicities
($Z\lesssim0.004$), the model \Ibar\ magnitudes flatten out for
$V-\Ic\lesssim1.1$.  This reflects the constant tip magnitude and small
color changes of the RGB as a function of metallicity at low [Fe/H],
which is seen in Galactic globular clusters \citep{1990AJ....100..162D}.
\citet{1997ApJ...475..399T} observed such a deviation in the \Ibar\ data
for the dwarf elliptical M~110 (NGC~205); regions with
$V-\Ic\approx0.8-1.0$ had \Ibar\ fainter than expected from
extrapolating the linear relation.

For comparison, we also show in Figure~\ref{ibar-compare} SBF
predictions by \citet{1994ApJS...95..107W} and \citet{wor99}.  We used
Worthey's on-line interpolation
engine\footnote{http://199.120.161.183/\~{}worthey/dial/dial\_a\_model.
Note that this version updates the models in \citet{1994ApJS...95..107W}
by including a more realistic treatment of the horizontal branch for
[Fe/H]$<-1.0$.}  to interpolate his results onto our metallicity grid.
One striking property of these models is the small spread in the model
locus, implying that variations in age and metallicity are very
degenerate.  The good agreement of the \citet{1994ApJS...95..107W}
[Fe/H]~=~--0.50 to +0.50 models with the observations was taken by
\citet{1997ApJ...475..399T} as validation of their empirical
calibration.  However, it is worth pointing out that the
\citet{1994ApJS...95..107W} results for \Ibar\ are too faint compared to
the data for $V-\Ic\lesssim1.1$, unlike our models. For
$V-\Ic\gtrsim1.1$, the \citet{1994ApJS...95..107W} models agree fairly
well with our models, though there are notable differences in the values
of \Ibar\ and $V-\Ic$ from the two set of models for any given age and
metallicity.  Also shown in Figure~\ref{ibar-compare} are the
\citet{wor99} models, which are based on the same Padova tracks as our
standard models for stars from the main-sequence through the early AGB.
The results from his updated models differ significantly from our own
and those of \citet{1994ApJS...95..107W}. Moreover, they do not match
the data very well.

\subsection{\Kpbar\ Measurements of Galaxies \label{section-kbar-compare}}

\begin{figure}
\centering
\vskip -0.2in
\includegraphics[width=4.5in,angle=0]{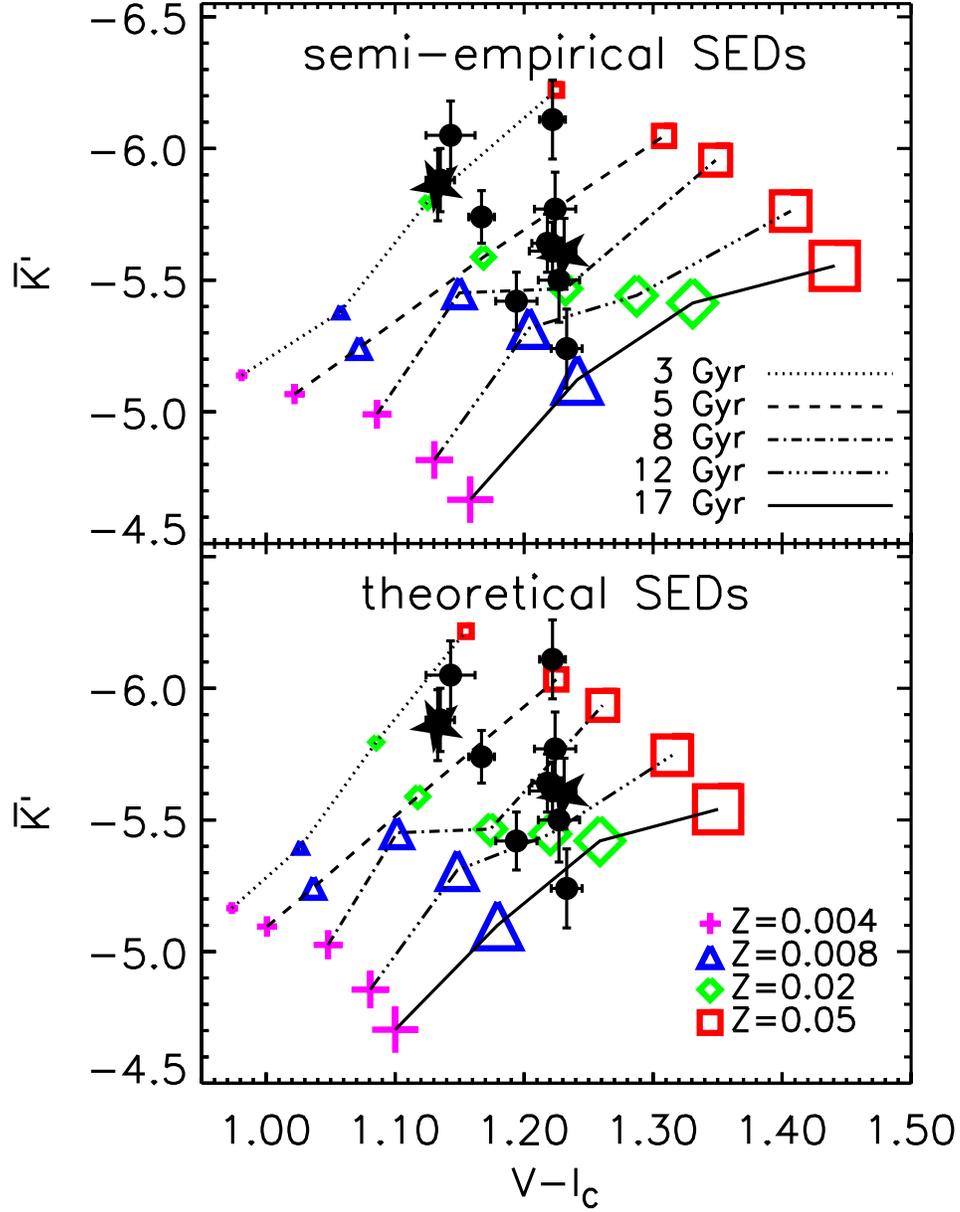}
\vskip 0.5in
\caption{\normalsize Comparison of different models with the \Kpbar\
observations. The models are computed using the Padova evolutionary
tracks with the semi-empirical SEDs (our standard models; {\sl top}) and
the theoretical SEDs ({\sl bottom}).  The high-S/N observations for
ellipticals in Virgo, Eridanus, and Fornax (Jensen \etal\ 1998) are
plotted as circles ($\bullet$), and those for M~32 and the bulge of M~31
(Luppino \& Tonry 1993) are plotted as stars ($\large \star$). The
colored model symbols are the same as in
Figure~\ref{ibar-compare}. Models with the same metallicity have the
same symbol, with increasing symbol size representing increasing ages.
Lines connect models of the same age at 3, 5, 8, 12, 17~Gyr, with the
oldest models having the reddest $V-\Ic$ colors. \label{kbar-compare}}
\end{figure}

In Figure~\ref{kbar-compare}, we compare our standard models to \Kpbar\
data from the literature.  The ranges of ages and metallicities in the
models are the same as in Figure~\ref{ibar-compare}. The observations
comprise the good S/N measurements of Fornax, Virgo, and Eridanus
galaxies from \citet{1998ApJ...505..111J}, using updated $I$-band SBF
distances to the galaxies from \citet{ton00}; this sample is composed
almost entirely of luminous galaxies ($M_B\lesssim-20$). We do not 
include the lower S/N data of Jensen \etal\ given the concerns about
systematic biases in SBF measurements at low S/N
\citep[see][]{1996ApJ...468..519J}. For the same reason, we do not
include data from \citet{1994ApJ...433..567P}, and also because these
authors did not correct their SBF measurements for globular cluster
contamination.  Finally, we include $\Kpbar$ data for M~31 and M~32 from
\citet{1993ApJ...410...81L}.

Unlike the \Ibar\ data in Figure~\ref{ibar-compare},
Figure~\ref{kbar-compare} shows that the \Kpbar\ data do not strongly
favor our standard models over the models with the theoretical SEDs.
Also in contrast to \Ibar, the effects of age and metallicity in the
\{$V-\Ic$, \Kpbar\} plane are more distinct, as can be seen by the
roughly orthogonal orientations between models with common ages and
those with common metallicities. (This effect was also noted by
\citealp{1993ApJ...409..530W}).  The \Kpbar\ models become slightly
fainter with increasing age. This is expected from the core
mass-luminosity relation for AGB stars.

The observations in Figure~\ref{kbar-compare} span a range of model ages
of about a factor of three, and the galaxies with the brightest \Kp-band
SBFs have SSP-equivalent ages near 3~Gyr.  The inferred metallicities
are roughly solar, which is somewhat higher than the value of
$\approx$~0.5\Zsun\ inferred by \citet{1998ApJ...505..111J} using the
\citet{1994ApJS...95..107W} models.\footnote{\citet{kun00} used
single-burst population synthesis models to interpret the optical
absorption-line strengths of three of the Fornax galaxies in
Figure~\ref{kbar-compare} (NGC~1379, 1399, and 1404). The ages and
metallicities of these galaxies inferred in Figure~\ref{kbar-compare}
from our models are comparable to his findings.}  At these ages and
metallicities, the brightest AGB stars are M-type stars.  The data in
Figure~\ref{kbar-compare} also show that the bluer ($V-\Ic\lesssim1.2$)
galaxies in the sample tend to have brighter \Kpbar. The implication
from the models is that the stellar populations dominating the
$\Kp$-band SBFs in the bluer galaxies are younger than those in the
redder galaxies; in addition, the bright \Kpbar\ of the bluer galaxies
suggests they also have roughly solar metallicities.

\subsection{Optical/IR Fluctuation Colors of Galaxies and Globular
Clusters \label{section-sbfcolors}}

SBF color measurements provide another interesting test of the
models. Since the fluctuations are almost entirely dominated by RGB and
AGB stars, SBF colors are a better tracer of the giant branch colors
than the integrated light.  In particular, unlike the integrated colors
used above, SBF colors are independent of model uncertainties in the
main sequence, subgiant branch, and horizontal branch stars. Also from a
practical standpoint, SBF colors are distance-independent (neglecting
the $k$-corrections). Thus they avoid any systematic errors from
uncertainties in the galaxy distances, which are inherent in using
absolute fluctuation magnitudes. However, the number of galaxies to date
with multi-color SBF data is relatively small.

\begin{figure}
\centering
\vskip -0.3in
\hskip -0.75in
\vbox{
\includegraphics[width=4in,angle=90]{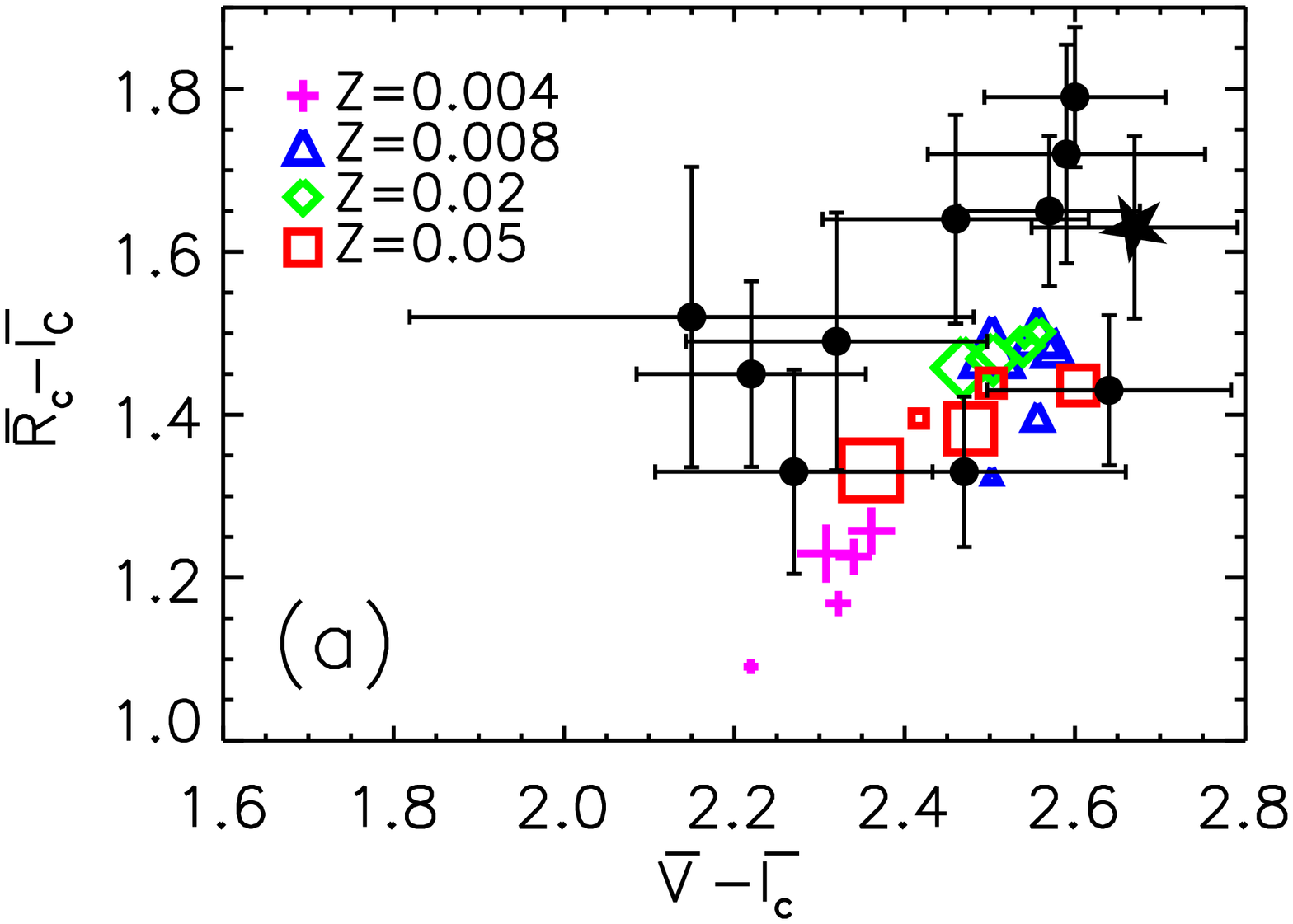}
\vskip -0.3in
\includegraphics[width=4in,angle=90]{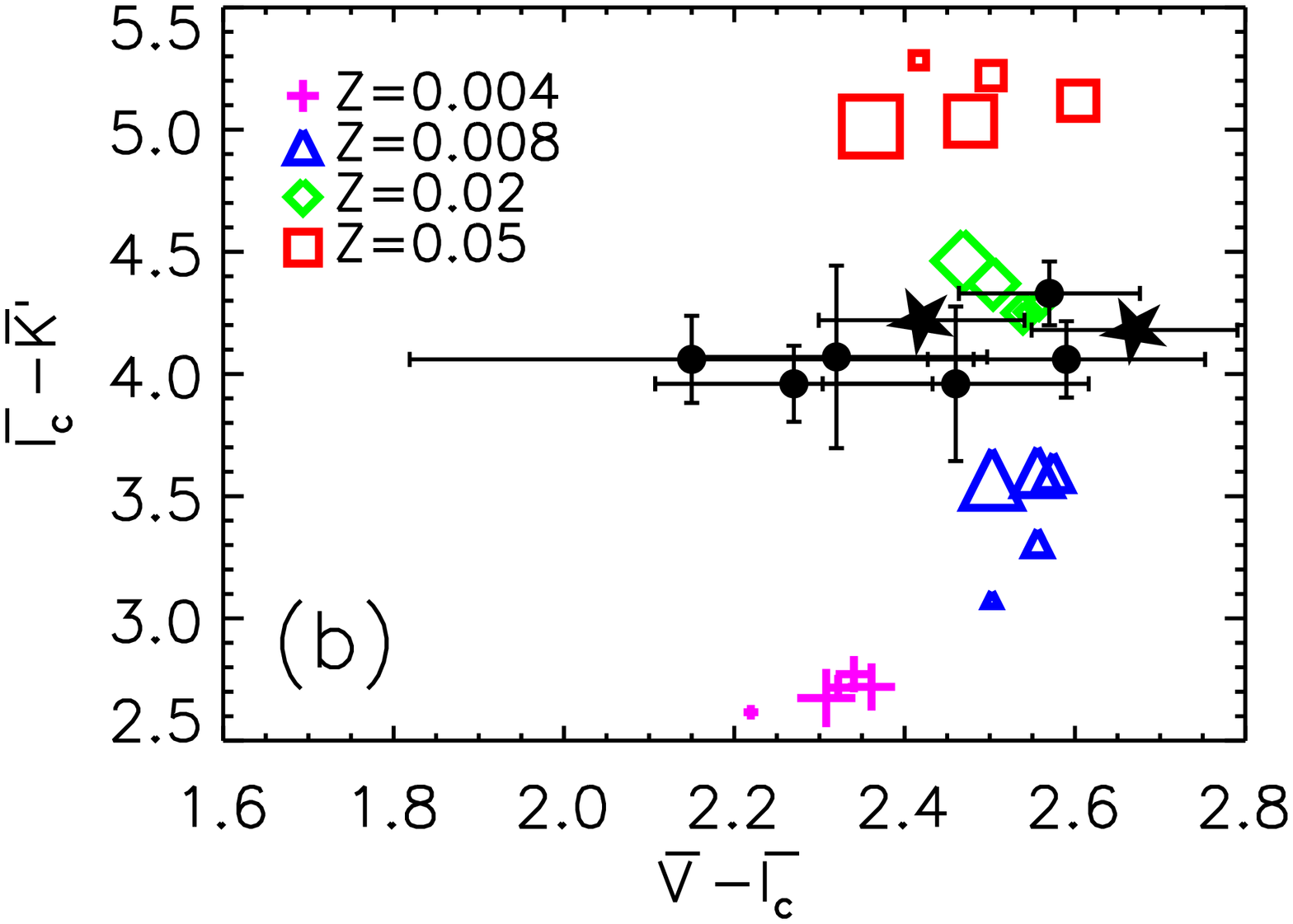}}
%\vskip -0.3in
\caption{\normalsize Comparison of our standard models with observed
optical/IR fluctuation colors. Virgo cluster galaxies are plotted as
circles ($\bullet$).  Local Group (M~32 in the top panel, M~31 and M~32
in the bottom panel) galaxies are plotted as stars ($\large
\star$). Models with the same metallicity have the same symbol, with
increasing symbol size representing increasing age \hbox{(3, 5, 8, 12,
17~Gyr)}. The galaxies with the bluest \Vbar--\Ibar\ color is NGC~4365.
\label{sbfcolors}}
\end{figure}

We compare our standard models with galaxies with good optical/IR SBF
color data in Figure~\ref{sbfcolors}.  The data are for M~31, M~32, and
Virgo cluster galaxies. The IR data are the same as in
Figure~\ref{kbar-compare}, and the optical data are from
\citet{1990AJ....100.1416T}, with some minor revisions from
\citet{ton00}.  For the \Vbar~\Rbar~\Ibar\ colors, most of the models
overlap each other (Figure~\ref{sbfcolors}a). This happens because of
the turnover of the tip of the giant branch and the increasing
contribution from core-He burning stars with increasing metallicity
\citep{1993ApJ...409..530W, 1994ApJ...429..557A}. Therefore, these
colors are largely degenerate to age and metallicity.  The optical/IR
(\Vbar~\Ibar~\Kpbar) model colors (Figure~\ref{sbfcolors}b) show the
same reversal as described above in the \Vbar--\Ibar\ color, but the
\Ibar--\Kpbar\ color increases monotonically with metallicity.  This
plot clearly demonstrates that the \Ibar--\Kpbar\ color is predicted to
be very metallicity-sensitive and highly age-insensitive. We return to
this point in \S~\ref{section-age-z}.

There is generally good agreement between our standard models and the
data.  The comparison with observed optical SBF colors in
Figure~\ref{sbfcolors}a indicates that our predictions for \Vbar--\Ibar\
are consistent with the data, while the predicted \Rbar--\Ibar\ color
may be $\approx$0.1~mag too blue.  However, Figure~\ref{sbfcolors}b
would instead suggest that the model \Vbar--\Ibar\ colors are too red by
$\approx$0.15~mag, an offset which is also suggested by comparisons with
globular cluster SBF colors (\S~\ref{globs}).  Thus, the models have
some difficulty in matching all the observed SBF colors simultaneously
to better than 0.1--0.15~mag.  \citet{1999phcc.conf..181B} did a similar
comparison using the \citet{1994ApJS...95..107W} models and found a
disagreement of about 0.3~mag between the \Vbar--\Ibar\ model colors and
the data, which they suggested was evidence for multi-metallicity
stellar populations in ellipticals.  Our models do not show this large a
discrepancy.  Furthermore, we point out that the relatively small
wavelength leverage of the \Vbar--\Ibar\ color is not a very strong
discriminant --- the \Bbar--\Ibar\ color should more sensitive to
populations of composite metallicity in ellipticals.  Past constraints
on such populations from \Vbar~\Rbar~\Ibar\ colors alone
\citep{1993A&A...275..433B, 1993ApJ...409..530W} could likely be
improved by expanding the wavelength range considered.

The agreement between observations and models is worsened if we use
theoretical SEDs in our models instead of semi-empirical ones. We find
that for \Rbar--\Ibar\ and \Vbar--\Ibar, the theoretical SEDs produce
model SBF colors $\approx$0.2--0.3~mag bluer relative to the
semi-empirical SEDs. In addition, the \Ibar--\Kpbar\ colors are
$\approx$0.2~mag too blue compared to the data.

Even though the number of available measurements is small, the SBF
colors have relatively little scatter from galaxy to galaxy.  The small
observed ranges of \Ibar--\Kpbar\ and \Vbar--\Ibar\ colors
(Figure~\ref{sbfcolors}b) suggests that the aperture-averaged
metallicities of the brightest giant stars are roughly consistent from
galaxy to galaxy. Also, the Local Group galaxies (M~31 and M~32) have
SBF colors comparable to those of the Virgo galaxies, despite the very
different physical apertures used in the measurements, since Virgo is
about 20 times farther away than M~31 and M~32.  This implies that the
effects of any SBF color gradients in these galaxies are small.

\begin{figure}
\centering
\hbox{\hskip -0.5in
\includegraphics[width=5in,angle=90]{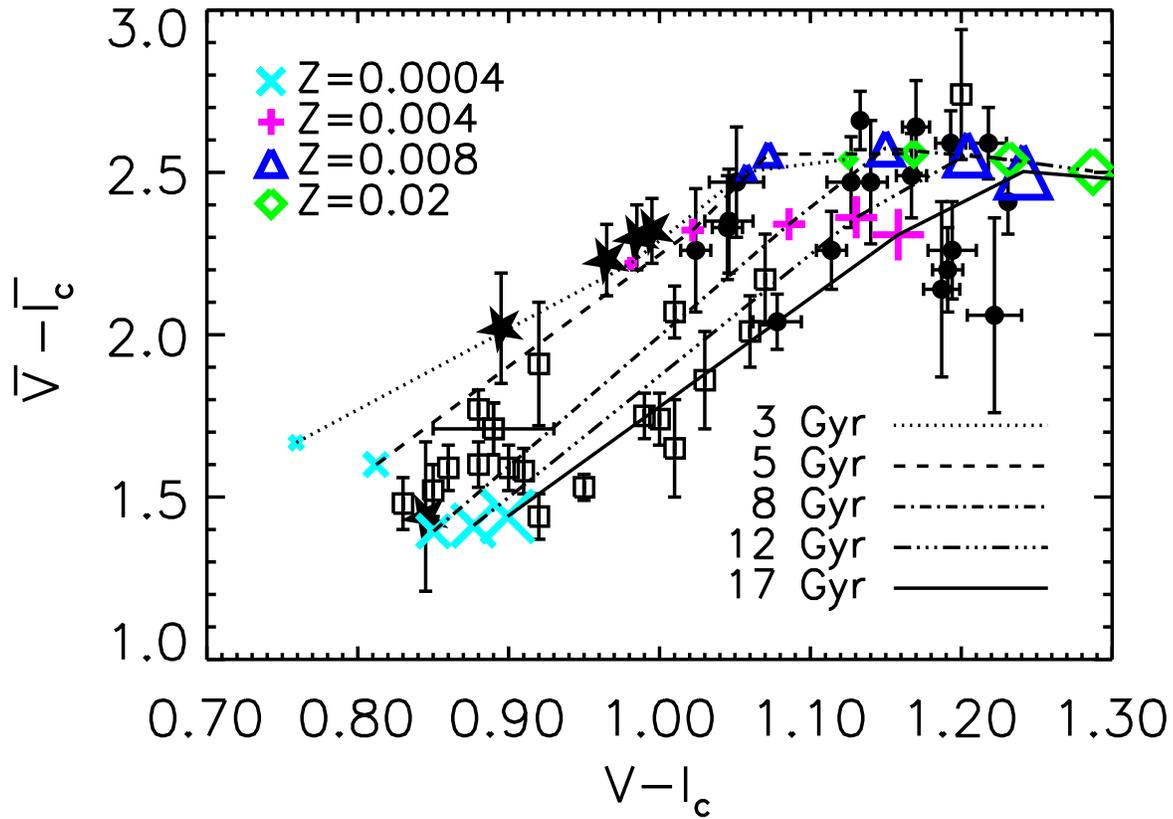}}
\caption{\normalsize Comparison of our standard models with observations
of \Vbar--\Ibar\ fluctuation color as a function of integrated $V-\Ic$
color.  Milky Way globular clusters are shown as squares ($\Box$) and
nearby early-type galaxies as circles ($\bullet$).  Different regions of
NGC~205 are shown as stars ($\star$).  For clarity, errors for the
globular cluster $V-\Ic$ colors are shown for only one cluster, chosen
to have the median error of the sample. Most of the galaxy sample is
from the Virgo cluster. \label{VIbar-v-VI}}
\end{figure}

In Figure~\ref{VIbar-v-VI}, we compare the \Vbar--\Ibar\ and $V-\Ic$
colors from our standard models with observations of both galaxies and
globular clusters. The globular cluster data are the same as in
\S~\ref{globs} and Figure~\ref{Mbar-v-feh}. The galaxy data are from
\citet{1994ApJ...429..557A} with some revisions from \citet{ton00}; the
galaxies have $M_B\approx-18.5$ to $-21.5$ and belong mostly to the
Virgo cluster.  The reddening corrections adopted for the galaxies have
been changed to those of \citet{1998ApJ...500..525S}.  Overall,
Figure~\ref{VIbar-v-VI} shows good agreement between our standard models
and the data and illustrates the expected metallicity dichotomy between
Galactic globular clusters and elliptical galaxies.  Interestingly, the
bulk of the galaxy data suggests a large range in age with a small range
in metallicity (mostly sub-solar), which is reminiscent of the trends
found in the smaller \Kpbar\ dataset. For the globular clusters, the
$V-\Ic$ measurement errors are much larger than for the galaxies, and
the observed color spread is consistent with little age dispersion.

\section{Discussion}

\subsection{Systematic Effects of Model Uncertainties}

In trying to extract information about the stellar populations of
galaxies using population synthesis models, it is important to explore
the effects of changing the model inputs. We have shown that our
standard models can reproduce most of the current observations, but some
slight offsets between the models and data (\eg,
Figs.~\ref{Mbar-v-feh} and~\ref{sbfcolors}) could be evidence either
for systematic errors in the model inputs or for real physical
differences between actual galaxy populations and SSPs.

The freedom of choice in the evolutionary tracks (Padova or Geneva) and
spectral libraries (empirical, semi-empirical and theoretical) of our
models allows us to explore the range of uncertainties in our SBF
predictions.  There are appreciable differences in the predicted
fluctuation magnitudes depending on the choice of model inputs, though
the differences do not follow obvious trends in metallicity, age or
bandpass.

\subsubsection{Multi-Metallicity Models: Choice of SEDs}

In comparing the results of multi-metallicity models from the Padova
tracks combined with either the semi-empirical or theoretical SEDs, the
most noticeable differences arise in the $I$-band region (\Ic\ and
$F814W$).  For all metallicities $Z\ge0.004$, \Ibar\ obtained from the
semi-empirical SEDs are about 0.1--0.2~mag brighter than those obtained
from the theoretical SEDs.  Similar differences occur over smaller
metallicity ranges for \Jbar\ ($Z\ge0.02$) and \Hbar\ ($Z=0.004-0.008$),
but with the semi-empirical predictions now being fainter than the ones
from the theoretical SEDs. Since the predicted SBF colors therefore
differ by 0.2--0.4~mag in this metallicity range, \Ibar--\Jbar\ and
\Ibar--\Hbar\ data could provide an independent test of the LCB97 SED
corrections (\S~\ref{bc2000}).  The SBF magnitudes from the two sets of
SEDs otherwise agree at the $\lesssim$0.1~mag level. Similarly, the
integrated colors and line indices obtained with both sets are
comparable, with the semi-empirical SEDs giving slightly redder $V-\Ic$
colors than the theoretical ones.

\subsubsection{Solar-Metallicity Models: Choice of Evolutionary Tracks
and SEDs} 

Table~\ref{table-solar-diff} provides the SBF magnitudes, integrated
colors, and line indices from the six possible types of
solar-metallicity models we can investigate from our two sets of
evolutionary tracks (Padova and Geneva) and three sets of SEDs
(theoretical, semi-empirical, and empirical).  Note that, {\em by
construction,} the theoretical SEDs should produce less accurate results
than the semi-empirical SEDs, which have been adjusted to match
observations of solar-neighborhood stars. There are considerable
($\gtrsim0.3$~mag) variations in the fluctuation magnitudes from
different choices of tracks and spectral libraries.  For a given set of
evolutionary tracks (Padova or Geneva), the predictions of the three
different sets of SEDs are relatively consistent, with differences at
the $\lesssim$0.15~mag level in all bandpasses (except for the $M$-band
SBFs from the empirical SEDs).

However, for a fixed set of SEDs, larger differences arise when changing
the stellar evolution prescription. The fluctuation magnitudes from the
Geneva models are up to 0.3~mag brighter in the optical than those from
the Padova models and $\approx0.2-0.5$~mag fainter in the infrared. This
is caused by the significant difference in the relative numbers of RGB
and AGB stars predicted by the two tracks.  (See also discussions in
\citealp{1996fstg.conf...14B} and \citealp{1996ApJ...457..625C}.) The
Geneva models have a much larger AGB contribution to the SBF signal at
solar-metallicity, especially from the early AGB phases.  The result is
that the Geneva tracks produce optical SBFs somewhat brighter and
near-IR SBFs somewhat fainter than the Padova tracks.  In contrast to
these large differences in SBF predictions, the integrated properties
from the Padova and Geneva models at solar metallicity are very
comparable.  \citet{1997AJ....114.1531B} examined the integrated spectra
of near solar-metallicity Galactic Bulge globular clusters and found
good agreement using semi-empirical SEDs with either the Padova or
Geneva tracks.

\begin{figure}
\centering
\hbox{\hskip -2in
\vbox{\includegraphics[width=3.0in,angle=90]{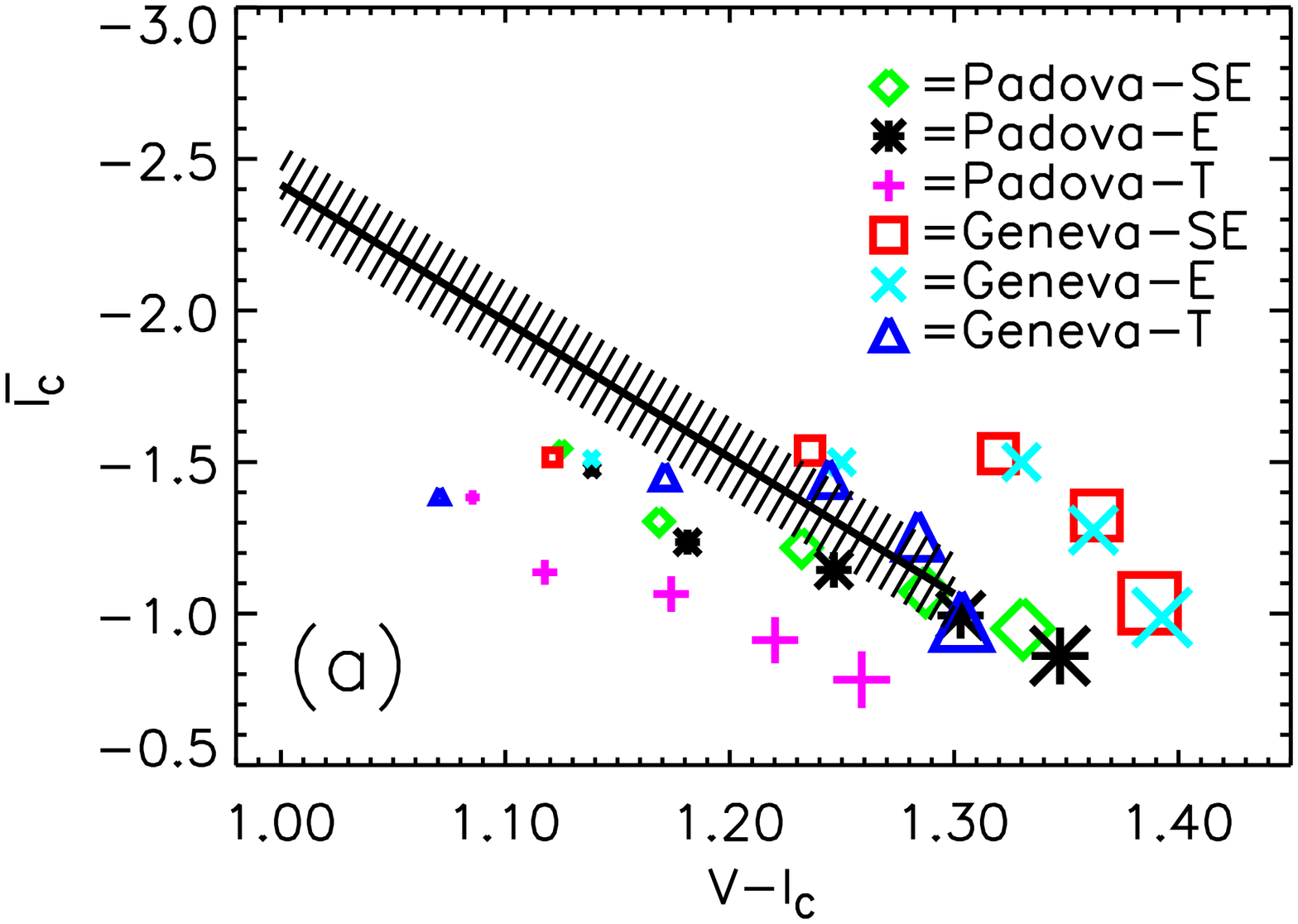}
\vskip -0.2in
\includegraphics[width=3.0in,angle=90]{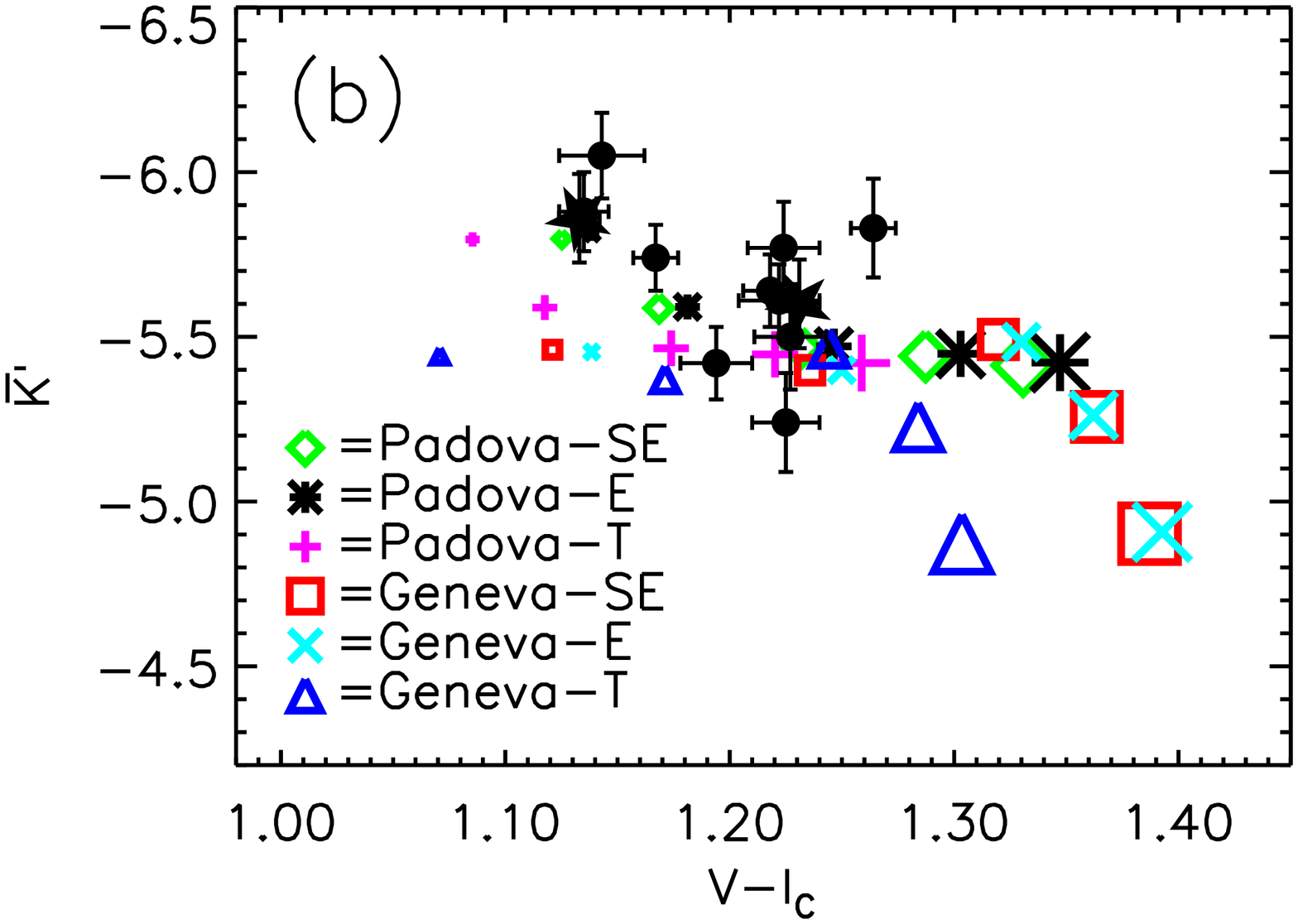}}
\hskip -3.4in
\vbox{\includegraphics[width=3.0in,angle=90]{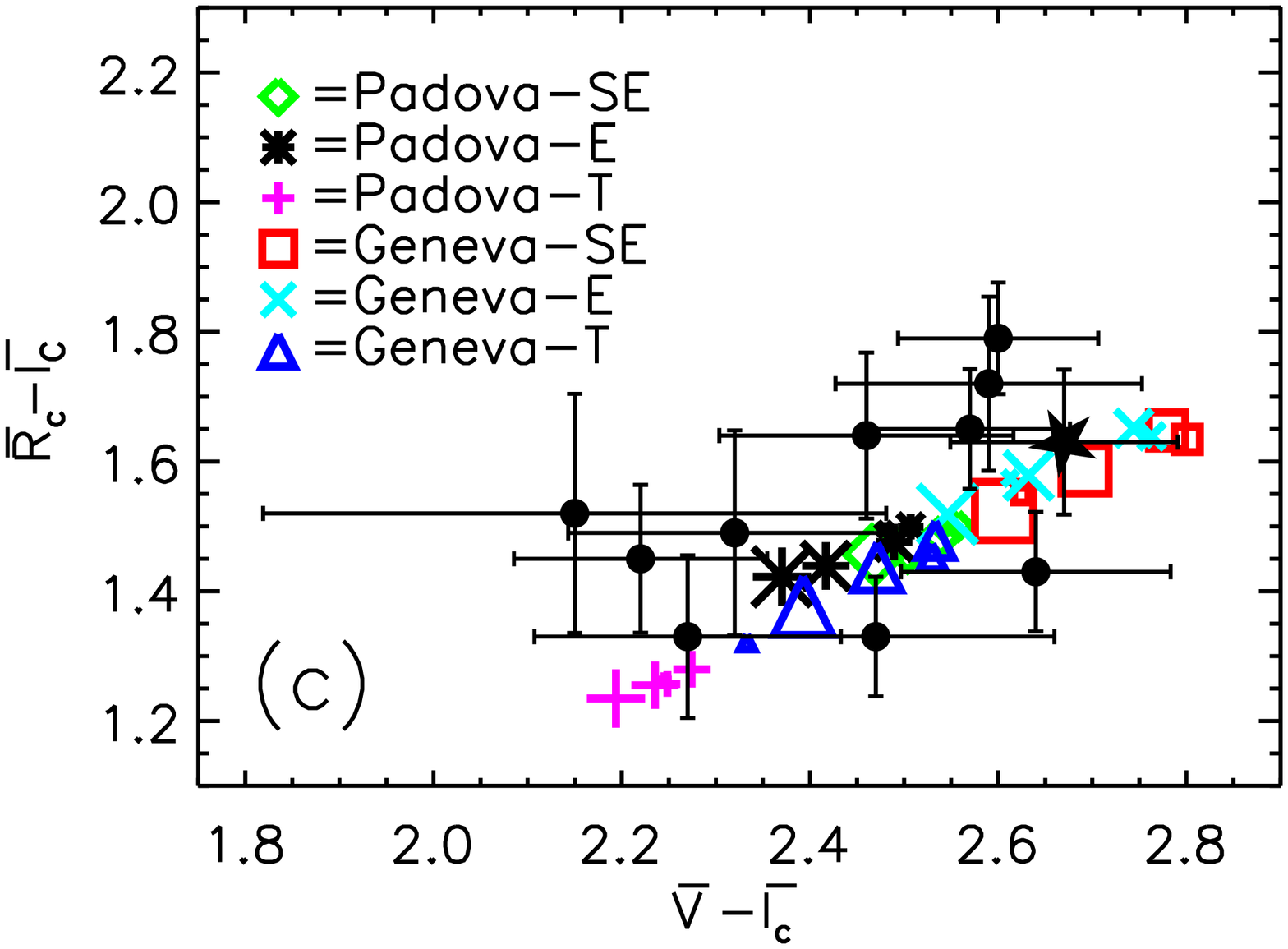}
\vskip -0.2in
\includegraphics[width=3.0in,angle=90]{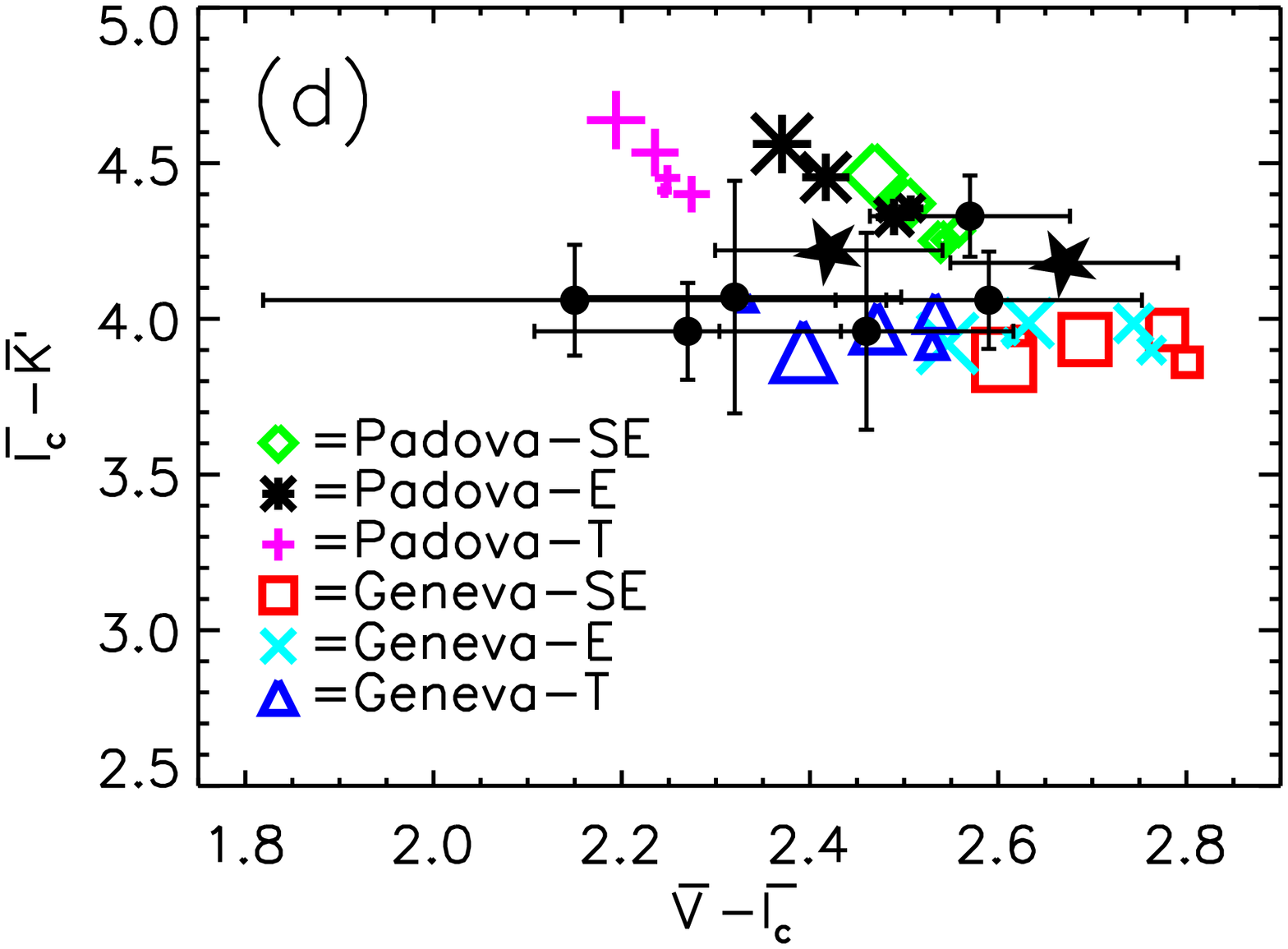}}}
\caption{\normalsize Comparison of our solar-metallicity models using
different combinations of stellar evolutionary tracks (Padova versus
Geneva) and spectral libraries (semi-empirical [SE], empirical [E], and
theoretical [T]) with the observations from Figures~\ref{ibar-compare},
\ref{kbar-compare}, and~\ref{sbfcolors}. All models are shown at ages of
3, 5, 8, 12, 17~Gyr, with increasing symbol size representing increasing
age.  As expected, for a fixed set of evolutionary tracks, the
semi-empirical SEDs give results similar to those from the empirical
SEDs and provide better agreement with the data than the theoretical
SEDs (see \S~\ref{bc2000}).
\label{solarmodels}}
\end{figure}

Figure~\ref{solarmodels} compares all six solar-metallicity models
against the data.  In general, for a fixed set of stellar evolutionary
tracks, the theoretical SEDs produce bluer integrated and SBF colors as
well as a slightly fainter \Ibar; the semi-empirical and empirical SEDs
give very comparable results, which is reassuring.  For a fixed choice
of input SEDs, the Geneva tracks produce slightly redder integrated and
optical SBF colors than the Padova tracks. The Geneva tracks also give
brighter \Ibar\ and fainter \Kbar\, and hence bluer \Ibar--\Kbar.

Overall, the SBF magnitudes (Figure~\ref{solarmodels}a and b) favor the
Padova models, especially the \Ibar\ data. The constraints from SBF
colors are less conclusive.  For the optical colors (\Vbar~\Rbar~\Ibar),
the choice of evolutionary tracks and SEDs are largely degenerate
(Figure~\ref{solarmodels}c), with the solar-metallicity models occupying
a locus similar to that of the multi-metallicity models in
Figure~\ref{sbfcolors}.  For the optical/near-IR colors
(\Vbar~\Ibar~\Kbar), the slight discrepancy noted in
\S~\ref{section-sbfcolors} between our standard models and the data is
not found when adopting the Geneva tracks with the theoretical SEDs
(Figure~\ref{solarmodels}d), though this combination is largely
disfavored by the other comparisons in Figure~\ref{solarmodels}.

\subsection{Constraints on the IMF and Stellar Evolutionary Lifetimes
\label{imf+lifetimes}}

As noted in \S~\ref{section-ibar-compare}, the tightness of the observed
correlation between \Ibar\ and $V-\Ic$ for elliptical galaxies reflects
a degeneracy between the effects of age and metallicity on these
observables. This degeneracy may be useful to constrain variations in
other parameters of the models, such as the IMF or the uncertainties in
the evolution of the cool stars.

\begin{figure}
\centering
\includegraphics[width=5in,angle=90]{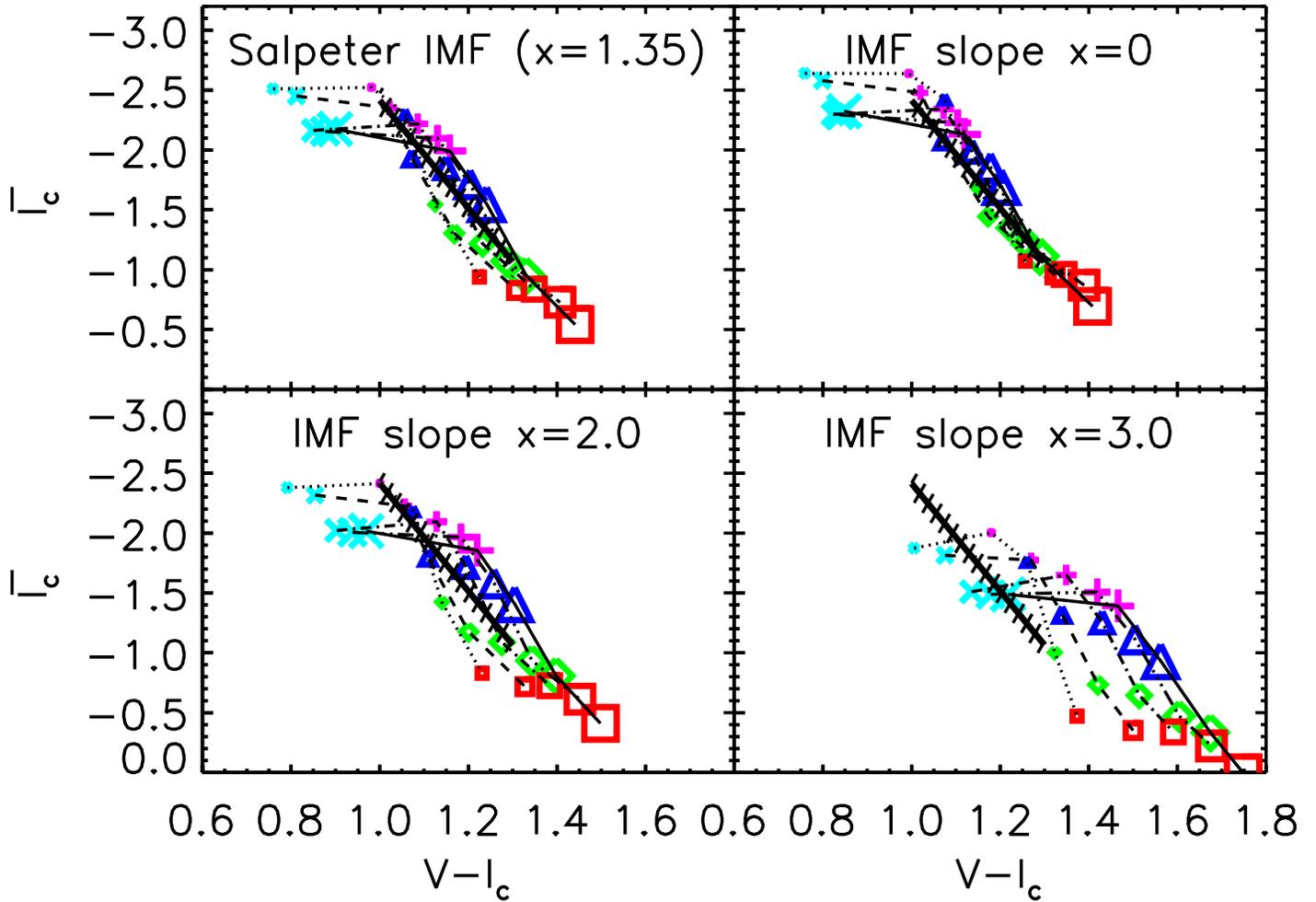}
\vskip 0.5in
\caption{\normalsize Effect of changing the slope of the IMF in our
standard models. The IMF is parameterized as $dN = M^{-(1+x)} dM$, with
$x=1.35$ for the Salpeter (1955) IMF.  The heavy black line and hatched
region represent the 1$\sigma$ spread in the empirical calibration. The
ages and metallicities of the models are the same as in
Figure~\ref{ibar-compare}.  The $I$-band SBF data disfavor an IMF much
steeper than Salpeter.
\label{compare-imf-ibar}}
\end{figure}

The tightness of the observed relation between \Ibar\ and $V-\Ic$ sets a
limit on how different the IMF of elliptical galaxies can be from that
in the solar neighborhood. Figure~\ref{compare-imf-ibar} shows the
effect of changing the IMF slope on the correlation between \Ibar\ and
$V-\Ic$. Lowering the IMF slope relative to Salpeter has only a small
effect; therefore the data are not very sensitive to IMFs flatter than
Salpeter.  A similar conclusion has been reached when purely using
integrated colors to constrain the IMF
\citep[e.g.][]{1978ApJ...222...14T, 1988ARA&A..26...51F}. However,
Figure~\ref{compare-imf-ibar} does suggests that the IMF in observed
elliptical galaxies cannot be much steeper (\ie, more dwarf-dominated)
than the IMF in the solar neighborhood. This interesting constraint from
SBF data complements those on the higher-mass end of the IMF from
chemical abundance and integrated light studies
\citep[e.g.,][]{1992ApJ...398...69W, 1998simf.conf....1K}.

\begin{figure}
\centering
\includegraphics[width=5in,angle=0]{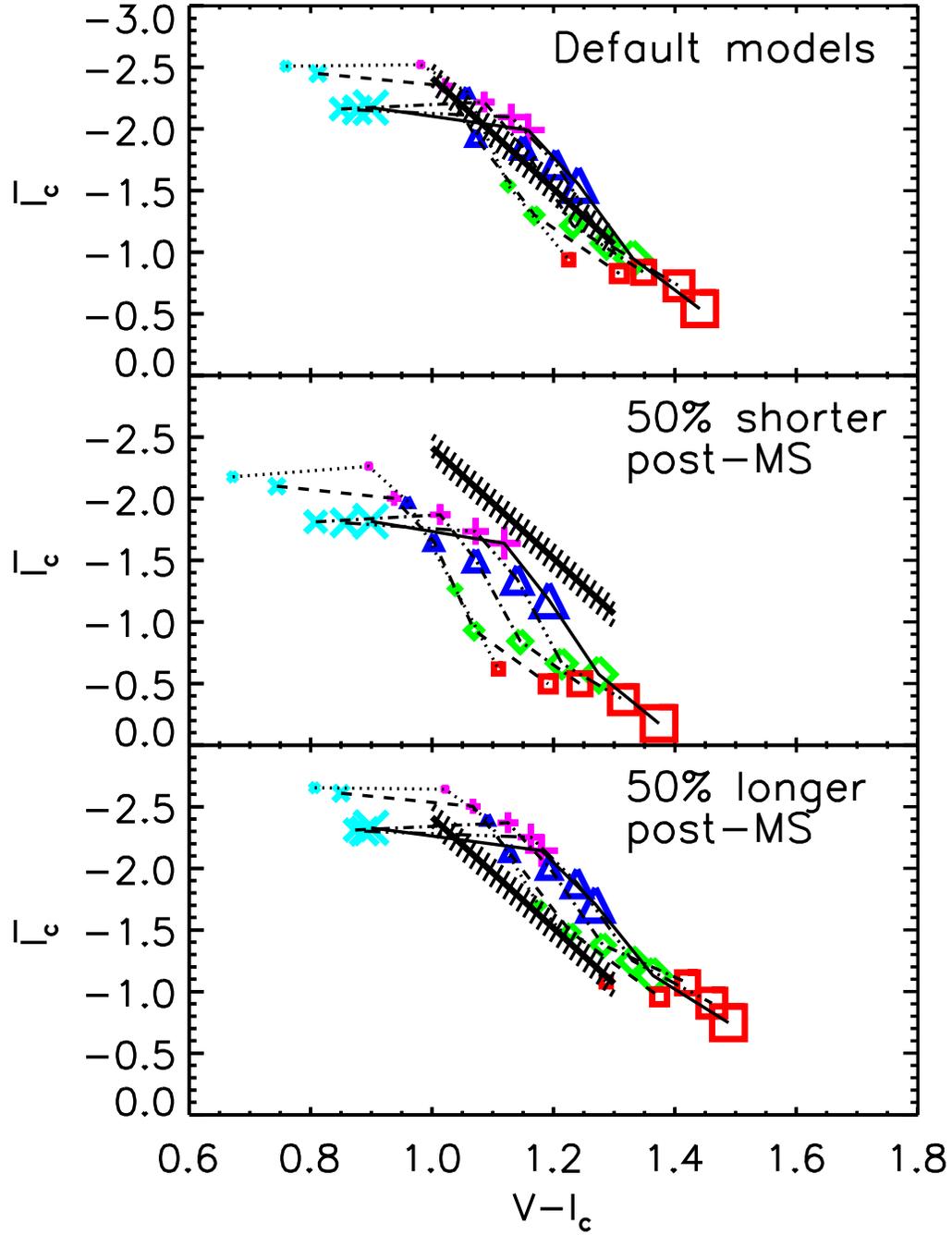}
\vskip 0.5in
\caption{\normalsize Effect of changing the lifetimes of the post-main
sequence evolutionary phases in our standard models.  The heavy black
line and hatched region represent the 1$\sigma$ spread in the empirical
calibration. The ages and metallicities of the models are the same as in
Figure~\ref{ibar-compare}.  The $I$-band SBF data indicate that the
post-main sequence evolutionary lifetimes in the models cannot be in
error by more than $\pm$50\% over wide ranges in age and
metallicity. \label{lifetimes-ibar}}
\end{figure}

Since the evolutionary lifetimes of low- and intermediate-mass stars are
still a major uncertainty in population synthesis models
\citep{1996ApJ...457..625C}, it is also revealing to investigate how
these are constrained by the tight correlation between \Ibar\ and
$V-\Ic$ color.  Figure~\ref{lifetimes-ibar} illustrates the effect of
changing the lifetimes of post-main sequence evolutionary phases (RGB,
HB, and AGB) by a global factor in our standard models.  The principal
effect is to change \Ibar, with small changes in the $V-\Ic$ color.
This comparison indicates that model uncertainties in the lifetimes of
the post-MS evolutionary phases are probably less than $\pm$50\%.
(Conversely, any large systematic change to the Cepheid distances which
set the zeropoint for the $I$-band SBF calibration would suggest that
the model lifetimes need to be adjusted.) This constraint is interesting
because it applies at the relatively high metallicities of elliptical
galaxies. For low-metallicity stellar populations, useful constraints on
the accuracy of the post-MS lifetimes can be derived from star counts in
Galactic globular clusters \citep[e.g.,][]{1988ARA&A..26..199R}.

\subsection{New Tools for Breaking the Age/Metallicity Degeneracy
\label{section-age-z}} 

Broad-band colors are largely degenerate to changes in age and
metallicity in old populations, with changes of \hbox{$S \equiv d({\rm
log}\ age)/d({\rm log}\ Z)\approx3/2$} approximately preserving the
colors (Worthey 1994).  Stellar absorption-line indices can be more
sensitive to either age or metallicity, \eg, $H\beta$ and $H\gamma$ are
age-sensitive ($S\lesssim1.0$) and Mg$_2$ and C4668 are
metallicity-sensitive ($S\approx 2-5$). SBFs are expected to depend
mostly on metallicity in old populations because they closely track the
temperature of the RGB and AGB, whose colors are governed by metallicity
(\S~\ref{obsframework}).  SBF magnitudes and colors of old populations
at near-IR wavelengths are predicted by our models and those of Worthey
(1994) to have \hbox{$d({\rm log}\ age)/d({\rm log}\ Z)\gtrsim5-6$}. This
strong sensitivity to metallicity suggests that SBF data, when combined
with age-sensitive observables, could effectively disentangle the
effects of age and metallicity in interpreting unresolved stellar
populations.

\begin{figure}
\centering
\vskip -0.3in
\hskip -0.75in
\vbox{
\includegraphics[width=4in,angle=90]{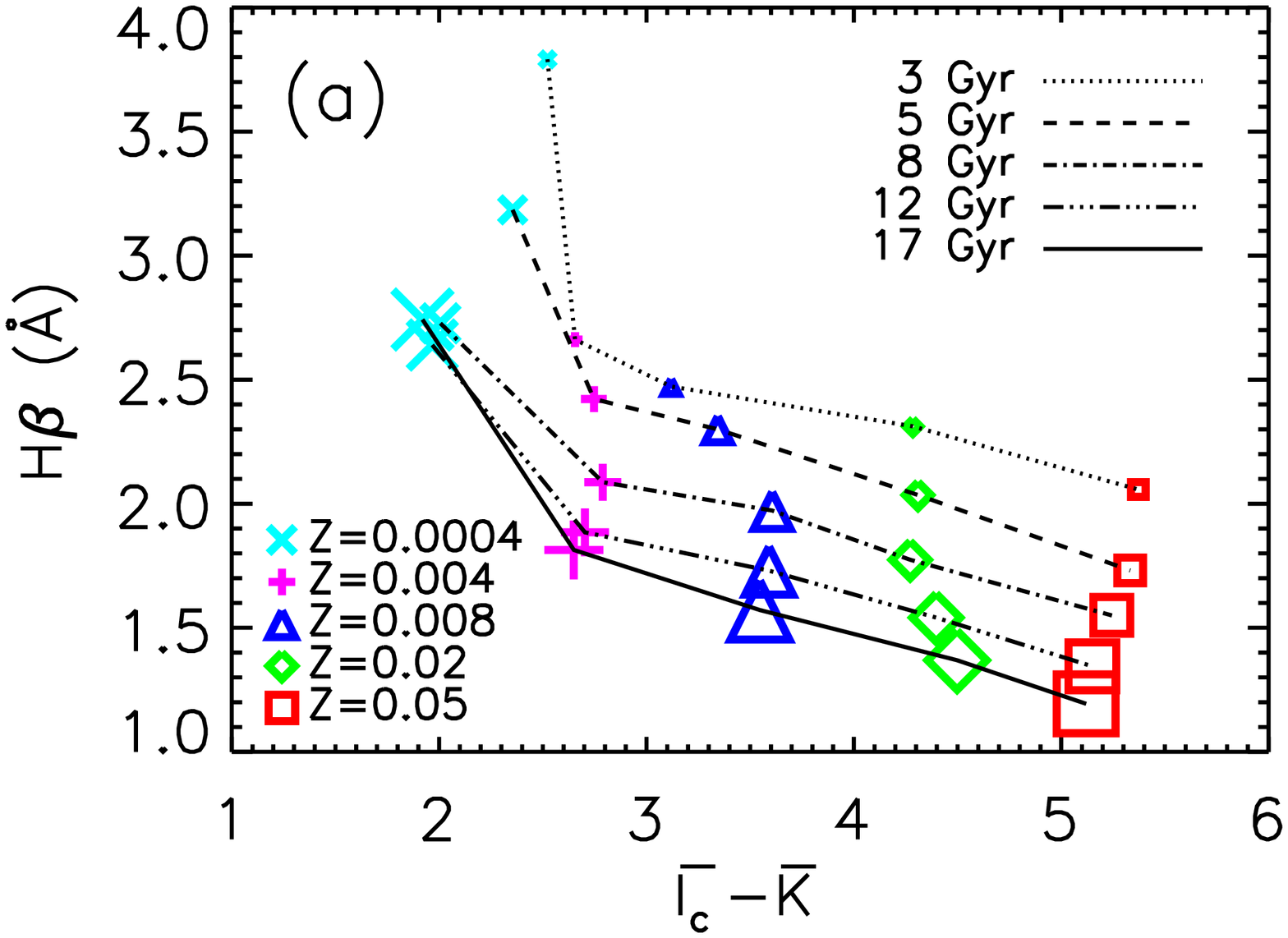}
\vskip -0.3in
\includegraphics[width=4in,angle=90]{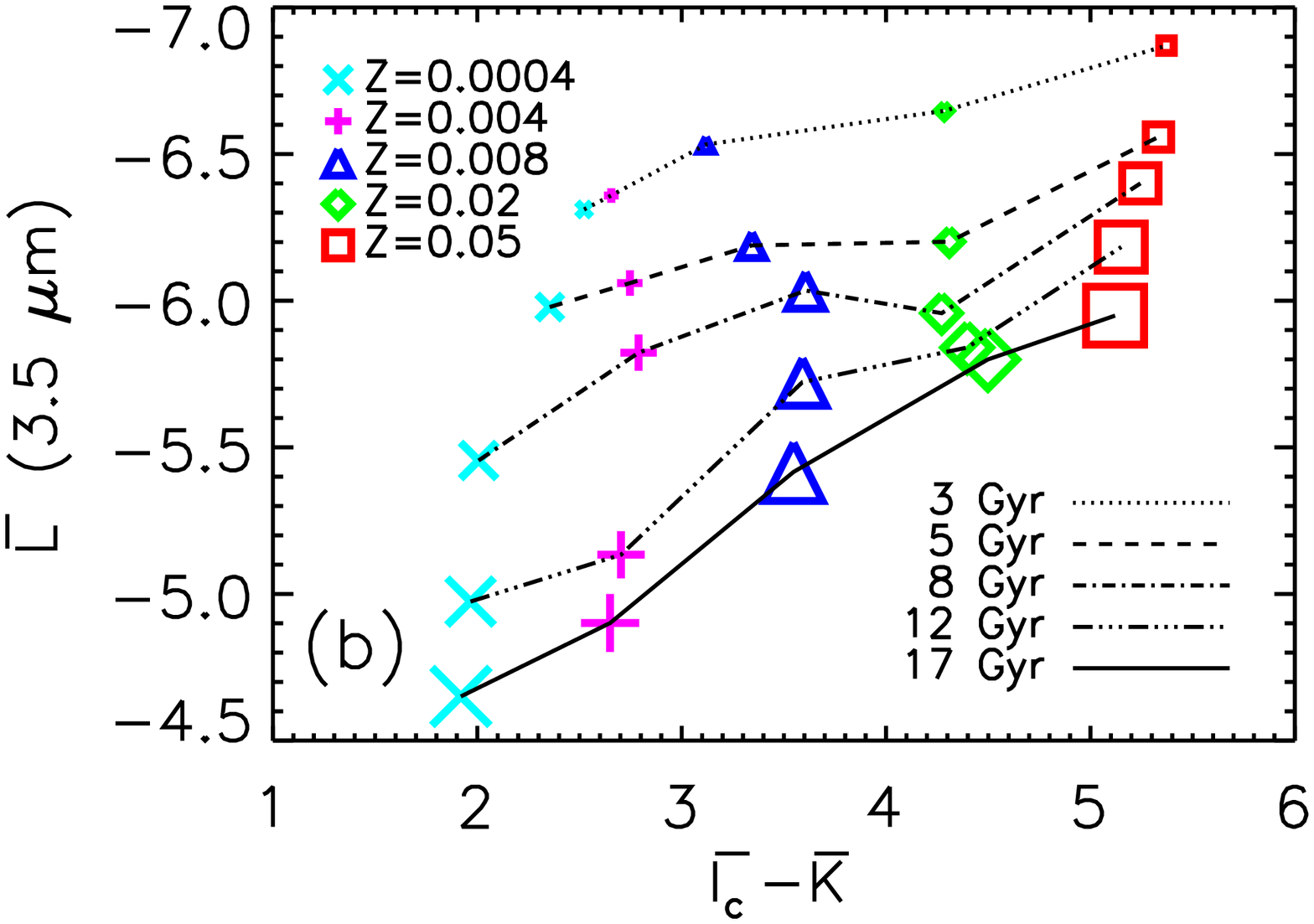}}
\caption{\normalsize Two examples of using SBF measurements to
disentangle age and metallicitiy effects in stellar populations
studies. {\bf Top:} The age-sensitive H$\beta$ absorption line versus
the metallicity-sensitive SBF \Ibar--\Kbar\ color. {\bf Bottom:} The
age-sensitive $L$-band (3.5~\micron) SBF magnitude versus the SBF
\Ibar--\Kbar\ color. \label{age-z}}
\end{figure}

A full investigation of using SBFs for stellar population studies is
beyond the scope of this paper; in addition, the existing datasets for
such analyses are limited.  Instead we highlight in Figure~\ref{age-z}
two possible methods to break the age-metallicity degeneracy, both of
which rely on the \Ibar--\Kbar\ color as a metallicity indicator.  Since
Balmer absorption lines are standard age indicators, the use of H$\beta$
in combination with \Ibar--\Kbar\ should be effective in distinguishing
age from metallicity (Figure~\ref{age-z}a). Similar results are obtained
when using H$\gamma_A$ as an age indicator. Our models also predict that
$L$ (3.5~\micron) and $M$-band (4.8~\micron) SBFs are very sensitive to
age (Figure~\ref{age-z}b), though this prediction should be treated with
caution (see \S~\ref{sbfmags}). Given the very high thermal background
of the Earth's atmosphere, $L$-band SBFs are measureable from the ground
only for the very closest galaxies such as M~31, while observing
$M$-band SBFs is presently impossible.  The future {\em Space Infrared
Telescope Facility} (\SIRTF) will enable SBF measurements in both of
these bands.

SBF colors such as \Ibar--\Kbar\ could offer some advantages over
metal-absorption lines as metallicity indicators for elliptical
galaxies. The interpretation of metal lines is generally hampered by
some possible complications: (1) the effect of selective
$\alpha$-element enhancement in elliptical galaxies which directly
affects indices such as Mg$_2$, and (2) the limited range in stellar
temperatures and gravities over which the analytic fitting functions
have been derived to parameterize line strengths
\citep[e.g.,][]{1994ApJS...94..687W}.  Also, absorption lines are
typically measured only in the central regions of galaxies, while SBF
measurements sample a much larger area. Finally, as seen in
Figure~\ref{age-z}, because of their weighting to the most luminous cool
stars, SBFs offer a much greater dynamic range than integrated spectral
properties; changes in metallicity should be more clearly detectable in
SBF measurements.

%- - - - - - - - - - - - - - - - - - - - - - - - - - - - - - - - - -%

\section{Conclusions \label{sec-conclusions}}

We have presented theoretical predictions for SBFs of single-burst
stellar populations (SSPs) spanning a wide range of ages (from 1 to
17~Gyr) and metallicities (from 1/200 to 2.5~times solar).  Our
calculations are based on the population synthesis models of Bruzual \&
Charlot (2000), in which the stellar evolution prescription and spectral
libraries are improved over the models used in previous SBF studies.  In
particular, our models have been optimized during the course of this
work by refining the prescription for the latest phases of stellar
evolution, which are important contributors to the optical and infrared
SBF signal. Our standard predictions are based on multi-metallicity
evolutionary tracks from the Padova school and semi-empirical stellar
spectra designed to match the observed color-temperature relations of
solar-neighborhood stars at solar metallicity (LCB97).

Using our models, we generate several {\bf basic predictions as a
function of age and metallicity}. 

%- - - - - - - - - - - - - - - %
\begin {enumerate}

\item We compute SBF magnitudes and integrated colors for a large set of
ground-based and space-based (\HST) optical and infrared
bandpasses. These are supplemented with the strengths of several optical
absorption-line indices on the Lick/IDS system.

\item We provide results for solar-metallicity models using several
combinations of stellar evolutionary tracks and spectral libraries.
These can be used to assess the systematic effects of different model
inputs on the results.

\item We predict the fractional contribution of different stellar
evolutionary phases to the SBFs. Since this information cannot be easily
derived from observations, the models provide insight into which phases
are important contributors to the SBF signal for a given bandpass.

\end{enumerate}
%- - - - - - - - - - - - - - - %

\noindent Our model results directly {\bf benefit SBF distance
determinations}, specifically:

%- - - - - - - - - - - - - - - %
\begin{enumerate}
\setcounter{enumi}{3}

\item We use the models to determine purely theoretical calibrations for
SBFs in many bandpasses. These are independent of any systematic errors
in Cepheid distances or reddening corrections, which affect only
empirical calibrations.

\item We tabulate $k$-corrections out to $cz\leq15,000$~\kms\
($z=0.04$), which are required for accurate determinations of \Ho.  We
find that the $k$-corrections are roughly linear in this redshift
range. Metallicity has a stronger effect on $k$-corrections in the
near-infrared than in optical, but the amplitudes of the corrections are
also generally smaller in the near-infrared. We conclude that systematic
errors from uncertainties in the $k$-corrections are not important
sources of error for $H_0$ determinations.

\item We suggest that the scatter in $I$-band SBF distances can be
further reduced by using the integrated $V-K$ galaxy color instead of
$V-\Ic$ to correct for stellar population variations between
galaxies. The reason for this improvement is that the $V-K$ color is
more sensitive to metallicity, which also drives the \Ibar\ signal.

\item Our models predict that the fluctuation magnitudes should be
independent of population age and metallicity around 1~\micron. A
similar conclusion was reached by \citet{1993ApJ...409..530W} using very
different population synthesis models --- this suggests that the
prediction is robust. Therefore, observations taken with a $Z$-band
filter from a large (8--10~m) ground-based telescope or with the $F110W$
filter in \HST\ NICMOS should allow SBF distance measurements which are
more robust against galaxian population variations.

\end{enumerate}
%- - - - - - - - - - - - - - - %

\noindent We have compared our model results with nearly all the SBF
measurements available to date.  Since SBFs are especially sensitive to
the cool, luminous stars on the upper RGB and AGB, they provide
important tests for population synthesis models.  The existing dataset
comprises Galactic globular clusters, M~31, M~32, and early-type
galaxies in nearby clusters. The $I$-band dataset is by far the most
extensive; there are some $K$-band SBF magnitudes and optical/IR SBF
colors, but more measurements are needed for further testing.

We find generally {\bf good agreement between models and data} and
also suggest some {\bf new tests for the models}. Specifically:

%- - - - - - - - - - - - - - - %
\begin{enumerate}
\setcounter{enumi}{7}

\item Our models reproduce \Vbar\ and \Ibar\ observations of Galactic
globular clusters. This test is complementary to those based on galaxy
data, since the globular clusters have much lower metallicities.  Models
with [Fe/H]~$\approx$~--0.7 might be $\approx$0.2~mag too red in
\Vbar--\Ibar, although more data are needed to verify this.

\item Our standard models provide the best agreement to date with the
tight empirical calibration of \Ibar\ over the entire observed range of
$V-\Ic$ galaxy color.  The zeropoint and slope of the calibration
predicted by our models agree remarkably well with those derived from
the data. Moreover, the models indicate a saturation of \Ibar\ for
$V-\Ic\lesssim1.0$, which is also seen in the observations. The reason
for this flattening is most likely the constancy of the $I$-band tip of
the RGB for metal-poor stellar populations. The small scatter in the
empirical calibration as a function of $V-\Ic$ galaxy color is also
reproduced by the models; this arises because of the partial
age/metallicity degeneracy in the \hbox{\{\Ibar, $V-\Ic$\}} parameter
space. This degeneracy is a boon for distance measurements.

\item Our standard models also agree with \Kpbar\ observations, although
this is based on a much smaller sample of galaxies. In the
\hbox{\{\Kpbar, $V-\Ic$\}} parameter space, changes in age and
metallicity are roughly orthogonal.  \Kpbar\ brightens for populations
of higher metallicities and younger ages, as expected from observations
of the RGB and AGB of star clusters in the Galaxy and the Magellanic
Clouds.

\item The optical/IR fluctuation colors predicted by our models agree
with the observations, although some discrepancies exist at the
$\approx0.1-0.2$~mag level. An advantage of testing the models against
measurements of SBF colors is that the data are immune to errors in the
galaxy distances.

\item The semi-empirical SEDs of LCB97 provide better agreement with SBF
observations at all metallicities than their theoretical
SEDs. Observations of \Ibar--\Jbar\ and \Ibar--\Hbar\ colors would help
to verify this result.  For solar metallicity, the results obtained from
the empirical spectral library of \citet{1998PASP..110..863P} agree
closely with those from the semi-empirical library of LCB97.

\item For solar metallicity, the Padova evolutionary tracks seem to
provide better agreement with SBF observations than the Geneva tracks.
The integrated spectral properties from the two sets of tracks are very
comparable at solar metallicities. However, the SBF data are a sensitive
test for deciding between the Geneva and Padova tracks, since the
differences are larger in the SBF predictions than in those for the
integrated spectra.

\item From the tightness of the empirical \Ibar\ calibration, we
conclude that the lifetimes of post-main sequence phases (RGB, core-He
burning, and AGB) in the evolutionary tracks are probably accurate to
within better than $\pm$50\%.

\end{enumerate}
%- - - - - - - - - - - - - - - %

\noindent By comparing our single-burst models with the available
dataset, mostly composed of luminous galaxies in nearby clusters, our
preliminary findings on {\bf the stellar populations dominating the
SBFs} are:

%- - - - - - - - - - - - - - - %
\begin{enumerate}
\setcounter{enumi}{14}

\item The metallicities inferred from SBF magnitudes and SBF
colors show little spread.  The metallicities favored by the optical/SBF
colors are slightly sub-solar, while those favored by the \Kpbar\ data
are around solar.

\item SBF color measurements show no obvious differences for galaxies
observed with different linear aperture sizes, though the available
dataset is small.  The implication is that SBF distance measurements
should be relatively insensitive to systematic errors due to aperture
effects.

\item The ages inferred from comparisons of both \Kpbar\ and
\Vbar--\Ibar\ with the $V-\Ic$ integrated color span a range of about a
factor of three, with the youngest ones near 3~Gyr. Note that estimates
based on combinations of SBF colors and integrated colors are
independent of the galaxy distances.

\item For old populations, the tightness of the empirical $I$-band SBF
calibration also indicates that the IMF in elliptical galaxies cannot
be significantly steeper than that in the solar neighborhood.

\end{enumerate}
%- - - - - - - - - - - - - - - %

\noindent Finally, we suggest that SBF measurements can offer {\bf
useful new tools for stellar population studies}:

%- - - - - - - - - - - - - - - %
\begin{enumerate}
\setcounter{enumi}{18}

\item In old populations, the SBF magnitudes and colors are predicted to
be very sensitive to metallicity, especially at near-IR ($JHK$)
wavelengths. This may offer a potent means of breaking the
age/metallicity degeneracy inherent in studies based on integrated
spectral properties.

\item We find that the \Ibar--\Kbar\ SBF color is very sensitive to
metallicity because of the decreasing temperature of the giant branch
with increasing metallicity. Thus, \Ibar--\Kbar\ might be used in
combination with age-sensitive observables such as Balmer absorption
lines to constrain the ages and metallicities of elliptical
galaxies. SBF colors may also present advantages over metal absorption
lines such as Mg$_2$ and C4668, which are affected by uncertainties in
the patterns of $\alpha$-element enhancement in elliptical galaxies.

\item Our models suggest that the $L$-band and $M$-band SBFs are 
very sensitive to age, although our predictions are not optimized
in this wavelength range. This potentially interesting result should
be further investigated using more appropriate models.

\item Observations of \Bbar--\Ibar\ with \Ibar--\Kbar\ may be useful to
identify stellar populations of different metallicities in elliptical
galaxies. 
\end{enumerate}
%- - - - - - - - - - - - - - - %

The single-burst models we have investigated can account for the full
observed ranges of SBF magnitudes, SBF colors, and integrated colors for
bright elliptical galaxies in nearby clusters.  It is important to
realize that, although the SBF observations can be most simply
reproduced by models with around solar metallicity and a significant
spread in age, a more refined analysis is required to interpret these
measurements in terms of the star formation history of elliptical
galaxies.  In particular, there are multiple lines of evidence that both
cluster and field elliptical galaxies have experienced more than one
episode of star formation \citep[e.g.,][]{1992AJ....104.1039S,
1996MNRAS.279....1B, 1999ApJ...518..576P}.  To constrain the ages and
metallicities of different stellar generations in elliptical galaxies,
we then require a combination of various age and metallicity
indicators. Unfortunately, the published SBF and absorption-line studies
contain few galaxies in common. In a future paper (M. Liu \etal, in
preparation), we exploit a more extensive set of new SBF measurements to
investigate the stellar content of elliptical galaxies.

\acknowledgements

It is pleasure to acknowledge useful discussions with Gustavo Bruzual,
Joe Jensen, Ivan King, Alvio Renzini, Mike Rich, Scott Trager, John
Tonry, and Guy Worthey, as well as the encouragement many of them
provided.  We are also grateful to Johns Blakeslee and Tonry for
providing us with updates to published optical SBF results, and to Steve
Lord for providing the ATRAN calculations. This research was supported
in part by the National Science Foundation through grant no.\
PHY94-07194 to the Institute of Theoretical Physics at UC Santa Barbara
and grant no.\ AST-9617173 to the authors.

%%%%%%%%%%%%%%%%%%%%%%%%%%%%%%%%%%%%%%%%%%%%%%%%%%%%%%%%%%%%%%%%%%%%%%




\clearpage
%\begin{thebibliography}{}
%\end{bibliography}

%-- BibTeX stuff --%
% ** need to hand-add the evol. track refs in Table 1 to *.bbl file **
%\bibliographystyle{apj}
%\bibliography{/rice/mliu/tex/bibtex/mliu}

%----------------------------------------------------------------------
%		FIGURE CAPTIONS
%----------------------------------------------------------------------

%% Generally speaking, only the figure captions, and not the figures
%% themselves, are included in electronic manuscript submissions.
%% Use \figcaption to format your figure captions. They should begin on a
%% new page.

%% No more than seven \figcaption commands are allowed per page,
%% so if you have more than seven captions, insert a \clearpage

%% There must be a \figcaption command for each legend. Key the text of the
%% legend and the optional \label in curly braces. If you wish, you may
%% include the name of the corresponding figure file in square brackets.
%% The label is for identification purposes only. It will not insert the
%% figures themselves into the document.
%% If you want to include your art in the paper, use \plotone.
%% Refer to the on-line documentation for details.

%\figcaption{}

\clearpage

%----------------------------------------%
%\input figures.tex
%----------------------------------------%

% Figure 1

% Figure 2

% Figure 3

% Figure 4

% Figure 5

% Figure 6

% Figure 7

% Figure 8

% Figure 9

% Figure 10

% Figure 11

% Figure 12

% Figure 13

% Figure 14

% Figure 15

%%----------------------------------------------------------------------
%%			TABLES
%%----------------------------------------------------------------------

%% Tables should be submitted one per page, so put a \clearpage before
%% Two options are available to the author for producing tables:  the
%% deluxetable environment provided by the AASTeX package or the LaTeX
%% table environment.  Use of deluxetable is preferred.

\clearpage
% description of the different BC2000 models

\begin{deluxetable}{llcl}  
\tablecolumns{3}  
\tablecaption{Model Options \label{table-models}}

\tablehead{\colhead{Inputs} & \colhead{Options} & 
\colhead{Metallicity Range\tablenotemark{a}} & \colhead{Reference}}

\startdata  
Evolutionary tracks & Geneva  & \Zsun\                         & Schaller \etal\ (1992)\tablenotemark{b} \\
                    & Padova  & $\{1/200-2.5\} \times \Zsun$   & Bressan \etal\ (1993)\tablenotemark{b} \\
&&&\\
Spectral libraries  & theoretical    & $\{1/200-2.5\} \times \Zsun$ & Lejeune \etal\ (1997, 1998) \\
                    & semi-empirical & $\{1/200-2.5\} \times \Zsun$ & Lejeune \etal\ (1997, 1998) \\
                    & empirical      & \Zsun\                       & Pickles (1998) \\
\enddata

\tablecomments{Our standard models use the Padova tracks with the
semi-empirical SEDs.}

\tablenotetext{a}{Solar metallicity is $\Zsun=0.02$ in these models.}

\tablenotetext{b}{More complete lists of references are available in
Charlot \etal\ (1996), Bruzual \etal\ (1997), and Bruzual \& Charlot
(2000).}

\end{deluxetable} 

\clearpage
% MAKETABLE.IDL: automatic LaTex table generation
% Fri Feb 18 15:12:13 2000  charlot@triscuit 
% input file = "TOTAL-salp.dat"

\thispagestyle{empty}
\begin{deluxetable}{ccccccccccccccccccccccccccc}
\rotate
\tablecaption{Predictions from our Standard Models\tablenotemark{a} \label{table-salp}}
\tablewidth{0pt}
\tabletypesize{\scriptsize}
\setlength{\tabcolsep}{0.02in}

\tablehead{ \colhead{${Z}$} & \colhead{Gyr} & \colhead{$\overline{B}$} & \colhead{$\overline{V}$} & \colhead{$\overline{R_c}$} & \colhead{$\overline{I_c}$} & \colhead{$\overline{F814W}$} & \colhead{$\overline{F110M}$} & \colhead{$\overline{F110W}$} & \colhead{$\overline{J}$} & \colhead{$\overline{F160W}$} & \colhead{$\overline{H}$} & \colhead{$\overline{K^{\prime}}$} & \colhead{$\overline{K_s}$} & \colhead{$\overline{K}$} & \colhead{$\overline{F222M}$} & \colhead{$\overline{L}$} & \colhead{$\overline{L^{\prime}}$} & \colhead{$\overline{M}$} & \colhead{$V$--$I_c$} & \colhead{$V$--$K$} & \colhead{$J$--$K$} & \colhead{H$\beta$} & \colhead{Mg$_2$} & \colhead{Mgb} & \colhead{H$\gamma_A$} & \colhead{C4668}}

%# SBFTABLE.IDL: compilation of SBF data files
%# Fri Feb 18 11:53:44 PST 2000
%# list of files processed:
%#   salp-z0001.sbf
%#   salp-z0004.sbf
%#   salp-z004.sbf
%#   salp-z008.sbf
%#   salp-z02.sbf
%#   salp-z05.sbf
%# SBF magnitudes generated by SBFCALC.PRO
%#   input file  = "all.inp"
%#   filter file = "filters.sbf"
%#
%#  Fri Feb 18 11:51:21 PST 2000
%#
%#    0       1       2       3       4       5       6       7       8       9      10      11      12      13      14      15      16      17      18      19      20      21      22      23      24      25      26      27
%#   Zmet  AgeGyr   phase       B       V      Rc      Ic   F814W   F110M   F110W       J   F160W       H      K'      Ks       K   F222M       L      L'       M   V-115    V-77   76-77   Hbeta     Mg2     Mgb  Hgam_A   C4668
%#
%#     Z   Gyr     V   R_c   I_c F814W F110M F110W     J F160W     H K^{\p   K_s     K F222M     L L^{\p     M $V$-- $V$-- $J$-- H\bet  Mg_2   Mgb H\gam C4668 C4668
%#

\startdata

0.0001 & $ 1$& $ 0.07$& $-0.56$& $-1.23$& $-2.10$& $-2.02$& $-3.00$& $-2.97$& $-3.37$& $-4.45$& $-4.60$& $-5.03$& $-5.05$& $-5.12$& $-5.02$& $-7.10$& $-7.12$& $-7.45$& $ 0.39$& $ 1.10$& $ 0.39$& $ 6.70$& $-0.05$& $-2.27$& \phs$ 8.83$& \phs$ 0.96$ \\
0.0001 & $ 2$& $ 0.38$& $-0.83$& $-1.68$& $-2.56$& $-2.48$& $-3.40$& $-3.37$& $-3.72$& $-4.69$& $-4.83$& $-5.18$& $-5.20$& $-5.24$& $-5.17$& $-6.60$& $-6.64$& $-6.91$& $ 0.66$& $ 1.68$& $ 0.54$& $ 4.76$& $-0.04$& $-1.34$& \phs$ 6.62$& \phs$ 0.40$ \\
0.0001 & $ 3$& $ 0.50$& $-0.73$& $-1.55$& $-2.41$& $-2.33$& $-3.22$& $-3.20$& $-3.55$& $-4.49$& $-4.63$& $-4.99$& $-5.00$& $-5.05$& $-4.98$& $-6.37$& $-6.41$& $-6.68$& $ 0.69$& $ 1.74$& $ 0.54$& $ 4.27$& $-0.03$& $-1.03$& \phs$ 6.22$& \phs$ 0.23$ \\
0.0001 & $ 5$& $ 0.50$& $-0.73$& $-1.50$& $-2.26$& $-2.19$& $-3.01$& $-2.97$& $-3.31$& $-4.16$& $-4.28$& $-4.64$& $-4.65$& $-4.71$& $-4.62$& $-6.00$& $-6.02$& $-6.31$& $ 0.75$& $ 1.80$& $ 0.53$& $ 3.54$& $-0.02$& $-0.53$& \phs$ 5.04$& $-0.03$ \\
0.0001 & $ 8$& $ 0.49$& $-0.68$& $-1.36$& $-1.99$& $-1.94$& $-2.61$& $-2.56$& $-2.88$& $-3.50$& $-3.59$& $-3.95$& $-3.97$& $-4.06$& $-3.94$& $-5.42$& $-5.42$& $-5.76$& $ 0.78$& $ 1.83$& $ 0.52$& $ 3.18$& $-0.00$& $-0.23$& \phs$ 4.14$& $-0.20$ \\
0.0001 & $12$& $ 0.54$& $-0.61$& $-1.28$& $-1.91$& $-1.86$& $-2.56$& $-2.51$& $-2.83$& $-3.46$& $-3.55$& $-3.89$& $-3.90$& $-3.98$& $-3.86$& $-4.96$& $-4.96$& $-5.27$& $ 0.82$& $ 1.90$& $ 0.53$& $ 3.02$& \phs$ 0.01$& $-0.04$& \phs$ 3.35$& $-0.26$ \\
0.0001 & $17$& $ 0.55$& $-0.61$& $-1.28$& $-1.90$& $-1.85$& $-2.54$& $-2.49$& $-2.81$& $-3.41$& $-3.49$& $-3.76$& $-3.77$& $-3.82$& $-3.74$& $-4.44$& $-4.44$& $-4.68$& $ 0.86$& $ 1.98$& $ 0.55$& $ 2.86$& \phs$ 0.02$& \phs$ 0.10$& \phs$ 2.70$& $-0.34$ \\[8pt]

0.0004 & $ 1$& $ 0.43$& $-0.65$& $-1.43$& $-2.25$& $-2.18$& $-3.07$& $-3.03$& $-3.39$& $-4.30$& $-4.42$& $-4.80$& $-4.82$& $-4.89$& $-4.81$& $-6.84$& $-6.87$& $-7.21$& $ 0.56$& $ 1.43$& $ 0.46$& $ 5.51$& $-0.02$& $-1.25$& \phs$ 6.55$& $-0.39$ \\
0.0004 & $ 2$& $ 0.35$& $-1.05$& $-1.89$& $-2.75$& $-2.67$& $-3.56$& $-3.53$& $-3.88$& $-4.76$& $-4.89$& $-5.21$& $-5.22$& $-5.26$& $-5.20$& $-6.55$& $-6.58$& $-6.86$& $ 0.74$& $ 1.83$& $ 0.56$& $ 4.11$& $-0.01$& $-0.54$& \phs$ 5.22$& $-0.46$ \\
0.0004 & $ 3$& $ 0.55$& $-0.84$& $-1.68$& $-2.51$& $-2.44$& $-3.32$& $-3.28$& $-3.63$& $-4.52$& $-4.64$& $-4.98$& $-4.99$& $-5.03$& $-4.97$& $-6.31$& $-6.34$& $-6.63$& $ 0.76$& $ 1.86$& $ 0.56$& $ 3.79$& $-0.00$& $-0.33$& \phs$ 4.91$& $-0.53$ \\
0.0004 & $ 5$& $ 0.58$& $-0.85$& $-1.66$& $-2.45$& $-2.39$& $-3.22$& $-3.18$& $-3.51$& $-4.32$& $-4.44$& $-4.75$& $-4.76$& $-4.81$& $-4.75$& $-5.98$& $-6.01$& $-6.28$& $ 0.81$& $ 1.92$& $ 0.56$& $ 3.18$& \phs$ 0.01$& \phs$ 0.10$& \phs$ 3.44$& $-0.56$ \\
0.0004 & $ 8$& $ 0.62$& $-0.77$& $-1.50$& $-2.16$& $-2.11$& $-2.84$& $-2.79$& $-3.10$& $-3.71$& $-3.79$& $-4.09$& $-4.11$& $-4.17$& $-4.09$& $-5.46$& $-5.46$& $-5.79$& $ 0.85$& $ 1.96$& $ 0.55$& $ 2.73$& \phs$ 0.03$& \phs$ 0.43$& \phs$ 2.01$& $-0.60$ \\
0.0004 & $12$& $ 0.62$& $-0.74$& $-1.48$& $-2.15$& $-2.10$& $-2.83$& $-2.78$& $-3.10$& $-3.72$& $-3.80$& $-4.07$& $-4.08$& $-4.12$& $-4.06$& $-4.97$& $-4.97$& $-5.23$& $ 0.88$& $ 2.02$& $ 0.56$& $ 2.64$& \phs$ 0.04$& \phs$ 0.55$& \phs$ 1.76$& $-0.62$ \\
0.0004 & $17$& $ 0.61$& $-0.73$& $-1.49$& $-2.17$& $-2.12$& $-2.85$& $-2.80$& $-3.12$& $-3.74$& $-3.82$& $-4.05$& $-4.06$& $-4.09$& $-4.05$& $-4.65$& $-4.65$& $-4.82$& $ 0.90$& $ 2.07$& $ 0.58$& $ 2.74$& \phs$ 0.05$& \phs$ 0.56$& \phs$ 2.07$& $-0.62$ \\[8pt]

0.0040 & $ 1$& $ 1.17$& $-0.36$& $-1.60$& $-2.80$& $-2.70$& $-3.78$& $-3.76$& $-4.13$& $-5.18$& $-5.32$& $-5.66$& $-5.66$& $-5.68$& $-5.65$& $-6.91$& $-6.94$& $-7.24$& $ 0.70$& $ 1.85$& $ 0.61$& $ 4.76$& \phs$ 0.05$& \phs$ 0.19$& \phs$ 6.58$& \phs$ 0.11$ \\
0.0040 & $ 2$& $ 1.33$& $-0.41$& $-1.56$& $-2.64$& $-2.56$& $-3.54$& $-3.50$& $-3.84$& $-4.79$& $-4.92$& $-5.27$& $-5.29$& $-5.32$& $-5.28$& $-6.63$& $-6.66$& $-7.02$& $ 0.89$& $ 2.19$& $ 0.68$& $ 3.32$& \phs$ 0.08$& \phs$ 1.12$& \phs$ 3.28$& \phs$ 0.79$ \\
0.0040 & $ 3$& $ 1.42$& $-0.30$& $-1.43$& $-2.52$& $-2.43$& $-3.41$& $-3.37$& $-3.71$& $-4.65$& $-4.77$& $-5.14$& $-5.15$& $-5.18$& $-5.14$& $-6.36$& $-6.38$& $-6.73$& $ 0.98$& $ 2.35$& $ 0.70$& $ 2.66$& \phs$ 0.11$& \phs$ 1.64$& \phs$ 0.58$& \phs$ 1.33$ \\
0.0040 & $ 5$& $ 1.66$& $-0.03$& $-1.18$& $-2.35$& $-2.26$& $-3.29$& $-3.24$& $-3.58$& $-4.55$& $-4.68$& $-5.07$& $-5.08$& $-5.10$& $-5.08$& $-6.06$& $-6.08$& $-6.37$& $ 1.02$& $ 2.42$& $ 0.72$& $ 2.42$& \phs$ 0.12$& \phs$ 1.90$& $-0.78$& \phs$ 1.42$ \\
0.0040 & $ 8$& $ 1.69$& \phs$ 0.12$& $-0.99$& $-2.22$& $-2.12$& $-3.23$& $-3.16$& $-3.49$& $-4.45$& $-4.58$& $-4.99$& $-5.00$& $-5.01$& $-5.01$& $-5.82$& $-5.83$& $-6.06$& $ 1.09$& $ 2.54$& $ 0.75$& $ 2.09$& \phs$ 0.14$& \phs$ 2.14$& $-2.52$& \phs$ 1.64$ \\
0.0040 & $12$& $ 1.78$& \phs$ 0.26$& $-0.84$& $-2.10$& $-1.99$& $-3.14$& $-3.06$& $-3.38$& $-4.32$& $-4.45$& $-4.82$& $-4.82$& $-4.80$& $-4.84$& $-5.13$& $-5.14$& $-4.98$& $ 1.13$& $ 2.61$& $ 0.76$& $ 1.88$& \phs$ 0.15$& \phs$ 2.40$& $-3.42$& \phs$ 1.82$ \\
0.0040 & $17$& $ 1.73$& \phs$ 0.32$& $-0.76$& $-1.99$& $-1.89$& $-3.01$& $-2.94$& $-3.26$& $-4.19$& $-4.31$& $-4.67$& $-4.67$& $-4.64$& $-4.69$& $-4.90$& $-4.91$& $-4.63$& $ 1.16$& $ 2.65$& $ 0.76$& $ 1.81$& \phs$ 0.16$& \phs$ 2.53$& $-3.63$& \phs$ 1.84$ \\[8pt]

0.0080 & $ 1$& $ 1.46$& $-0.02$& $-1.35$& $-2.68$& $-2.57$& $-3.75$& $-3.71$& $-4.09$& $-5.17$& $-5.32$& $-5.70$& $-5.71$& $-5.74$& $-5.70$& $-7.07$& $-7.10$& $-7.45$& $ 0.75$& $ 1.98$& $ 0.65$& $ 4.35$& \phs$ 0.08$& \phs$ 0.81$& \phs$ 5.87$& \phs$ 0.92$ \\
0.0080 & $ 2$& $ 1.79$& \phs$ 0.06$& $-1.14$& $-2.45$& $-2.34$& $-3.60$& $-3.52$& $-3.88$& $-4.89$& $-5.02$& $-5.47$& $-5.48$& $-5.51$& $-5.48$& $-6.74$& $-6.76$& $-7.13$& $ 0.98$& $ 2.44$& $ 0.76$& $ 2.98$& \phs$ 0.13$& \phs$ 1.85$& \phs$ 1.50$& \phs$ 2.10$ \\
0.0080 & $ 3$& $ 1.92$& \phs$ 0.20$& $-0.98$& $-2.30$& $-2.19$& $-3.50$& $-3.41$& $-3.77$& $-4.78$& $-4.92$& $-5.38$& $-5.39$& $-5.42$& $-5.39$& $-6.53$& $-6.55$& $-6.91$& $ 1.06$& $ 2.59$& $ 0.78$& $ 2.47$& \phs$ 0.15$& \phs$ 2.30$& $-1.07$& \phs$ 2.68$ \\
0.0080 & $ 5$& $ 2.24$& \phs$ 0.62$& $-0.53$& $-1.93$& $-1.82$& $-3.26$& $-3.17$& $-3.57$& $-4.61$& $-4.76$& $-5.24$& $-5.25$& $-5.28$& $-5.25$& $-6.19$& $-6.20$& $-6.53$& $ 1.07$& $ 2.62$& $ 0.79$& $ 2.30$& \phs$ 0.16$& \phs$ 2.53$& $-2.37$& \phs$ 2.71$ \\
0.0080 & $ 8$& $ 2.18$& \phs$ 0.72$& $-0.37$& $-1.85$& $-1.74$& $-3.45$& $-3.32$& $-3.76$& $-4.81$& $-4.97$& $-5.45$& $-5.47$& $-5.46$& $-5.50$& $-6.03$& $-6.06$& $-6.30$& $ 1.15$& $ 2.81$& $ 0.83$& $ 1.97$& \phs$ 0.18$& \phs$ 2.88$& $-3.93$& \phs$ 3.10$ \\
0.0080 & $12$& $ 2.20$& \phs$ 0.84$& $-0.22$& $-1.72$& $-1.60$& $-3.35$& $-3.22$& $-3.67$& $-4.70$& $-4.86$& $-5.32$& $-5.33$& $-5.31$& $-5.37$& $-5.72$& $-5.75$& $-5.92$& $ 1.20$& $ 2.91$& $ 0.85$& $ 1.73$& \phs$ 0.20$& \phs$ 3.15$& $-5.07$& \phs$ 3.36$ \\
0.0080 & $17$& $ 2.24$& \phs$ 0.95$& $-0.07$& $-1.55$& $-1.44$& $-3.16$& $-3.04$& $-3.49$& $-4.52$& $-4.68$& $-5.12$& $-5.13$& $-5.10$& $-5.18$& $-5.42$& $-5.44$& $-5.57$& $ 1.24$& $ 2.96$& $ 0.85$& $ 1.57$& \phs$ 0.21$& \phs$ 3.34$& $-5.69$& \phs$ 3.46$ \\[8pt]

0.0200 & $ 1$& $ 1.81$& \phs$ 0.46$& $-0.77$& $-2.22$& $-2.10$& $-3.59$& $-3.51$& $-3.91$& $-5.01$& $-5.16$& $-5.64$& $-5.65$& $-5.69$& $-5.64$& $-7.09$& $-7.12$& $-7.50$& $ 0.82$& $ 2.19$& $ 0.70$& $ 3.86$& \phs$ 0.12$& \phs$ 1.66$& \phs$ 4.42$& \phs$ 2.86$ \\
0.0200 & $ 2$& $ 2.36$& \phs$ 0.78$& $-0.35$& $-1.81$& $-1.71$& $-3.64$& $-3.48$& $-3.99$& $-5.05$& $-5.22$& $-5.78$& $-5.80$& $-5.83$& $-5.82$& $-6.89$& $-6.92$& $-7.31$& $ 1.04$& $ 2.72$& $ 0.84$& $ 2.79$& \phs$ 0.17$& \phs$ 2.50$& $-0.30$& \phs$ 3.97$ \\
0.0200 & $ 3$& $ 2.60$& \phs$ 1.00$& $-0.05$& $-1.54$& $-1.45$& $-3.72$& $-3.51$& $-4.06$& $-5.09$& $-5.27$& $-5.80$& $-5.82$& $-5.83$& $-5.88$& $-6.65$& $-6.70$& $-7.05$& $ 1.13$& $ 2.94$& $ 0.88$& $ 2.31$& \phs$ 0.20$& \phs$ 3.03$& $-3.16$& \phs$ 4.73$ \\
0.0200 & $ 5$& $ 2.83$& \phs$ 1.25$& \phs$ 0.20$& $-1.30$& $-1.22$& $-3.53$& $-3.31$& $-3.87$& $-4.89$& $-5.07$& $-5.59$& $-5.62$& $-5.61$& $-5.68$& $-6.20$& $-6.28$& $-6.59$& $ 1.17$& $ 2.99$& $ 0.87$& $ 2.04$& \phs$ 0.22$& \phs$ 3.36$& $-4.52$& \phs$ 5.04$ \\
0.0200 & $ 8$& $ 2.61$& \phs$ 1.32$& \phs$ 0.27$& $-1.22$& $-1.14$& $-3.44$& $-3.23$& $-3.79$& $-4.79$& $-4.96$& $-5.47$& $-5.50$& $-5.49$& $-5.56$& $-5.96$& $-6.05$& $-6.33$& $ 1.23$& $ 3.12$& $ 0.90$& $ 1.77$& \phs$ 0.25$& \phs$ 3.72$& $-5.70$& \phs$ 5.47$ \\
0.0200 & $12$& $ 2.48$& \phs$ 1.43$& \phs$ 0.40$& $-1.07$& $-1.00$& $-3.44$& $-3.20$& $-3.77$& $-4.76$& $-4.95$& $-5.44$& $-5.48$& $-5.47$& $-5.58$& $-5.84$& $-5.99$& $-6.26$& $ 1.29$& $ 3.25$& $ 0.92$& $ 1.54$& \phs$ 0.27$& \phs$ 4.04$& $-6.71$& \phs$ 5.85$ \\
0.0200 & $17$& $ 2.51$& \phs$ 1.52$& \phs$ 0.51$& $-0.95$& $-0.89$& $-3.41$& $-3.15$& $-3.74$& $-4.73$& $-4.93$& $-5.41$& $-5.46$& $-5.45$& $-5.58$& $-5.80$& $-5.98$& $-6.26$& $ 1.33$& $ 3.34$& $ 0.93$& $ 1.37$& \phs$ 0.29$& \phs$ 4.24$& $-7.44$& \phs$ 6.14$ \\[8pt]

0.0500 & $ 1$& $ 2.26$& \phs$ 1.00$& $-0.32$& $-2.10$& $-1.98$& $-4.67$& $-4.43$& $-5.17$& $-6.33$& $-6.55$& $-7.14$& $-7.19$& $-7.21$& $-7.29$& $-7.68$& $-7.84$& $-8.11$& $ 0.96$& $ 2.91$& $ 0.98$& $ 3.29$& \phs$ 0.18$& \phs$ 2.53$& \phs$ 1.88$& \phs$ 5.68$ \\
0.0500 & $ 2$& $ 2.86$& \phs$ 1.32$& \phs$ 0.23$& $-1.28$& $-1.21$& $-4.25$& $-3.97$& $-4.77$& $-5.82$& $-6.07$& $-6.63$& $-6.72$& $-6.74$& $-6.91$& $-7.30$& $-7.53$& $-7.86$& $ 1.14$& $ 3.22$& $ 0.98$& $ 2.45$& \phs$ 0.23$& \phs$ 3.34$& $-3.54$& \phs$ 7.01$ \\
0.0500 & $ 3$& $ 3.02$& \phs$ 1.48$& \phs$ 0.46$& $-0.94$& $-0.88$& $-3.94$& $-3.66$& $-4.43$& $-5.44$& $-5.68$& $-6.22$& $-6.31$& $-6.31$& $-6.49$& $-6.87$& $-7.09$& $-7.42$& $ 1.23$& $ 3.35$& $ 0.98$& $ 2.06$& \phs$ 0.27$& \phs$ 3.83$& $-5.41$& \phs$ 7.85$ \\
0.0500 & $ 5$& $ 2.78$& \phs$ 1.67$& \phs$ 0.61$& $-0.83$& $-0.77$& $-3.78$& $-3.51$& $-4.30$& $-5.29$& $-5.54$& $-6.05$& $-6.15$& $-6.16$& $-6.39$& $-6.56$& $-6.89$& $-7.21$& $ 1.31$& $ 3.51$& $ 1.00$& $ 1.73$& \phs$ 0.30$& \phs$ 4.35$& $-6.91$& \phs$ 8.58$ \\
0.0500 & $ 8$& $ 2.61$& \phs$ 1.77$& \phs$ 0.60$& $-0.84$& $-0.77$& $-3.68$& $-3.41$& $-4.21$& $-5.19$& $-5.45$& $-5.96$& $-6.08$& $-6.08$& $-6.34$& $-6.40$& $-6.80$& $-7.12$& $ 1.35$& $ 3.57$& $ 1.00$& $ 1.55$& \phs$ 0.32$& \phs$ 4.66$& $-7.64$& \phs$ 8.97$ \\
0.0500 & $12$& $ 2.39$& \phs$ 1.75$& \phs$ 0.66$& $-0.73$& $-0.67$& $-3.51$& $-3.25$& $-4.03$& $-5.01$& $-5.26$& $-5.76$& $-5.88$& $-5.88$& $-6.12$& $-6.18$& $-6.56$& $-6.89$& $ 1.41$& $ 3.67$& $ 1.01$& $ 1.34$& \phs$ 0.36$& \phs$ 5.02$& $-8.49$& \phs$ 9.72$ \\
0.0500 & $17$& $ 2.35$& \phs$ 1.82$& \phs$ 0.79$& $-0.54$& $-0.48$& $-3.31$& $-3.04$& $-3.82$& $-4.80$& $-5.06$& $-5.55$& $-5.66$& $-5.66$& $-5.90$& $-5.95$& $-6.32$& $-6.66$& $ 1.44$& $ 3.71$& $ 1.00$& $ 1.19$& \phs$ 0.38$& \phs$ 5.28$& $-9.00$& $10.33$ \\
\enddata

\tablenotetext{a}{Our standard models use the Padova evolutionary
tracks, the semi-empirical SEDs, and a Salpeter IMF.}

\end{deluxetable}

\clearpage
% << Coefficients of robust linear fit to SBF mag vs. galaxy property >>
%     SBF mag = c0 + c1*(galaxy property)
%
% Sat Dec 18 23:40:03 PST 1999
%
%   FILTER FILE = ../sbfdata/filters.sbf
%   DATA FILES =  ../sbfdata/salp-z004.sbf ../sbfdata/salp-z008.sbf ../sbfdata/salp-z02.sbf
%  ../sbfdata/salp-z05.sbf

\thispagestyle{empty}
\begin{deluxetable}{lcccccccccccccccccccc}  
%\begin{deluxetable}{lccccccccccc}  

\rotate
\tablewidth{0pt}
%\tabletypesize{\small}
\tabletypesize{\footnotesize}
\setlength{\tabcolsep}{0.075in}

\tablecolumns{21}  
\tablecaption{Coefficients of Robust Linear Fits to SBF Magnitudes \label{table-calcobs}}

\tablehead{  
\colhead{}  &  \multicolumn{3}{c}{$\overline{I_c}$} &
\colhead{}  &  \multicolumn{3}{c}{$\overline{F814W}$} &
\colhead{}  &  \multicolumn{3}{c}{$\Jbar$} &
\colhead{}  &  \multicolumn{3}{c}{$\overline{F160W}$} &
\colhead{}  &  \multicolumn{3}{c}{$\Kbar$} \\
\cline{2-4} \cline{6-8} \cline{10-12} \cline{14-16} \cline{18-20}
%\cline{2-4} \cline{6-8} \cline{10-12} \\  
\colhead{} & \colhead{$c_0$} & \colhead{$c_1$} & \colhead{$\sigma$} & 
\colhead{} & \colhead{$c_0$} & \colhead{$c_1$} & \colhead{$\sigma$} & 
\colhead{} & \colhead{$c_0$} & \colhead{$c_1$} & \colhead{$\sigma$} & 
\colhead{} & \colhead{$c_0$} & \colhead{$c_1$} & \colhead{$\sigma$} & 
\colhead{} & \colhead{$c_0$} & \colhead{$c_1$} & \colhead{$\sigma$}}

\startdata  
as a function of $V-\Ic$     &  $-$7.03 & \phs4.56 & \phs0.19   &&   $-$6.77 & \phs4.42 &  0.18  &&   $-$1.95 &  $-$1.45 &  0.22   &&   $-$2.96 &  $-$1.44 &  0.23   &&  $-$2.54 &  $-$2.34 &  0.31 \\
as a function of $V-K$       &  $-$5.59 & \phs1.36 & \phs0.10   &&   $-$5.38 & \phs1.32 &  0.10  &&   $-$2.14 &  $-$0.52 &  0.17   &&   $-$3.12 &  $-$0.53 &  0.19   &&  $-$2.90 &  $-$0.83 &  0.24 \\
as a function of $J-K$       &  $-$6.50 & \phs5.80 & \phs0.12   &&   $-$6.26 & \phs5.62 &  0.11  &&   $-$1.63 &  $-$2.41 &  0.15   &&   $-$2.58 &  $-$2.46 &  0.16   &&  $-$2.12 &  $-$3.78 &  0.20 \\
as a function of H$\beta$    & \phs1.05 &  $-$1.41 & \phs0.33   &&  \phs1.07 &  $-$1.37 &  0.32  &&   $-$4.29 & \phs0.31 &  0.26   &&   $-$5.26 & \phs0.30 &  0.27   &&  $-$6.34 & \phs0.53 &  0.38 \\
as a function of Mg$_2$      &  $-$3.10 & \phs7.13 & \phs0.11   &&   $-$2.96 & \phs6.91 &  0.11  &&   $-$3.14 &  $-$2.58 &  0.19   &&   $-$4.13 &  $-$2.58 &  0.20   &&  $-$4.45 &  $-$4.13 &  0.26 \\
as a function of Mgb         &  $-$3.37 & \phs0.55 & \phs0.09   &&   $-$3.22 & \phs0.54 &  0.09  &&   $-$3.04 &  $-$0.20 &  0.19   &&   $-$4.02 &  $-$0.20 &  0.20   &&  $-$4.29 &  $-$0.32 &  0.26 \\
as a function of H$\gamma_A$ &  $-$2.69 &  $-$0.24 & \phs0.17   &&   $-$2.57 &  $-$0.23 &  0.16  &&   $-$3.33 & \phs0.08 &  0.21   &&   $-$4.32 & \phs0.08 &  0.23   &&  $-$4.76 & \phs0.12 &  0.31 \\
as a function of C4668       &  $-$2.37 & \phs0.19 & \phs0.15   &&   $-$2.25 & \phs0.18 &  0.15  &&   $-$3.35 &  $-$0.08 &  0.16   &&   $-$4.34 &  $-$0.08 &  0.17   &&  $-$4.81 &  $-$0.12 &  0.21 \\
\enddata

%                                   F814W_bar                          F160W_bar        
%                                   c0    c1   rms                     c0    c1   rms   
% as a function of V-Ic          &   $-$6.77 & \phs4.42 &  0.18  &&  $-$2.96 &  $-$1.44 &  0.23 \\
% as a function of V-K           &   $-$5.38 & \phs1.32 &  0.10  &&  $-$3.12 &  $-$0.53 &  0.19 \\
% as a function of J-K           &   $-$6.26 & \phs5.62 &  0.11  &&  $-$2.58 &  $-$2.46 &  0.16 \\
% as a function of Hbeta         &  \phs1.07 &  $-$1.37 &  0.32  &&  $-$5.26 & \phs0.30 &  0.27 \\
% as a function of Mg2           &   $-$2.96 & \phs6.91 &  0.11  &&  $-$4.13 &  $-$2.58 &  0.20 \\
% as a function of Mgb           &   $-$3.22 & \phs0.54 &  0.09  &&  $-$4.02 &  $-$0.20 &  0.20 \\
% as a function of Hgam_A        &   $-$2.57 &  $-$0.23 &  0.16  &&  $-$4.32 & \phs0.08 &  0.23 \\
% as a function of C4668         &   $-$2.25 & \phs0.18 &  0.15  &&  $-$4.34 &  $-$0.08 &  0.17 \\

\tablecomments{The fit is $\Mbar = c_0 + c_1 \times {\rm (galaxy\
property)}$ for models with $Z=\{0.004, 0.008, 0.02, 0.05\}$ and ages of
5, 8, 12, 17~Gyr. The rms scatter about the fitted line is $\sigma$,
given in magnitudes.}

\end{deluxetable} 

\clearpage
% ALLCALC-KCORR.IDL: automatic LaTex table generation
% Tue Dec 21 07:05:14 1999  charlot@triscuit 

\thispagestyle{empty}
\begin{deluxetable}{lcccccccccc}
\rotate
\tablecaption{Linear coefficients for SBF $k$-corrections 
as a Function of Model Metallicity ($Z$) for $cz\le15,000$~\kms\label{table-kcorr}}
\tablewidth{0pt}
\tabletypesize{\footnotesize}
%\tabletypesize{\small}

%# ALLCALC-KCORR.IDL: Tue Dec 21 07:05:15 1999  charlot@triscuit 
%# linear fits to K-corrections as a function of metallicity
%# automatically generated LaTex file
%#
%#  list of Z's = "z004, z008, z02, z05,"
%#  list of input prefixes = cz0-salp-, cz3000-salp-, cz6000-salp-, cz9000-salp-, cz12000-salp-,"
%#
%#  list of redshifts =  0.0000, 0.010000, 0.020000, 0.030000, 0.040000
%#  list of model age used =  5, 8, 12, 17
%#

\tablehead{ \colhead{${Z}$} & \colhead{$\overline{R_c}$} & \colhead{$\overline{I_c}$} & \colhead{$\overline{F814W}$} & \colhead{$\overline{J}$} & \colhead{$\overline{F160W}$} & \colhead{$\overline{K'}$} & \colhead{$\overline{K}$} & \colhead{${V-\Ic}$} & \colhead{${V-K}$} & \colhead{$J-K$}}

\startdata

  0.004 &  \phs$ 3.53\ (0.08)$&  \phs$ 3.79\ (0.30)$&  \phs$ 3.24\ (0.33)$&  $-0.05\ (0.28)$&  \phs$ 0.74\ (0.14)$&  $-2.45\ (0.14)$&  $-2.89\ (0.54)$&  \phs$ 1.07\ (0.02)$&  \phs$ 4.32\ (0.19)$&  \phs$ 2.11\ (0.06)$\\
  0.008 &  \phs$ 3.62\ (0.16)$&  \phs$ 6.20\ (0.50)$&  \phs$ 5.23\ (0.52)$&  \phs$ 0.33\ (0.12)$&  \phs$ 1.31\ (0.13)$&  $-2.17\ (0.17)$&  $-2.85\ (0.49)$&  \phs$ 0.90\ (0.03)$&  \phs$ 4.59\ (0.23)$&  \phs$ 2.27\ (0.09)$\\
   0.02 &  \phs$ 3.40\ (0.07)$&  \phs$ 6.91\ (0.30)$&  \phs$ 6.24\ (0.20)$&  \phs$ 0.25\ (0.17)$&  \phs$ 1.90\ (0.21)$&  $-0.90\ (0.51)$&  $-2.04\ (0.27)$&  \phs$ 0.88\ (0.06)$&  \phs$ 4.88\ (0.20)$&  \phs$ 2.27\ (0.10)$\\
   0.05 &  \phs$ 3.65\ (0.29)$&  \phs$ 6.10\ (0.50)$&  \phs$ 5.33\ (0.55)$&  \phs$ 1.02\ (0.48)$&  \phs$ 2.24\ (0.54)$&  \phs$ 3.52\ (0.45)$&  \phs$ 0.61\ (0.47)$&  \phs$ 1.10\ (0.08)$&  \phs$ 4.96\ (0.29)$&  \phs$ 1.97\ (0.09)$\\

\enddata

\tablecomments{The $k(z)$ term is given by $a \times z$ where $z$ is the
redshift and $a$ is the number tabulated above, for different
observables (SBF magnitude or galaxy color) as a function model
metallicity. The convention for the correction is $k(z) = X(z) - X(z=0)$
where $X$ is an SBF magnitude or galaxy color.  The number in
parenthesis is the rms scatter for the linear coefficient averaged over
models of different ages (see \S~\ref{kcorr}).}

\end{deluxetable}

\clearpage
% MAKETABLE.IDL: automatic LaTex table generation
% Fri Feb 18 15:15:19 2000  charlot@triscuit 
% input file = "TOTAL-solar-diff.dat"

\thispagestyle{empty}
\begin{deluxetable}{ccccccccccccccccccccccccccc}
\rotate
\tablecaption{Solar Metallicity Models \label{table-solar-diff}}
\tablewidth{0pt}
\tabletypesize{\scriptsize}
\setlength{\tabcolsep}{0.02in}
 
\tablehead{ \colhead{${Z}$} & \colhead{Gyr} & \colhead{$\overline{B}$} & \colhead{$\overline{V}$} & \colhead{$\overline{R_c}$} & \colhead{$\overline{I_c}$} & \colhead{$\overline{F814W}$} & \colhead{$\overline{F110M}$} & \colhead{$\overline{F110W}$} & \colhead{$\overline{J}$} & \colhead{$\overline{F160W}$} & \colhead{$\overline{H}$} & \colhead{$\overline{K^{\prime}}$} & \colhead{$\overline{K_s}$} & \colhead{$\overline{K}$} & \colhead{$\overline{F222M}$} & \colhead{$\overline{L}$} & \colhead{$\overline{L^{\prime}}$} & \colhead{$\overline{M}$} & \colhead{$V$--$I_c$} & \colhead{$V$--$K$} & \colhead{$J$--$K$} & \colhead{H$\beta$} & \colhead{Mg$_2$} & \colhead{Mgb} & \colhead{H$\gamma_A$} & \colhead{C4668}}

%# SBFSOLARTABLE.IDL: compilation of SBF data files for solar Z
%# first file is listed, for all other files give difference of (that file) - (1st file)
%# Fri Feb 18 11:55:12 PST 2000
%# *FILE = salp-z02.sbf
%# SBF magnitudes generated by SBFCALC.PRO
%#   input file  = "all.inp"
%#   filter file = "filters.sbf"
%#
%#  Fri Feb 18 11:51:34 PST 2000
%#
%#    0       1       2       3       4       5       6       7       8       9      10      11      12      13      14      15      16      17      18      19      20      21      22      23      24      25      26      27
%#   Zmet  AgeGyr   phase       B       V      Rc      Ic   F814W   F110M   F110W       J   F160W       H      K'      Ks       K   F222M       L      L'       M   V-115    V-77   76-77   Hbeta     Mg2     Mgb  Hgam_A   C4668
%#
%# *FILE = salp-z02.sbf
%# *FILE = salp-th-z02.sbf
%# *FILE = salp-pickles.sbf
%# *FILE = geneva-salp-z02.sbf
%# *FILE = geneva-salp-th-z02.sbf
%# *FILE = geneva-salp-pickles.sbf
%#                                                                                                                                                                  
%#

\startdata

\cutinhead{Padova Evolutionary Tracks, Semi-Empirical SEDs ({\em Standard Model})}
0.02 & $ 1$& \phs$ 1.81$& \phs$ 0.46$& $-0.77$& $-2.22$& $-2.10$& $-3.59$& $-3.51$& $-3.91$& $-5.01$& $-5.16$& $-5.64$& $-5.65$& $-5.69$& $-5.64$& $-7.09$& $-7.12$& $-7.50$& \phs$ 0.82$& \phs$ 2.19$& \phs$ 0.70$& \phs$ 3.86$& \phs$ 0.12$& \phs$ 1.66$& \phs$ 4.42$& \phs$ 2.86$ \\
0.02 & $ 2$& \phs$ 2.36$& \phs$ 0.78$& $-0.35$& $-1.81$& $-1.71$& $-3.64$& $-3.48$& $-3.99$& $-5.05$& $-5.22$& $-5.78$& $-5.80$& $-5.83$& $-5.82$& $-6.89$& $-6.92$& $-7.31$& \phs$ 1.04$& \phs$ 2.72$& \phs$ 0.84$& \phs$ 2.79$& \phs$ 0.17$& \phs$ 2.50$& $-0.30$& \phs$ 3.97$ \\
0.02 & $ 3$& \phs$ 2.60$& \phs$ 1.00$& $-0.05$& $-1.54$& $-1.45$& $-3.72$& $-3.51$& $-4.06$& $-5.09$& $-5.27$& $-5.80$& $-5.82$& $-5.83$& $-5.88$& $-6.65$& $-6.70$& $-7.05$& \phs$ 1.13$& \phs$ 2.94$& \phs$ 0.88$& \phs$ 2.31$& \phs$ 0.20$& \phs$ 3.03$& $-3.16$& \phs$ 4.73$ \\
0.02 & $ 5$& \phs$ 2.83$& \phs$ 1.25$& \phs$ 0.20$& $-1.30$& $-1.22$& $-3.53$& $-3.31$& $-3.87$& $-4.89$& $-5.07$& $-5.59$& $-5.62$& $-5.61$& $-5.68$& $-6.20$& $-6.28$& $-6.59$& \phs$ 1.17$& \phs$ 2.99$& \phs$ 0.87$& \phs$ 2.04$& \phs$ 0.22$& \phs$ 3.36$& $-4.52$& \phs$ 5.04$ \\
0.02 & $ 8$& \phs$ 2.61$& \phs$ 1.32$& \phs$ 0.27$& $-1.22$& $-1.14$& $-3.44$& $-3.23$& $-3.79$& $-4.79$& $-4.96$& $-5.47$& $-5.50$& $-5.49$& $-5.56$& $-5.96$& $-6.05$& $-6.33$& \phs$ 1.23$& \phs$ 3.12$& \phs$ 0.90$& \phs$ 1.77$& \phs$ 0.25$& \phs$ 3.72$& $-5.70$& \phs$ 5.47$ \\
0.02 & $12$& \phs$ 2.48$& \phs$ 1.43$& \phs$ 0.40$& $-1.07$& $-1.00$& $-3.44$& $-3.20$& $-3.77$& $-4.76$& $-4.95$& $-5.44$& $-5.48$& $-5.47$& $-5.58$& $-5.84$& $-5.99$& $-6.26$& \phs$ 1.29$& \phs$ 3.25$& \phs$ 0.92$& \phs$ 1.54$& \phs$ 0.27$& \phs$ 4.04$& $-6.71$& \phs$ 5.85$ \\
0.02 & $17$& \phs$ 2.51$& \phs$ 1.52$& \phs$ 0.51$& $-0.95$& $-0.89$& $-3.41$& $-3.15$& $-3.74$& $-4.73$& $-4.93$& $-5.41$& $-5.46$& $-5.45$& $-5.58$& $-5.80$& $-5.98$& $-6.26$& \phs$ 1.33$& \phs$ 3.34$& \phs$ 0.93$& \phs$ 1.37$& \phs$ 0.29$& \phs$ 4.24$& $-7.44$& \phs$ 6.14$ \\

\cutinhead{Padova Evolutionary Tracks, Theoretical SEDs ({\em Difference from Standard Model})}
0.02 & $ 1$& $-0.01$& $-0.07$& $-0.05$& \phs$ 0.07$& \phs$ 0.06$& \phs$ 0.12$& \phs$ 0.08$& \phs$ 0.01$& $-0.06$& $-0.06$& $-0.01$& $-0.01$& \phs$ 0.00$& \phs$ 0.00$& $-0.01$& $-0.01$& $-0.01$& \phs$ 0.01$& \phs$ 0.02$& \phs$ 0.01$& $-0.05$& \phs$ 0.00$& \phs$ 0.05$& $-0.15$& \phs$ 0.12$ \\
0.02 & $ 2$& $-0.07$& $-0.13$& $-0.07$& \phs$ 0.15$& \phs$ 0.14$& \phs$ 0.09$& \phs$ 0.06$& $-0.05$& $-0.07$& $-0.07$& \phs$ 0.00$& \phs$ 0.00$& \phs$ 0.01$& \phs$ 0.01$& \phs$ 0.01$& \phs$ 0.01$& \phs$ 0.00$& $-0.03$& \phs$ 0.01$& \phs$ 0.01$& $-0.02$& \phs$ 0.00$& \phs$ 0.06$& $-0.14$& \phs$ 0.14$ \\
0.02 & $ 3$& $-0.09$& $-0.13$& $-0.07$& \phs$ 0.16$& \phs$ 0.16$& \phs$ 0.09$& \phs$ 0.05$& $-0.08$& $-0.08$& $-0.07$& \phs$ 0.00$& \phs$ 0.00$& \phs$ 0.01$& \phs$ 0.01$& \phs$ 0.03$& \phs$ 0.03$& \phs$ 0.01$& $-0.04$& \phs$ 0.00$& \phs$ 0.00$& \phs$ 0.00$& \phs$ 0.00$& \phs$ 0.07$& $-0.10$& \phs$ 0.15$ \\
0.02 & $ 5$& $-0.09$& $-0.14$& $-0.08$& \phs$ 0.17$& \phs$ 0.16$& \phs$ 0.09$& \phs$ 0.05$& $-0.08$& $-0.09$& $-0.07$& \phs$ 0.00$& \phs$ 0.00$& \phs$ 0.00$& \phs$ 0.00$& \phs$ 0.04$& \phs$ 0.04$& \phs$ 0.02$& $-0.05$& \phs$ 0.00$& \phs$ 0.00$& \phs$ 0.01$& \phs$ 0.00$& \phs$ 0.06$& $-0.08$& \phs$ 0.14$ \\
0.02 & $ 8$& $-0.06$& $-0.11$& $-0.05$& \phs$ 0.15$& \phs$ 0.15$& \phs$ 0.07$& \phs$ 0.03$& $-0.09$& $-0.08$& $-0.06$& \phs$ 0.00$& \phs$ 0.00$& \phs$ 0.00$& \phs$ 0.00$& \phs$ 0.05$& \phs$ 0.05$& \phs$ 0.02$& $-0.06$& \phs$ 0.00$& \phs$ 0.00$& \phs$ 0.01$& \phs$ 0.00$& \phs$ 0.05$& $-0.06$& \phs$ 0.13$ \\
0.02 & $12$& $-0.04$& $-0.11$& $-0.05$& \phs$ 0.16$& \phs$ 0.16$& \phs$ 0.07$& \phs$ 0.03$& $-0.11$& $-0.08$& $-0.06$& \phs$ 0.00$& $-0.01$& $-0.01$& $-0.01$& \phs$ 0.06$& \phs$ 0.05$& \phs$ 0.02$& $-0.07$& \phs$ 0.01$& \phs$ 0.00$& \phs$ 0.01$& \phs$ 0.00$& \phs$ 0.05$& $-0.04$& \phs$ 0.12$ \\
0.02 & $17$& $-0.04$& $-0.11$& $-0.05$& \phs$ 0.17$& \phs$ 0.17$& \phs$ 0.07$& \phs$ 0.02$& $-0.13$& $-0.08$& $-0.06$& $-0.01$& $-0.01$& $-0.01$& $-0.02$& \phs$ 0.06$& \phs$ 0.05$& \phs$ 0.02$& $-0.07$& \phs$ 0.01$& $-0.01$& \phs$ 0.01$& \phs$ 0.00$& \phs$ 0.05$& $-0.01$& \phs$ 0.12$ \\

\cutinhead{Padova Evolutionary Tracks, Empirical SEDs ({\em Difference from Standard Model})}
0.02 & $ 1$& $-0.04$& \phs$ 0.00$& \phs$ 0.03$& \phs$ 0.03$& \phs$ 0.04$& \phs$ 0.06$& \phs$ 0.02$& $-0.06$& $-0.10$& $-0.10$& $-0.03$& $-0.04$& $-0.04$& $-0.05$& $-0.02$& $-0.02$& \phs$ 0.00$& \phs$ 0.04$& \phs$ 0.02$& \phs$ 0.01$& $-0.06$& \phs$ 0.00$& \phs$ 0.10$& $-0.51$& \phs$ 0.19$ \\
0.02 & $ 2$& $-0.11$& \phs$ 0.01$& \phs$ 0.06$& \phs$ 0.04$& \phs$ 0.06$& \phs$ 0.10$& \phs$ 0.04$& $-0.07$& $-0.14$& $-0.14$& $-0.01$& $-0.01$& $-0.01$& $-0.03$& \phs$ 0.02$& \phs$ 0.02$& \phs$ 0.08$& \phs$ 0.02$& \phs$ 0.03$& \phs$ 0.01$& $-0.02$& \phs$ 0.00$& \phs$ 0.11$& $-0.42$& \phs$ 0.19$ \\
0.02 & $ 3$& $-0.13$& \phs$ 0.01$& \phs$ 0.07$& \phs$ 0.07$& \phs$ 0.08$& \phs$ 0.12$& \phs$ 0.05$& $-0.09$& $-0.17$& $-0.17$& $-0.02$& $-0.02$& $-0.01$& $-0.03$& \phs$ 0.05$& \phs$ 0.09$& \phs$ 0.16$& \phs$ 0.01$& \phs$ 0.02$& \phs$ 0.00$& $-0.01$& \phs$ 0.00$& \phs$ 0.10$& $-0.30$& \phs$ 0.18$ \\
0.02 & $ 5$& $-0.12$& \phs$ 0.02$& \phs$ 0.07$& \phs$ 0.07$& \phs$ 0.09$& \phs$ 0.13$& \phs$ 0.06$& $-0.09$& $-0.16$& $-0.15$& \phs$ 0.00$& $-0.01$& \phs$ 0.00$& $-0.02$& \phs$ 0.10$& \phs$ 0.16$& \phs$ 0.27$& \phs$ 0.01$& \phs$ 0.04$& \phs$ 0.00$& \phs$ 0.02$& \phs$ 0.00$& \phs$ 0.07$& $-0.20$& \phs$ 0.10$ \\
0.02 & $ 8$& $-0.06$& \phs$ 0.02$& \phs$ 0.06$& \phs$ 0.07$& \phs$ 0.10$& \phs$ 0.12$& \phs$ 0.05$& $-0.09$& $-0.15$& $-0.15$& \phs$ 0.00$& $-0.01$& \phs$ 0.00$& $-0.02$& \phs$ 0.12$& \phs$ 0.18$& \phs$ 0.33$& \phs$ 0.01$& \phs$ 0.05$& \phs$ 0.01$& \phs$ 0.01$& \phs$ 0.00$& \phs$ 0.06$& $-0.18$& \phs$ 0.10$ \\
0.02 & $12$& $-0.14$& $-0.01$& \phs$ 0.05$& \phs$ 0.08$& \phs$ 0.11$& \phs$ 0.16$& \phs$ 0.06$& $-0.11$& $-0.16$& $-0.15$& $-0.01$& $-0.01$& \phs$ 0.00$& $-0.01$& \phs$ 0.13$& \phs$ 0.24$& \phs$ 0.46$& \phs$ 0.02$& \phs$ 0.04$& \phs$ 0.00$& \phs$ 0.01$& \phs$ 0.00$& \phs$ 0.06$& $-0.14$& \phs$ 0.10$ \\
0.02 & $17$& $-0.18$& $-0.01$& \phs$ 0.05$& \phs$ 0.09$& \phs$ 0.12$& \phs$ 0.18$& \phs$ 0.07$& $-0.12$& $-0.16$& $-0.15$& $-0.01$& $-0.01$& \phs$ 0.00$& $-0.01$& \phs$ 0.12$& \phs$ 0.25$& \phs$ 0.50$& \phs$ 0.02$& \phs$ 0.03$& \phs$ 0.00$& \phs$ 0.00$& \phs$ 0.00$& \phs$ 0.05$& $-0.11$& \phs$ 0.08$ \\

\cutinhead{Geneva Evolutionary Tracks, Semi-Empirical SEDs ({\em Difference from Standard Model})}
0.02 & $ 1$& $-0.10$& $-0.15$& $-0.08$& \phs$ 0.16$& \phs$ 0.14$& \phs$ 0.47$& \phs$ 0.41$& \phs$ 0.42$& \phs$ 0.41$& \phs$ 0.41$& \phs$ 0.38$& \phs$ 0.38$& \phs$ 0.35$& \phs$ 0.40$& \phs$ 0.22$& \phs$ 0.22$& \phs$ 0.16$& \phs$ 0.06$& \phs$ 0.07$& \phs$ 0.01$& $-0.01$& \phs$ 0.02$& \phs$ 0.01$& \phs$ 0.40$& \phs$ 0.01$ \\
0.02 & $ 2$& \phs$ 0.20$& \phs$ 0.22$& \phs$ 0.17$& \phs$ 0.03$& \phs$ 0.05$& \phs$ 0.11$& \phs$ 0.08$& \phs$ 0.11$& \phs$ 0.13$& \phs$ 0.15$& \phs$ 0.16$& \phs$ 0.17$& \phs$ 0.16$& \phs$ 0.20$& \phs$ 0.12$& \phs$ 0.14$& \phs$ 0.12$& \phs$ 0.00$& \phs$ 0.02$& \phs$ 0.01$& \phs$ 0.26$& $-0.01$& $-0.18$& \phs$ 1.00$& $-0.47$ \\
0.02 & $ 3$& \phs$ 0.09$& \phs$ 0.11$& \phs$ 0.08$& \phs$ 0.03$& \phs$ 0.04$& \phs$ 0.30$& \phs$ 0.24$& \phs$ 0.29$& \phs$ 0.32$& \phs$ 0.34$& \phs$ 0.34$& \phs$ 0.35$& \phs$ 0.34$& \phs$ 0.40$& \phs$ 0.20$& \phs$ 0.24$& \phs$ 0.20$& \phs$ 0.00$& $-0.06$& $-0.02$& \phs$ 0.26$& $-0.01$& $-0.23$& \phs$ 1.33$& $-0.46$ \\
0.02 & $ 5$& \phs$ 0.09$& \phs$ 0.01$& $-0.10$& $-0.23$& $-0.21$& \phs$ 0.09$& \phs$ 0.02$& \phs$ 0.07$& \phs$ 0.12$& \phs$ 0.15$& \phs$ 0.19$& \phs$ 0.21$& \phs$ 0.21$& \phs$ 0.25$& \phs$ 0.29$& \phs$ 0.35$& \phs$ 0.38$& \phs$ 0.07$& \phs$ 0.12$& \phs$ 0.03$& \phs$ 0.14$& \phs$ 0.00$& $-0.01$& \phs$ 0.74$& $-0.13$ \\
0.02 & $ 8$& $-0.04$& $-0.07$& $-0.15$& $-0.31$& $-0.29$& $-0.10$& $-0.16$& $-0.12$& $-0.07$& $-0.04$& $-0.02$& $-0.01$& \phs$ 0.00$& \phs$ 0.03$& \phs$ 0.04$& \phs$ 0.11$& \phs$ 0.14$& \phs$ 0.09$& \phs$ 0.16$& \phs$ 0.03$& \phs$ 0.06$& \phs$ 0.01$& \phs$ 0.09$& \phs$ 0.29$& \phs$ 0.09$ \\
0.02 & $12$& \phs$ 0.06$& $-0.06$& $-0.13$& $-0.25$& $-0.22$& \phs$ 0.13$& \phs$ 0.04$& \phs$ 0.10$& \phs$ 0.14$& \phs$ 0.19$& \phs$ 0.18$& \phs$ 0.21$& \phs$ 0.22$& \phs$ 0.28$& \phs$ 0.20$& \phs$ 0.33$& \phs$ 0.34$& \phs$ 0.08$& \phs$ 0.08$& \phs$ 0.00$& \phs$ 0.02$& \phs$ 0.01$& \phs$ 0.10$& \phs$ 0.07$& \phs$ 0.10$ \\
0.02 & $17$& \phs$ 0.15$& \phs$ 0.05$& $-0.02$& $-0.08$& $-0.05$& \phs$ 0.47$& \phs$ 0.36$& \phs$ 0.44$& \phs$ 0.46$& \phs$ 0.52$& \phs$ 0.51$& \phs$ 0.55$& \phs$ 0.55$& \phs$ 0.64$& \phs$ 0.49$& \phs$ 0.64$& \phs$ 0.64$& \phs$ 0.06$& $-0.02$& $-0.03$& \phs$ 0.00$& \phs$ 0.01$& \phs$ 0.11$& \phs$ 0.11$& $-0.01$ \\

\cutinhead{Geneva Evolutionary Tracks, Theoretical SEDs ({\em Difference from Standard Model})}
0.02 & $ 1$& $-0.10$& $-0.16$& $-0.07$& \phs$ 0.19$& \phs$ 0.17$& \phs$ 0.49$& \phs$ 0.42$& \phs$ 0.42$& \phs$ 0.41$& \phs$ 0.42$& \phs$ 0.40$& \phs$ 0.39$& \phs$ 0.36$& \phs$ 0.41$& \phs$ 0.22$& \phs$ 0.22$& \phs$ 0.16$& \phs$ 0.06$& \phs$ 0.11$& \phs$ 0.02$& $-0.07$& \phs$ 0.02$& \phs$ 0.05$& \phs$ 0.21$& \phs$ 0.13$ \\
0.02 & $ 2$& \phs$ 0.12$& \phs$ 0.04$& \phs$ 0.07$& \phs$ 0.13$& \phs$ 0.14$& \phs$ 0.23$& \phs$ 0.16$& \phs$ 0.11$& \phs$ 0.07$& \phs$ 0.09$& \phs$ 0.18$& \phs$ 0.20$& \phs$ 0.19$& \phs$ 0.23$& \phs$ 0.14$& \phs$ 0.15$& \phs$ 0.13$& $-0.04$& \phs$ 0.02$& \phs$ 0.02$& \phs$ 0.24$& \phs$ 0.00$& $-0.06$& \phs$ 0.86$& $-0.26$ \\
0.02 & $ 3$& $-0.01$& $-0.05$& $-0.02$& \phs$ 0.15$& \phs$ 0.16$& \phs$ 0.41$& \phs$ 0.31$& \phs$ 0.28$& \phs$ 0.24$& \phs$ 0.28$& \phs$ 0.35$& \phs$ 0.37$& \phs$ 0.36$& \phs$ 0.43$& \phs$ 0.22$& \phs$ 0.26$& \phs$ 0.21$& $-0.05$& $-0.07$& $-0.01$& \phs$ 0.25$& $-0.01$& $-0.12$& \phs$ 1.21$& $-0.25$ \\
0.02 & $ 5$& $-0.05$& $-0.18$& $-0.20$& $-0.15$& $-0.13$& \phs$ 0.17$& \phs$ 0.06$& \phs$ 0.05$& \phs$ 0.06$& \phs$ 0.10$& \phs$ 0.22$& \phs$ 0.24$& \phs$ 0.25$& \phs$ 0.29$& \phs$ 0.34$& \phs$ 0.41$& \phs$ 0.42$& \phs$ 0.00$& \phs$ 0.10$& \phs$ 0.03$& \phs$ 0.15$& \phs$ 0.01$& \phs$ 0.11$& \phs$ 0.65$& \phs$ 0.10$ \\
0.02 & $ 8$& $-0.13$& $-0.23$& $-0.23$& $-0.23$& $-0.21$& $-0.04$& $-0.13$& $-0.15$& $-0.13$& $-0.09$& \phs$ 0.00$& \phs$ 0.02$& \phs$ 0.03$& \phs$ 0.07$& \phs$ 0.11$& \phs$ 0.17$& \phs$ 0.18$& \phs$ 0.01$& \phs$ 0.14$& \phs$ 0.03$& \phs$ 0.08$& \phs$ 0.01$& \phs$ 0.21$& \phs$ 0.23$& \phs$ 0.31$ \\
0.02 & $12$& $-0.02$& $-0.22$& $-0.22$& $-0.19$& $-0.16$& \phs$ 0.19$& \phs$ 0.07$& \phs$ 0.07$& \phs$ 0.09$& \phs$ 0.14$& \phs$ 0.21$& \phs$ 0.25$& \phs$ 0.25$& \phs$ 0.32$& \phs$ 0.28$& \phs$ 0.40$& \phs$ 0.39$& \phs$ 0.00$& \phs$ 0.06$& \phs$ 0.00$& \phs$ 0.05$& \phs$ 0.01$& \phs$ 0.23$& \phs$ 0.05$& \phs$ 0.32$ \\
0.02 & $17$& \phs$ 0.08$& $-0.10$& $-0.11$& $-0.03$& \phs$ 0.01$& \phs$ 0.53$& \phs$ 0.39$& \phs$ 0.41$& \phs$ 0.41$& \phs$ 0.47$& \phs$ 0.54$& \phs$ 0.58$& \phs$ 0.58$& \phs$ 0.68$& \phs$ 0.57$& \phs$ 0.72$& \phs$ 0.70$& $-0.03$& $-0.03$& $-0.03$& \phs$ 0.04$& \phs$ 0.01$& \phs$ 0.24$& \phs$ 0.12$& \phs$ 0.21$ \\

\cutinhead{Geneva Evolutionary Tracks, Empirical SEDs ({\em Difference from Standard Model})}
0.02 & $ 1$& $-0.15$& $-0.14$& $-0.03$& \phs$ 0.18$& \phs$ 0.18$& \phs$ 0.46$& \phs$ 0.39$& \phs$ 0.38$& \phs$ 0.39$& \phs$ 0.39$& \phs$ 0.41$& \phs$ 0.40$& \phs$ 0.37$& \phs$ 0.41$& \phs$ 0.22$& \phs$ 0.21$& \phs$ 0.21$& \phs$ 0.07$& \phs$ 0.15$& \phs$ 0.05$& $-0.08$& \phs$ 0.02$& \phs$ 0.13$& $-0.19$& \phs$ 0.24$ \\
0.02 & $ 2$& \phs$ 0.14$& \phs$ 0.24$& \phs$ 0.23$& \phs$ 0.06$& \phs$ 0.08$& \phs$ 0.17$& \phs$ 0.11$& \phs$ 0.06$& \phs$ 0.02$& \phs$ 0.03$& \phs$ 0.16$& \phs$ 0.17$& \phs$ 0.16$& \phs$ 0.18$& \phs$ 0.14$& \phs$ 0.15$& \phs$ 0.19$& \phs$ 0.02$& \phs$ 0.07$& \phs$ 0.03$& \phs$ 0.27$& $-0.01$& $-0.04$& \phs$ 0.73$& $-0.31$ \\
0.02 & $ 3$& $-0.02$& \phs$ 0.11$& \phs$ 0.11$& \phs$ 0.03$& \phs$ 0.06$& \phs$ 0.37$& \phs$ 0.28$& \phs$ 0.25$& \phs$ 0.18$& \phs$ 0.21$& \phs$ 0.34$& \phs$ 0.36$& \phs$ 0.35$& \phs$ 0.40$& \phs$ 0.25$& \phs$ 0.30$& \phs$ 0.31$& \phs$ 0.01$& $-0.02$& $-0.01$& \phs$ 0.26$& $-0.01$& $-0.08$& \phs$ 1.03$& $-0.26$ \\
0.02 & $ 5$& $-0.05$& \phs$ 0.01$& $-0.06$& $-0.20$& $-0.17$& \phs$ 0.14$& \phs$ 0.05$& \phs$ 0.02$& $-0.02$& \phs$ 0.01$& \phs$ 0.19$& \phs$ 0.21$& \phs$ 0.22$& \phs$ 0.24$& \phs$ 0.42$& \phs$ 0.50$& \phs$ 0.62$& \phs$ 0.08$& \phs$ 0.17$& \phs$ 0.03$& \phs$ 0.15$& \phs$ 0.00$& \phs$ 0.12$& \phs$ 0.49$& \phs$ 0.07$ \\
0.02 & $ 8$& $-0.16$& $-0.08$& $-0.12$& $-0.28$& $-0.25$& $-0.04$& $-0.13$& $-0.17$& $-0.22$& $-0.19$& $-0.02$& \phs$ 0.00$& \phs$ 0.01$& \phs$ 0.03$& \phs$ 0.21$& \phs$ 0.30$& \phs$ 0.46$& \phs$ 0.10$& \phs$ 0.21$& \phs$ 0.04$& \phs$ 0.07$& \phs$ 0.01$& \phs$ 0.21$& \phs$ 0.07$& \phs$ 0.27$ \\
0.02 & $12$& $-0.13$& $-0.07$& $-0.09$& $-0.20$& $-0.16$& \phs$ 0.18$& \phs$ 0.07$& \phs$ 0.04$& $-0.02$& \phs$ 0.02$& \phs$ 0.18$& \phs$ 0.20$& \phs$ 0.21$& \phs$ 0.26$& \phs$ 0.37$& \phs$ 0.52$& \phs$ 0.70$& \phs$ 0.07$& \phs$ 0.11$& \phs$ 0.01$& \phs$ 0.04$& \phs$ 0.01$& \phs$ 0.19$& $-0.02$& \phs$ 0.22$ \\
0.02 & $17$& $-0.04$& \phs$ 0.04$& \phs$ 0.02$& $-0.04$& \phs$ 0.01$& \phs$ 0.52$& \phs$ 0.38$& \phs$ 0.36$& \phs$ 0.31$& \phs$ 0.36$& \phs$ 0.51$& \phs$ 0.53$& \phs$ 0.54$& \phs$ 0.61$& \phs$ 0.64$& \phs$ 0.81$& \phs$ 0.97$& \phs$ 0.06$& \phs$ 0.01$& $-0.02$& \phs$ 0.02$& \phs$ 0.01$& \phs$ 0.19$& \phs$ 0.05$& \phs$ 0.09$ \\

\enddata

\tablecomments{The top set of results gives the model output for our
standard solar-metallicity model (Padova evolutionary tracks with
semi-empirical SEDs). For all other models, the differences from our
standard models are tabulated, \eg, \{Geneva model, Pickles SED\} --
\{Padova model, semi-empirical SEDs\}.}

\end{deluxetable}

\end{document}